\documentclass[usegraphicx,usenatbib,useAMS]{mn2e}
\newcommand{\be}{\begin{equation}}
\newcommand{\ba}{\begin{eqnarray}}
\newcommand{\ee}{\end{equation}}
\newcommand{\ea}{\end{eqnarray}}

\newcommand{\Del}{\Delta}

\newcommand{\pa}{\partial}

\def\simless{\mathbin{\lower 3pt\hbox
   {$\rlap{\raise 5pt\hbox{$\char'074$}}\mathchar"7218$}}}   
\def\simgreat{\mathbin{\lower 3pt\hbox
   {$\rlap{\raise 5pt\hbox{$\char'076$}}\mathchar"7218$}}}   

\title[Photoevaporation of Cosmological Minihalos during Reionization]
{Photoevaporation of Cosmological Minihalos during Reionization}
\author[P.~R.~Shapiro et al.]{Paul~R.~Shapiro$^1$, Ilian~T.~Iliev,$^2$ 
and Alejandro~C.~Raga$^3$\\
$^1$ Department of Astronomy, University of Texas, Austin, TX 78712-1083\\
$^2$ Osservatorio Astrofisico di Arcetri, Largo Enrico Fermi 5,
50125 Firenze, Italy\\
$^3$ Instituto de Ciencias Nucleares,
	Universidad Nacional Autonoma de M\'exico (UNAM),
	Apdo. Postal 70-543, 04510 M\'exico, 
\\ 
D. F., M\'exico}

\begin{document}

\maketitle

\label{firstpage}

\begin{abstract}
Energy released by a small fraction of the baryons in the universe, which
condensed out while the IGM was cold, dark, and neutral, reheated and
reionized it by redshift 6, exposing other baryons already condensed into
dwarf-galaxy minihalos to the glare of ionizing radiation.  We present
the first gas dynamical simulations of the photoevaporation of cosmological
minihalos overtaken by the ionization fronts which swept through the IGM
during the reionization epoch in the currently-favored $\Lambda$CDM universe,
including the effects of radiative transfer.  These simulations demonstrate
the phenomenon of I-front trapping inside minihalos, in which the weak, R-type
fronts which traveled supersonically across the IGM decelerated when they
encountered the dense, neutral gas inside minihalos, and were thereby
transformed into D-type I-fronts, preceded by shock waves.  For a  minihalo
with virial temperature below $10^4$K, the I-front gradually burned its
way through the minihalo which trapped it, removing all of its baryonic gas 
by causing a supersonic, evaporative wind to blow backwards into the IGM,
away from the exposed layers
of minihalo gas just behind the advancing I-front.  We describe this
process in detail, along with some of its observable consequences, for the
illustrative case of a minihalo of total mass $10^7 M_\odot$, exposed
to a distant source of ionizing radiation with either a stellar or
quasar-like spectrum, after it was overtaken at redshift $z = 9$ by the
weak, R-type I-front which ionized the IGM surrounding the source.
For a source at $z=9$ which emits $10^{56}$ ionizing photons 
per second at 1 Mpc (or, equivalently, $10^{52}$ ionizing photons per 
second at 10 kpc), the photoevaporation of this minihalo takes about 
100--150 Myrs, depending on the source spectrum, ending at about 
$z = $7--7.5.

Such hitherto neglected feedback effects were widespread during the
reionization epoch.  N-body simulations and analytical estimates of
halo formation in the $\Lambda$CDM model suggest that sub-kpc minihalos
such as these, with virial temperatures below $10^4$K, were so common
as to cover the sky around larger-mass source halos and possibly 
dominate the absorption of ionizing photons during reionization.
This means that previous estimates of the number of ionizing photons
per H atom required to complete reionization which neglected this effect
may be too low.  Regardless of their effect on the progress of
reionization, however, the minihalos were so abundant that random lines of
sight thru the high-$z$ universe should encounter many of them, which suggests
that it may be possible to observe the processes described here
in the absorption spectra of distant sources.  
\end{abstract}


\section{Introduction}
\label{intro}
\subsection{The reionization epoch}
Observations of the intergalactic medium (IGM) indicate that the universe was
reionized by $z\approx 6$ but that reionization began much before this.
The absence of a Gunn-Peterson (GP) trough in the spectra of high-redshift 
quasars at observed wavelengths $\lambda_\alpha(1+z)$ due to Ly$\alpha$ 
resonance scattering by H atoms in the IGM at $z<6$ indicates that the IGM
was very highly ionized at all $z<6$ \citep{F_etal00}. Since the universe 
recombined at $z\sim1000$, something must have occurred to reionize it
by $z\sim6$.  

The detection of GP troughs in the spectra of SDSS quasars at $z\geq6$ places a 
lower limit on the mean H~I density in the IGM at $z=6$ 
large enough to suggest that reionization might only just have 
ended at $z\approx6$ \citep{Betal01,DCSM01,F_etal03}.
Limits on the GP effect at $z<6$, that is,
imply a hydrogen neutral fraction significantly {\it smaller} 
than the value at $z=6$, which can be explained by a photoionized IGM only if
the ionizing background radiation increased dramatically in a short interval of 
cosmic time at this epoch \citep{F_etal02,CMcD02}.
Since reionization is believed to have proceeded inhomogeneously,
with isolated patches of ionized gas which eventually grew to fill 
all of space,  
the epoch of overlap would have been accompanied by the required rapid increase 
in the ionizing radiation background, as the absorption mean free path for this 
radiation suddenly increased.
For this reason, the GP detections at $z=6$ and limits
at $z<6$ are thought to indicate that the epoch of
overlap -- the end of reionization --
occurred at a redshift close to $z=6$. 

The recent discovery by the WMAP satellite of 
polarization of the CMB which fluctuates on an angular scale of $10^\circ$, on the other 
hand, amounts to a detection of a foreground electron scattering optical depth 
through the IGM of $\tau_{es}\simeq0.17\pm0.04$ \citep{Ketal03,Spetal03}. This suggests 
that the IGM was either mostly ionized by a redshift of $z\simeq17\pm5$, or else was at 
least partially ionized starting at even higher $z$ \citep{Spetal03}. 
The GP detection at $z=6$ described above argues against the simplest interpretation
of the WMAP polarization result in which the universe was fully reionized by redshift
$z\simgreat15$ and stayed ionized thereafter. Another argument against that simple 
interpretation is based upon the
fact that, after the IGM was reheated by its reionization, it would gradually have 
cooled by adiabatic expansion even though it continued to be photoionized by background 
radiation. If reionization finished as early as implied by the simplest interpretation
of the WMAP results, this would have lowered the temperature of the IGM at redshifts $z\leq4$ 
below the level deduced from the Ly $\alpha$ forest \citep{HH03}.

In short, reionization is currently believed to have begun by
redshift $z\ga15$ to accomodate the WMAP results
but to extend until $z=6$ to accomodate observations of the 
GP effect and of the Ly $\alpha$ 
forest at $z\leq6$. When and how this 
reionization occurred and what its consequences were for the subsequent 
epochs is one of the major unsolved problems of cosmology.

\subsection{Ionization fronts in the IGM}

The neutral, opaque IGM out of which the first bound objects 
condensed was dramatically 
reheated and reionized at some time between a redshift $z\approx30$ and 
$ z\approx6$ by
the radiation released by some of these objects. 
When the first sources turned on, they 
ionized their surroundings by propagating weak, R-type ionization 
fronts which moved 
outward supersonically with respect to both the neutral gas ahead 
of and the ionized gas 
behind the front, racing ahead of the hydrodynamical response of 
the IGM, as first 
described by \citet{S86} and \citet{SG87}. These authors solved 
the problem of the time-varying
radius of a spherical I-front which surrounds an isolated source 
in a cosmologically-expanding 
IGM analytically, generalizing the I-front jump condition and radiative
transfer to cosmological conditions. They applied these solutions to determine 
when and how fast the I-fronts 
surrounding isolated sources would grow to overlap and, thereby, 
complete the reionization
of the universe\footnote{This provided the first indication that 
quasars alone did not reionize 
the universe; the observed quasar luminosity 
function extrapolated to higher redshift 
was not sufficient to do so in time to satisfy the GP constraint, 
suggesting that some other 
source was responsible, such as starlight.}. 
The effect of density inhomogeneity on the rate of the I-front propagation
was described by a mean ``clumping factor'' $C\equiv\langle n^2\rangle/\langle n\rangle^2>1$, 
which slowed the I-fronts by increasing 
the average recombination rate per atom inside clumps. This 
suffices to describe the
rate of I-front propagation as long as the clumps are either not 
self-shielding or, if
so, only absorb a small fraction of the ionizing photons emitted 
by the central source.

Numerical radiative transfer methods are currently under development 
to solve this problem 
in 3-D for the inhomogeneous density distribution which arises as 
cosmic structure forms,
so far primarily limited to the transfer of ionizing radiation 
through pre-computed density 
fields generated by cosmological simulations which do not include 
radiative transfer, so 
there is no back-reaction of the radiative transfer 
calculation on the density field or any aspect of the gas dynamics 
\citep{RS99,ANM99,CFMR01,NUS01,Soetal01,C02,HN02,CSW03}. Approximate 
approaches which 
incorporate some of the important effects of radiative transport 
during reionization
within the context of cosmological gas dynamics simulation have also 
been developed 
\citep{G00,GA01,RGS02a}. These recent attempts to model 
inhomogeneous reionization 
numerically are generally handicapped by their limited spatial resolution,
which prevents them from fully
resolving the most important (sub-kpc-sized) density inhomogeneities. 
The questions of 
what dynamical effect the I-front had on the density inhomogeneities 
it encountered and how 
the presence of these inhomogeneities affected the I-fronts and 
reionization, therefore, require further analysis. 

Towards this end, we have developed a radiation-hydrodynamics 
code which incorporates radiative transfer and focused our attention on
properly resolving 
this small-scale structure. 
Here we shall present in some detail our results of the first 
radiation-hydrodynamical simulations of the back-reaction 
of a cosmological I-front on a 
gravitationally-bound density inhomogeneity it encounters -- a 
$10^7M_\odot$ dwarf galaxy
minihalo -- during reionization, along with some general considerations which 
apply to other halos as well. 
A second paper to follow will present the results for a much wider 
range of cases, with different halo masses and I-front encounter epochs, 
to quantify just how
the process described here varies with these properties. 
As it turns out, the photoevaporation
of sub-kpc-sized objects like these may be the dominant process
by which ionizing photons are
absorbed during reionization, so this problem is of critical importance. 
In addition, 
observations of the absorption spectra of high redshift sources 
like those which reionized 
the universe should reveal the presence of these photoevaporative 
flows and provide a useful 
diagnostic of the reionization process.

\subsection{The photoevaporation of dwarf galaxy minihaloes overtaken by cosmological 
ionization fronts}

The effect which small-scale
clumpiness had on reionization depended upon the sizes, densities, and spatial
distribution of the clumps overtaken by the I-fronts during reionization.
For the currently-favored $\Lambda$CDM model $(\Omega_0=1-\lambda_0=0.3$, 
$h=0.7$, $\Omega_bh^2=0.02$, primordial power spectrum index $n_p=1$; COBE-normalized),
which is the model we adopt throughout this paper, the universe  at $z>6$
was already filled with dwarf galaxies capable of 
trapping a piece of the global, intergalactic I-fronts which reionized the
universe and photoevaporating their gaseous baryons back into the IGM. 
Prior to their encounter with these I-fronts,``minihalos''
with $T_{\rm vir}\leq10^4\rm K$ were neutral and optically thick to hydrogen
ionizing radiation, as long as their total mass exceeded the Jeans mass $M_J$ in 
the unperturbed background IGM prior to reionization (see \S~\ref{analyt_sect}), 
as was required to enable baryons to collapse into the halo along with dark matter. Their
``Str\"omgren numbers'' $L_S\equiv2R_{\rm halo}/\ell_S$, the ratio of a halo's 
diameter to its Str\"omgren length $\ell_S$ inside the halo (the length
of a column of gas within which the unshielded arrival rate of ionizing
photons just balances the total recombination rate), were large.
For a uniform gas of H density $n_{\rm H}$, located a distance $r_{\rm Mpc}$
(in Mpc) from a UV source emitting $N_{\rm ph,56}$ ionizing photons (in units
of $10^{56}\rm s^{-1}$), the Str\"omgren length is only 
$\ell_S\approx(100\,{\rm pc})(N_{\rm ph,56}/r_{\rm Mpc}^2)
(n_{\rm H}/0.1\,{\rm cm}^{-3})^{-2}$, so $L_S\gg1$ for most of the range of
halo masses and sources of interest, as we will discuss in \S~\ref{strom_sect}. 
In that case, the intergalactic,
weak, R-type I-front which entered each minihalo during reionization would
have decelerated to about twice the sound speed of the ionized gas before
it could exit the other side, thereby transforming itself into a D-type
front, preceded by a shock. Typically, the side facing the source would
then have expelled a supersonic wind backwards toward the source, which
shocked the surrounding IGM as the minihalo photoevaporated.

The importance of this photoevaporation process has long been recognized in the study 
of interstellar clouds exposed to ionizing starlight 
\citep{OS55,S78,B89,BM90,LL94,BD95,Letal96,GH01}.   
Radiation-hydrodynamical simulations were performed in 2-D of a stellar I-front overtaking 
a clump inside a molecular cloud \citep{KSW82,SWK83}. More recently, 2-D simulations for the 
case of a circumstellar cloud ionized by a single nearby star have also been performed
\citep{MRCLBSN98}. In the cosmological context, however, the importance of this process
has only recently been fully appreciated. 

In proposing the expanding minihalo model to 
explain Lyman $\alpha$ forest (``LF'') quasar absorption lines, \citet{BSS88} discussed
how gas originally confined by the gravity of dark minihalos in the CDM model with
virial temperatures below $10^4$ K would have been 
expelled by pressure forces if photoionization by ionizing background radiation suddenly 
heated all the gas to an isothermal condition at $T\approx10^4$ K, a correct description 
only in the optically-thin limit. The first realistic discussion of the photoevaporation of 
such minihaloes by cosmological I-fronts, including the first radiation-hydrodynamical 
simulations of this process, was by \citet{SRM97,SRM98}.
This work demonstrated the importance of taking proper
account of optical depth and self-shielding simultaneously with hydrodynamics, in order
to derive the correct dynamical flow, ionization structure and the resulting observational
characteristics. \citet{BL99} subsequently confirmed the relative importance of this 
process as a feedback effect on dwarf galaxy minihaloes during reionization, using static models 
of uniformly-illuminated spherical clouds in thermal and ionization equilibrium, without
accounting for gas dynamics. They took H atom self-shielding into account and assumed that all 
gas which is instantaneously heated above the minihalo virial temperature must be evaporated. 
They concluded that 50\%-90\% of the gas in gravitationally-bound objects when  
reionization occurred should have been evaporated. However, by neglecting the time-dependent
gas dynamical nature of this phenomenon, their model failed to capture some essential physics,
leading to incorrect results, as we will show below. 

Further results of radiation-hydrodynamical simulations of a cosmological
minihalo overrun
by a weak, R-type I-front in the surrounding IGM
were summarized in \citet{SR00a,SR00b,SR01} and \citet{S01}. These results were
preliminary versions of the final, more advanced results which we shall present 
in this paper in some detail.  Before we do, it is worth summarizing why  
we believe that the particular problem simulated here is not
only generic to the minihalos
which formed prior to the end of the reionization epoch, but likely to
have affected the global outcome of cosmic reionization, as well.

According to the currently prevalent paradigm of cosmic structure 
formation -- the flat, cold-dark matter universe with cosmological constant 
($\Lambda$CDM) -- cosmological structures formed hierarchically, with small mass
haloes forming first and merging over time to form ever larger halos. The first 
baryonic structures to emerge in this way were the ``minihalos'' which began forming
at $z>20-30$, with masses in the range from about $10^4M_\odot$ to $10^8M_\odot$, where 
the minimum is set by the Jeans mass in the cold and neutral IGM prior to its reionization.
The term ``minihalo'' used here specifically refers to halos at any epoch with
virial temperatures $T_{vir}\leq10^4$ K. In such halos, the thermal collisional ionization rate 
alone was not sufficient to ionize the H and He atoms significantly, so the minihalo gas was almost 
entirely neutral, in the absence of ionizing radiation. Moreover, collisional line excitation 
of H atoms is exponentially suppressed below $10^4$ K and, thus, a purely atomic gas of 
H and He could not radiatively cool effectively below that temperature. Such minihalo gas was able 
to cool below $10^4$ K only if $\rm H_2$ molecules formed in sufficient abundance to cool it by 
rotational-vibrational line excitation, to temperatures as low as $T\simless100$ K. This $\rm H_2$ 
formation, catalyzed by the tiny residual ionized fraction, was essential if the minihalo gas
was to cool and, thereby, compress relative to the dark matter, enough to become self-gravitating
and form stars. Simulations suggest that minihalos of mass $M\simgreat10^6M_\odot$ did just 
that, thereby forming the first metal-free (``Pop. III'') stars of mass
$M_*\simgreat100M_\odot$ at $z\simless30$ \citep{BCL02,ABN02}, perhaps leading to the first 
miniquasars, as well.    

While this process of star formation in minihalos is important as the 
origin of the first 
sources of ionizing radiation in the universe, minihalos are not likely 
to have been the dominant 
source of cosmic reionization. The $\rm H_2$ formation required for 
minihalos to form stars would have
been suppressed easily by photodissociation in the Lyman-Werner bands 
by the background of UV 
radiation created by the very first minihalos which formed stars, when 
the ionizing 
radiation background was still much too low to cause reionization 
\citep{HAR00}. If so,
most minihalos must not have formed stars and must, instead, have 
remained ``sterile'', 
dark-matter-dominated halos filled with stable, neutral atomic gas 
at the halo virial temperature 
$T_{\rm vir}\leq10^4$ K. In that case, when some other source of 
radiation ionized the universe
and drove an I-front into the minihalo, which photoheated the gas 
to $T>10^4$ K, minihalo 
gravity would have been unable to confine the ionized gas and, 
eventually, all of it was 
expelled. A caveat 
to this description above of the negative feedback on the $\rm H_2$ 
in minihalos due to the 
first UV sources is the possibility that the presence of a weak X-ray 
background might have 
counterbalanced this $\rm H_2$ photodissociation by enhancing 
the $\rm H_2$ formation rate in places where 
X-ray ionization increased the small ionized fraction inside the 
halos \citep{HAR00}.\footnote{The net effect of these early X-rays on
the $\rm H_2$ in minihalos may have beem more negative than positive,
however, if they raised the entropy floor of the IGM from which the
minihalos condensed \citep{OH03}. Another study suggests that the near
environment of early-star forming halos might also have experienced some
{\it positive} feedback on the $\rm H_2$, which partially offset
the negative feedback due to $\rm H_2$ dissociation by UV sources,
although not necessarily {\it inside} the pre-existing minihalos
(e.g. \citealt{RGS02b}); this study did not include an X-ray background.}
Apart from 
such caveats, however, the primary source of reionizing photons 
is believed to have been 
the halos with $T_{\rm vir}\geq10^4$ K. For such halos, radiative 
cooling by H atom line 
excitation alone was efficient enough to compress the gas to enable 
self-gravity and 
self-shielding, leading to star formation. For the purposes of this 
paper, we shall 
henceforth accept the premise that the minihalos were predominantly 
``sterile'' targets of source halos with $T_{\rm vir}\geq10^4$ K, of 
higher mass.

In that case, it has been shown, source halos would generally have found their
skies covered by the minihalos between them and neighboring source halos,
until the ionization fronts created by those sources overtook the minihalos
and photoevaporated them, possibly consuming a significant fraction of the
ionizing photons released by the sources in the process
(\citealt{HAM01}, \citealt{S01}, \citealt{BL02} and \citealt{Setal03}). 
According to
these authors, the phenomenon of minihalo photoevaporation during reionization
identified by \cite{SRM97,SRM98} 
not only was quite common and had an important negative feedback effect
on the minihalos, it also had an important effect on reionization, itself,
by screening the ionizing sources and increasing the number of ionizing
photons required per baryon to complete reionization.   
While these results demonstrate the importance of minihalos to the cosmic
reionization story, we will show that a 
correct quantitative estimate of 
these effects requires detailed gas dynamics and radiative transfer simulations.
In addition, as pointed out by 
\cite{S01}, there is a very large number of minihalos per unit redshift
along an arbitrary line-of-sight (LOS) through the universe at $z\geq6$, so the
process of 
their photoevaporation is potentially observable in the absorption
spectra of very high-$z$ 
sources.

This paper is organized as follows. In \S~\ref{analyt_sect}, we
discuss our minihalo model and 
the statistics of minihalos in the $\Lambda$CDM universe, as well
as the analytical expectations 
for how the photoevaporation of cosmological minihalos proceeds and
what its outcome is.
In \S~\ref{calc_sect}, we outline the basic equations used to model
the gas dynamics of 
photoevaporation, the microphysical processes considered and the
initial conditions for our simulations. In \S~\ref{num_methods_sect},
we discuss our 
numerical method and tests, and the parameters of our simulations. 
In \S~\ref{results_sect},   
we describe the photoevaporation process in detail for the illustrative 
case of a minihalo of 
$10^7M_\odot$ at $z=9$ exposed to  ionizing radiation from one of three
types of sources:
(1) starlight with a $50,000$ K black-body spectrum (hereafter,
referred to as ``BB 5e4''), 
representative of massive Pop. II stars, (2) starlight with a $10^5$ K black-body spectrum
(hereafter, ``BB 1e5''), as expected for massive Pop. III stars, 
and (3) QSO-like, with a power-law 
spectrum $F_\nu\propto\nu^{-1.8}$ ($\nu>\nu_H$ ) (hereafter, ``QSO'').
We show our results for the structure of the global I-front 
propagating in the IGM during its weak, R-type phase (\S~\ref{I-front_IGM_sect}), the encounter between this global I-front and
a minihalo along its path and the penetration
of the minihalo by the front in its weak, R-type phase (\S~5.2), 
the 
trapping of the I-front inside the minihalo and its transition to 
D-type (\S~5.3),
the structure of the photoevaporative flow 
during the D-type phase (\S~\ref{flow_sect}), 
the evolution of temperature structure (\S~5.5) and I-front
location and speed (\S~5.6),
the evaporation times
(\S~\ref{evap_time_sect}), the consumption of ionizing photons (\S~\ref{phot_consump_sect}),
and some observational diagnostics (\S~\ref{observ_sect}). Finally, in \S~\ref{summary_sect}, we 
give our summary and conclusions.

\section{Analytical considerations}
\label{analyt_sect}
\subsection{Cosmological minihalos during reionization in a $\Lambda$CDM Universe}
\subsubsection{Minihalo model}
Our initial minihalo model consists of a virialized region in hydrostatic equilibrium 
embedded in an exterior region of cosmological infall. For the former, we have previously 
developed an analytical model for the postcollapse equilibrium structure of virialized 
objects that condense out of a cosmological background universe, either matter-dominated or
flat with a cosmological constant \citep{SIR99,IS01}. This Truncated Isothermal Sphere, or 
TIS, model assumes that cosmological halos form from the collapse and virialization of 
``top-hat'' density perturbations and are spherical, isotropic, and isothermal. This leads 
to a unique, nonsingular TIS, a particular solution of the Lane-Emden equation (suitably 
modified when $\Lambda\neq0$). The size $r_t$ and velocity dispersion $\sigma_V$ are unique 
functions of the mass $M$ and formation redshift $z_{\rm coll}$ of the object for a given 
background universe. The TIS profile flattens to a constant central value, $\rho_0$, which
is roughly proportional to the critical density of the universe at the epoch of collapse,
$\rho_c(z_{\rm coll})$,
with a small core radius $r_0\approx r_t/30$
[where $\sigma_V^2=4\pi G\rho_0 r_0^2$ and 
$r_0\equiv r_{\rm King}/3$ for the ``King radius'' $r_{\rm King}$, defined by \citet{BT87}, 
p. 228]. While the TIS model does not produce the central cusp at very small radii in the dark 
matter density profile of halos predicted by numerical CDM N-body simulations, it does 
reproduce many of the average properties of these halos remarkably well, suggesting that
it is a useful approximation for halos which result from more realistic initial conditions
\citep{SIR99,IS01,IS02}. In particular, the TIS mass profile agrees 
well with the fit by NFW to numerical simulations (i.e. fractional deviation $\simless20\%$)
at all radii outside of a few TIS core radii (i.e. outside a King radius or so). Our 
application here of the TIS model is not sensitive to the small difference between the TIS 
and NFW dark-matter density profiles at small radii. In both cases, the baryonic profile is
nonsingular.

The TIS halo is uniquely specified by the central density $\rho_0$ 
and core radius $r_0$. Since 
the central density is proportional to $\rho_c(z_{\rm coll})$, it 
is also possible to specify
the profile by the two parameters, total mass and collapse redshift, 
$(M,z_{\rm coll})$, 
which is equivalent to specifying the pair $(r_0,\rho_0)$. This makes 
the TIS model extremely
useful in combination with other analytical methods which determine
the typical epochs of
collapse for halos of different masses, such as the well-known 
Press-Schechter approximation.
We note that it is customary elsewhere to define the total mass 
of profiles used to model CDM halos as
$M_{\rm 200}$, the mass inside a sphere of radius $r_{\rm 200}$ 
with a mean density which 
is $200 \rho_c(z)$ at some redshift $z$. For the sake of comparison 
with the TIS profile, if 
we were to fix $M_{\rm 200}$ and  $r_{\rm 200}$ for halos of 
different profiles, which amounts 
to fixing $M$ and $z_{\rm coll}$ for the TIS halo, then 
$M=1.167M_{\rm 200}=772.6\rho_0r_0^3$,
$\rho_0=18,000\rho_c(z_{\rm coll})$ and $r_{\rm 200}=24.2r_0$ 
if $z=z_{\rm coll}$.

As applied here to model minihalos,  
the TIS is embedded in a self-similar, spherical, cosmological 
infall, using the solution for the latter in an Einstein-de Sitter universe 
derived by \citet{B85}, generalized by us to apply it to  
a low-density background universe at the early times considered here 
(Iliev \& Shapiro 2001, Appendix A). 
As discussed in detail in \citet{SIR99} and \citet{IS01}, the truncation 
radius of the TIS solution coincides almost exactly with the location of 
the shock in the 
self-similar infall solution, allowing a natural match of the two models. The 
procedure which we used to accomplish that is discussed in detail in the Appendix.

For the calculations in this paper, our fiducial case is a minihalo of mass 
$M=10^7M_\odot$ which collapses at 
$z=9$. According to the TIS model, this minihalo has a virial 
radius $r_t=0.76$ kpc, a virial temperature $T_{\rm vir}=4000$ K, and 
a DM velocity 
dispersion $\sigma_V=5.2\rm\,km\, s^{-1}$. In units of these fiducial values,
we define $M_7\equiv M/(10^7 M_\odot)$ and $(1+z)_{10}\equiv (1+z)/10$. We 
shall express temperature in terms of $T_4=T/(10^4\,\rm K)$. 

\subsubsection{Statistical properties of minihalos}

This TIS model for the internal structure of an individual minihalo of a given mass 
$M$ which forms at a given collapse epoch $z_{\rm coll}$ can be combined with the 
Press-Schechter (PS) approximation to determine the statistical expectations for the
properties of halos of different masses at each epoch which result from the Gaussian-random 
initial fluctuations of the $\Lambda$CDM universe.

\begin{figure}
\includegraphics[width=3.3in]{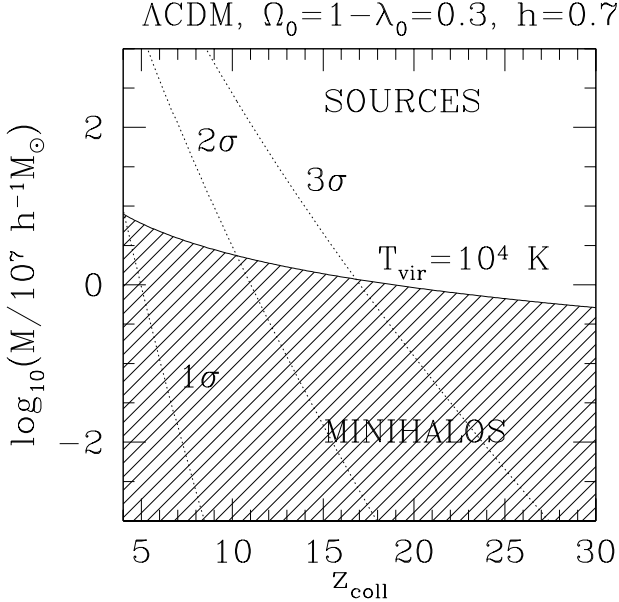}
\caption{How common are minihalos at high redshift? 
Plot of minihalo mass vs. redshift. Shaded region represents minihalos: halos 
with virial temperature $T_{\rm vir}<10^4$ K. Dotted curves represent $\nu-\sigma$ 
fluctuations with $\nu=1,$ 2 and 3, predicted by the PS approximation, as labeled.}
\label{M_zcoll_T}
\end{figure}

In Figure~\ref{M_zcoll_T}, we plot the curves in the halo $M-z_{\rm coll}$ 
plane which correspond 
to halos which collapse from $\nu$-$\sigma$ fluctuations for $\nu=1,2,$ 
and 3. Here  
$\nu\equiv{\delta_{\rm crit}}/\sigma(M)$, $\sigma(M)$ is the rms 
fluctuation in the dark 
matter density field, filtered on mass scale $M$, as predicted by 
linear theory, and 
$\delta_{\rm crit}$ is the overdensity of a top-hat perturbation 
given by extrapolating 
the linear solution to the time of infinite collapse in the exact 
nonlinear solution.
The value $\nu=1$ corresponds to the most common halos, while high values of 
$\nu$ correspond to rare density peaks.  

The minihalos span a mass range from $M_{\rm min}$ to 
$M_{\rm max}$ which varies with redshift, with $M_{\rm min}$ close to 
the Jeans mass of the uncollapsed IGM prior to reionization,
\ba
M_J\!\!\!\!&=&\!\!\!\!
5.7\times10^3M_\odot\left(\frac{\Omega_0h^2}{0.15}\right)^{-1/2}
\left(\frac{\Omega_bh^2}{0.02}\right)^{-3/5}\nonumber\\
&&\times\left(1+z\right)_{10}^{3/2},
\ea
while 
\be
M_{\rm max}=3.95\times10^7M_\odot 
\left(\frac{\Omega_0h^2}{0.15}\right)^{-1/2}\left(1+z\right)_{10}^{-3/2}
\ee
is the mass for which $T_{\rm vir}=10^4$ K according to the TIS model 
(Iliev \& Shapiro 2001). 
This maximum minihalo 
mass at each epoch (i.e. the
curve of constant $T_{\rm vir} =10^4$ K) is shown in Figure~\ref{M_zcoll_T}.
Figure~\ref{M_zcoll_T} demonstrates 
that, below redshift $z\sim10$, all minihalos result from
fluctuations with $\nu<2$, so they are quite typical and fairly common. 
Integrating over the entire mass function of minihalos from $M_{\rm min}$ to
$M_{\rm max}$, we find that
the total collapsed mass fraction in minihalos is (36\%,28\%,10\%,3\%) at
$z=(6,9,15,20)$, respectively.

\subsection{Ionizing flux}
\label{ion_flux_sect}
We shall assume that each minihalo is photoevaporated by an
isotropic point source located at a proper
distance $r$ from the minihalo centre, which emits $N_{\rm ph}$ H ionizing
photons per second with luminosity $L_\nu (\rm erg\,s^{-1}Hz^{-1})$, so that
\be
N_{\rm ph}=\int^\infty_{\nu_H}d\nu\frac{L_\nu}{h\nu}.
\ee
The unattenuated flux at the location of the minihalo centre is just $F_\nu=L_\nu/(4\pi r^2)$,
since $r\ll c/H(z)$, the horizon size during minihalo photoevaporation. We shall, henceforth,
parametrize the flux to which a given halo is exposed in terms of the dimensionless 
frequency-integrated photon flux, $F_0\equiv N_{\rm ph,56}/r^2_{\rm Mpc}$. For example, 
a source which emits $10^{56}$ (or $10^{52}$) ionizing photons per 
second at a distance of 1 Mpc (or 10 kpc) corresponds to $F_0=1$, or, in physical 
units, a flux of $8.356\times10^5\rm cm^{-2}s^{-1}$. What values of $F_0$ are relevant to 
the encounter between a minihalo and an I-front during cosmic reionization?

To get a sense of how large this fiducial flux is in terms of an 
equivalent isotropic radiation
background, we can estimate the mean ionizing photon intensity
$J_\gamma(\rm cm^{-2}s^{-1}ster^{-1})$ required to irradiate an atom 
on the minihalo surface 
with the same flux, by equating this flux to $2\pi J_{\gamma}$. 
For a QSO-like ionizing background with a power-law
energy spectrum, $J_\nu\propto\nu^{-\alpha}$, $\alpha>1$, 
for example, the integrated photon 
intensity is $J_\gamma=1.509\times10^5J_{-21}(1+\alpha)^{-1}$, 
where $J_{-21}$ is the mean 
intensity at the H Lyman limit in units of $10^{-21}\rm erg\, cm^{-2}s^{-1}Hz^{-1}ster^{-1}$.
The equivalent dimensionless flux parameter is then given by 
\be
F_{0,\rm equ}=1.13(1+\alpha)^{-1}J_{-21}, 
\ee
so $F_0=1$ is roughly equivalent to $J_{-21}\sim1$.

We can estimate the appropriate range to consider for $F_0$ as follows.
Suppose the luminosity of a dwarf galaxy is produced by a  
super-massive star cluster (SSC). \cite{TM01} estimate the ionizing
luminosity of such an SSC, using the observed stellar mass function in the R136 cluster 
in 30 Doradus as input to the STARBURST99 code of \citet{Leietal99},
finding (for $0.1\leq M_*/M_\odot\leq100$):
\be
N_{\rm ph,52}=3.94M_{*,6},
\label{ssc_lum}
\ee 
where $M_*$ is the total mass in stars, $M_{p}\equiv M/(10^pM_\odot)$ and 
$N_{\rm ph,q}\equiv N_{\rm ph}/(10^q \,\rm s^{-1})$ is the flux of ionizing 
photons. 

Alternatively, \cite{MR00} have estimated that, for a standard Salpeter IMF, approximately 
$3000-4000$ ionizing photons are emitted for every stellar baryon. Hence, if 
the galaxy gas mass turned into stars is $M_*$, the total number
of ionizing photons emitted during the lifetime of those stars is 
\be
N_{\rm tot}=\int N_{\rm ph}dt\approx10^{66.6}M_{*,6}.
\ee
Assuming an average lifetime of $\tau_{\rm lifetime}\approx10^7$ yrs, we obtain
\be
N_{\rm ph,52}\approx 1.6 M_{*,6},
\ee
similar to the result in equation~(\ref{ssc_lum}). On the other hand, 
\cite{BKL01} estimate that the number of photons produced per baryon by 
Pop. III stars can be about 10-20 times higher than this. 

If we let $f_b=\Omega_b/\Omega_0=0.136$ be the cosmic mean baryonic mass fraction
and $f_*$ be the fraction of these baryons which are converted into stars in
the dwarf galaxy halo, then  
$M_*=f_*f_bM_{\rm halo}$. 
\citet{TM01} find $f_*=0.5$ for their SSC.
 If the target minihalo mass $M_{\rm target}\ll M_{\rm halo}$,
the mean distance between source and halo is $r\approx d_{\rm sep}/2$, while if 
$M_{\rm target}\approx M_{\rm halo}$, it is $r=d_{\rm sep}$. For source halos of 
mass $10^8M_\odot$ at $z=9$, $d_{\rm sep}\approx 50$ kpc \citep{S01}, so
$r_{\rm Mpc}=(0.5-1)\times d_{\rm sep,Mpc}\approx 0.025-0.05$. The characteristic 
dimensionless flux $F_0$ for such sources at this epoch is, therefore, given by 
\be
F_0=\frac{N_{\rm ph,56}}{r_{\rm Mpc}^2}
	=(1-4)\times3.94\times10^{-2}
		\frac{f_*f_bM_{\rm halo,8}}{d_{\rm sep,Mpc}^2}\approx 1-4,
\ee
if we adopt the luminosity per gas mass  estimated  by \cite{TM01}.
For halos in the dwarf galaxy mass range, the PS approximation for the halo mass 
function in $\Lambda$CDM yields $dn_{\rm halo}/dM\propto M^{-2}$ \citep{S01}, so the 
mean separation of dwarf galaxy sources of mass M is 
$d_{\rm sep}=(Mdn_{\rm halo}/dM)^{-1/3}\propto M^{1/3}$. If the mass-to-light ratio $M/L$ 
is independent of halo mass, then the flux incident on a minihalo due to one of 
these halos scales only weakly with source halo mass, $F_0\propto M^{1/3}$, 
up to the galaxy mass 
scale above which the PS mass function cuts off exponentially, at which point $F_0$ drops 
more rapidly with increasing mass. 

Finally, we can get a simple, direct estimate of $F_0$ based on the mean flux required to 
reionize the universe, as follows. Suppose the universe was filled with 
sources that supplied $\bar{\xi}$ ionizing photons per H atom steadily over the time 
between the source turn-on epoch $z_{\rm on}$ and the epoch of reionization 
overlap at $z_{\rm ov}$. Assuming that all photons remained energetic enough 
to ionize H atoms despite their continuous redshifting, and neglecting the
absorption of these photons during the extended reionization epoch, the mean 
comoving number density of ionizing photons at any redshift in the range 
$z_{\rm on}\leq z \leq z_{\rm ov}$ would be just 
\be
\bar{n}_\gamma=\bar{\xi}{\bar{n}_{\rm H}}\frac{t(z)-t_{\rm on}}{t_{\rm ov}-t_{\rm on}},
\ee
where $t$ is the age of the universe at each redshift, as labelled, and $\bar{n}_{\rm H}$
here refers to the mean IGM H density. The mean photon intensity $J_\gamma$ of this 
isotropic background of ionizing radiation would then be given by 
\be
J_\gamma=\frac{c}{4\pi}\bar{n}_\gamma
   =\frac{\bar{\xi}{\bar{n}_{\rm H}}c}{4\pi}\frac{t(z)-t_{\rm on}}{t_{\rm ov}-t_{\rm on}}
\ee
This corresponds to an effective photon flux $F_{\rm eff}$ given by 
\be
F_{\rm eff}=2\pi J_\gamma
   =\frac{\bar{\xi}{\bar{n}_{\rm H}}c}{2}\frac{t(z)-t_{\rm on}}{t_{\rm ov}-t_{\rm on}},
\ee
or a dimensionless flux 
\be
F_{0,\rm eff}
   = 3\bar{\xi}\left(1+z\right)_{10}^3\left(\frac{\Omega_bh^2}{0.02}\right) 
	\left[\frac{t(z)-t_{\rm on}}{t_{\rm ov}-t_{\rm on}}\right],
\ee
For $t_{\rm on}\ll t_{\rm ov}$ the quantity in square brackets can be replaced by 
$t(z)/t_{\rm ov}$ to yield an upper limit,
\be
F_{0,\rm eff}\simless3\bar{\xi}\left(1+z\right)_{10}^3\left(\frac{\Omega_bh^2}{0.02}\right)
	\left(\frac{1+z}{1+z_{\rm ov}}\right)^{-3/2}, 
\ee 
or, if $z_{\rm ov}=6$ and $\Omega_bh^2=0.02$,
\be
F_{0,\rm eff}\simless1.8\bar{\xi}\left(1+z\right)_{10}^{3/2}.
\ee
This suggests that $F_0\sim 1$ is a reasonable fiducial choice for the flux to which 
minihalos were exposed during reionization before overlap. In this paper we present 
results for $F_0=1$ only. We defer the study of the effects of varying $F_0$ to the 
companion paper.  

\subsection{Minihalo Str\"omgren numbers and I-front trapping}
\label{strom_sect}
\subsubsection{I-front trapping}

For a spherical H~II region in a uniform interstellar gas, 
the early expansion phase of 
the weak, R-type I-front ends when the I-front decelerates 
to twice the sound speed of the 
ionized gas, at which point it makes the transition from 
R-critical to D-critical, 
preceded by a shock front \citep{S78}. 
This transition occurs when the 
radius of the I-front approaches that of the Str\"omgren sphere. By analogy, we 
expect our intergalactic, weak, R-type I-front to make a similar transition to D-type
after it enters a minihalo when it slows to the R-critical speed as it travels up the 
density gradient inside the minihalo, a distance comparable to the Str\"omgren length
along its path through the halo. In order to ``trap'' the I-front, therefore, the minihalo 
size must exceed this length. We can estimate the ``trapping'' condition as follows.

The Str\"omgren length $\ell_S(r)$ at impact parameter $r$ is given by 
\be
F=\int_0^{\ell_S(r)}d\ell n_e n_H\alpha_H^{(2)}
\label{strom}
\ee
i.e. by balancing the number of recombinations with the number of 
ionizing photons arriving along a given LOS.\footnote{We have adapted
cylindrical coordinates $(r,x)$ in which the $x$-axis is the line between the
source and the minihalo center, the axis of symmetry of this problem. In
writing equation~(\ref{strom}), we have assumed that the distance
between the source and the minihalo is much greater than the size of the
minihalo, so we can approximate the LOS as parallel to the $x$-axis.}
The integration in equation~(\ref{strom})
is done along the LOS, $n_H$ is the H number density, $n_e$
is the electron density,
$F$ is the flux of 
ionizing photons, $\alpha_H^{(2)}$ is the Case B recombination coefficient 
for hydrogen, and we have assumed for simplicity
that only the H in the gas is ionized. 
We then define the ``Str\"omgren number'' for the
halo as $L_S\equiv2r_t/\ell_S(0)$, where $r_t$ is the radius of the halo and 
$\ell_S(0)$ is the Str\"omgren length for zero impact parameter.
If $L_S>1$, then the halo is able to trap the I-front, while if $L_S<1$, 
the halo would be ionized quickly by the passage of the weak, 
R-type I-front across
it, which will not slow down enough to be trapped. 
In the latter case, if $L_S\ll 1$, 
the halo gas would be uniformly photoheated to $T\geq10^4$ K before any mass motions could 
occur, and the pressure gradient would blow the gas apart isotropically, long after the
I-front had exited the halo.

\begin{figure}
\includegraphics[width=3.4in]{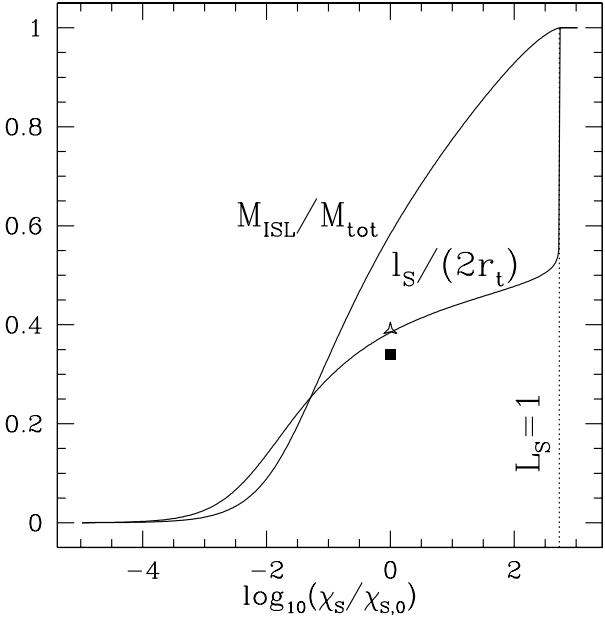} 
\caption{Inverse Str\"omgren Layer. The Str\"omgren length along the 
axis, $l_S(0)$, in units of the halo diameter $2r_t$ and ISL mass $M_{\rm ISL}$
in units of the total mass $M_{\rm tot}$ vs. $\chi_S/\chi_{S,0}$, where
$\chi_S\equiv F/(\rho_0^2r_0)$, $\chi_{S,0}$ is $\chi_S$ for our fiducial halo 
and $F_0=1$. Symbols indicate the (roughly) corresponding simulation 
results for 
$M_{\rm ISL}/M_{\rm tot}$ (star) and $\ell_S(0)/(2r_t)$ (square) at 
the moment of 
I-front transition from R-type to D-type (see \S~\ref{trapping_sect}). 
Dotted line
indicates the value $\chi_{\rm S,crit}$ for which $L_S=1$.
We assume $T_4=2$.}
\label{ISL}
\end{figure}

The threshold condition $L_S=1$ is equivalent to
\be
F=\frac{2\alpha_H^{(2)}f_b^2}{(\mu_Hm_H)^2}\rho_0^2r_0
\int_0^{\zeta_t}\tilde{\rho}^2(\zeta)d\zeta,
\label{strom2}
\ee
where $\tilde{\rho}\equiv\rho/\rho_0$, $\zeta\equiv r/r_0$, $f_b=\Omega_b/\Omega_0$ 
is the baryonic mass fraction and $\mu_Hm_H$ is the 
mean gas mass per H atom. The dependences on the source flux and 
on the minihalo mass and size at 
different redshift are combined if we define the parameter
\be
\label{chis}
\chi_S\equiv\frac{F}{\rho_0^2r_0}=\chi_{S,0}F_0M_7^{-1/3}
\left({1+z}\right)_{10}^{-5}\left(\frac{\Omega_0h^2}{0.15}\right)^{-5/3},
\ee
where 
$M_7\equiv M_{\rm tot}/(10^7M_{\odot})$ and
$\chi_{S,0}\equiv4.1\times10^{30}\mbox{g$^{-2}$cm$^3$s$^{-1}$}$. We have 
used the fact that $I\equiv2\int_0^{\zeta_t}\tilde{\rho}^2(\zeta)d\zeta=3.49$ and the 
scalings of $\rho_0$ and $r_0$ with the mass of the halo and its redshift of collapse 
according to the TIS model at high redshift \citep{IS01}. There is a direct correspondence 
between $L_S$ and $\chi_S$, and the condition $L_S=1$ is equivalent to 
$\chi_S=\chi_{S,\rm crit}$ where

\ba
\chi_{S,\rm crit}\!\!\!\!&=&\!\!\!\!3.3\times10^{33}T_4^{-3/4}
\left({\Omega_bh^2\over0.02}\right)^2\nonumber\\
&&\times\left({\Omega_0h^2\over0.15}\right)^{-2}{\rm g^{-2}cm^3s^{-1}},
\ea

\noindent 
(where $T_4$ refers to the temperature of the ionized gas)
while 
$L_S>1$ and $L_S<1$ correspond to $\chi_S>\chi_{S,\rm crit}$ and
$\chi_S<\chi_{S,\rm crit}$, respectively. We plot $\ell_S(0)$ vs. $\chi_S$ in 
Figure~\ref{ISL}. For our fiducial case,
$\chi_S=\chi_{S,0}\ll\chi_{S,\rm crit}$, so we expect the I-front to be easily
trapped by the minihalo in this case. In general, the critical minihalo
mass which is just large enough to trap the I-front by making $L_S=1$
is given by setting $\chi_S=\chi_{S,\rm crit}$ in equation~(\ref{chis}) to
yield

\ba
M_{7,\rm crit}\!\!\!\!&=&\!\!\!\!1.8\times10^{-9}
F_0^3\left[(1+z)_{10}\right]^{-15}T_4^{9/4}
\left({\Omega_0h^2\over0.15}\right)
\nonumber\\&&\times\left({\Omega_bh^2\over0.02}\right)^{-6}\,.
\ea

\noindent For $F_0\la10$, even the smallest minihalos which formed
at the last possible moment before the I-front overtook them at the end
of the reionization epoch at $z_{\rm ov}=6$ would
have been able to trap the I-front.

For a uniform halo with gas number density $\langle{n}_H\rangle$, equation~(\ref{strom}) reduces to
\be
\ell_S=\frac{F}{\alpha_H^{(2)}\langle n_H\rangle^2}.
\label{l_S_uniform}
\ee
In terms of the I-front trapping condition ($L_S=1$), a TIS halo with mean density
$\bar{\rho}_{\rm TIS}$ is equivalent to a uniform halo of the same size 
but with mean density
given by
\be
\bar{\rho}=\rho_0\left(\frac{I}{2\zeta_t}\right)^{1/2}
	=0.24\rho_0=34\bar{\rho}_{\rm TIS},
\label{equiv_SUS}
\ee
where $\bar{\rho}_{\rm TIS}$ is the mean density of the TIS.
Hence, a uniform halo of the same size as the TIS minihalo which is just
large enough to trap the I-front will do so only if it is 34 times more
massive than that TIS halo.

One important implication of this result is that it would be very incorrect
to use the volume-average mean density of a minihalo instead of this
much larger density in equation~(\ref{equiv_SUS}) to determine whether
a minihalo of a given mass will trap the I-front. In particular, if we
replace the realistic TIS profile of the centrally-concentrated
minihalo by a uniform-density minihalo with the same mass and radius, instead, 
the Str\"omgren number, $L_{S,\rm uniform}$, 
of this uniform halo will be very much
smaller than $L_S$ for the TIS minihalo, according to

\ba
L_{S,\rm uniform}\!\!\!\!&=&\!\!\!\!
8.9\times10^{-4}(\chi_{S,\rm crit}/\chi_S)
=0.72M_7^{1/3}(1+z)_{10}^5
\nonumber\\&&\times
F_0^{-1}T_4^{-3/4}
\left({\Omega_bh^2\over0.02}\right)^2\left({\Omega_0h^2\over0.15}\right)^{-1/3}
\,.
\ea

\noindent 
Hence, when a TIS minihalo of a given mass is just marginally able to trap the
I-front (i.e. $\chi_S=\chi_{S,\rm crit}$), a uniform halo with the
same mass and radius has a Str\"omgren number $L_S=8.9\times10^{-4}$,
far below the minimum required to trap the I-front!
This means that the phenomenon of I-front trapping can be
missed entirely by numerical simulations of reionization which
do not fully resolve the internal structure of individual
minihalos, even when their resolution in mass and length are,
in principle, high enough to identify minihalos of the correct mass and
mean density. For this reason, the simulations we shall
report here are high enough in resolution to
guarantee that the internal structure of
individual minihalos is fully resolved.

\begin{figure}
\includegraphics[width=3.6in]{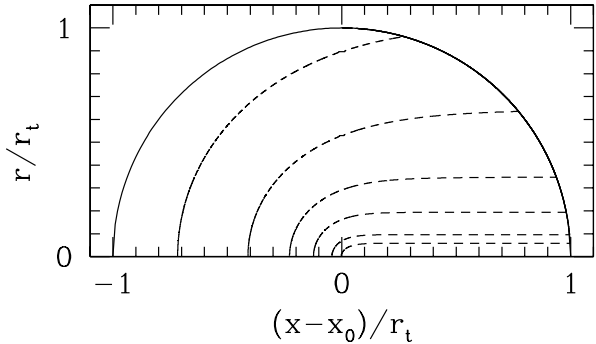}
\vspace{-4cm}
\caption{The ISL shape inside minihalo (halo boundary indicated by the solid line), 
according to the ISL approximation for $\chi/\chi_{S,0}=0.01$, 0.1, 1, 10, 100, 
and 250 (dashed-line curves from left to right). $x_0$ is the position of the
centre of the halo along the $x$-axis. Both the impact parameter $r$ 
and axis coordinate $x-x_0$ are in units of the halo radius $r_t$. Source is located
outside the box, to the left, at $x=0$, where $x_0\gg r_t$.
We assume $T_4=2$.}
\label{ISL_shape}
\end{figure}

\subsubsection{Inverse Str\"omgren Layer (ISL) in minihalo}

We define the Inverse Str\"omgren Layer as the region inside the 
minihalo which is the 
$\rm H\,II$ region predicted by the Str\"omgren approximation. This 
region is bounded by the 
minihalo surface nearest to the source and by the surface defined at 
each impact parameter
$r$ by $\ell_S(r)$ in equation~(\ref{strom}). This ISL provides an 
estimate of the minihalo 
region which will be ionized first, during the weak, R-type I-front 
phase, and of the 
approximate location of the surface inside the minihalo where the transition to D-type will
occur as the R-type front approaches it. In Figure~\ref{ISL}, 
we show the fraction of the halo mass which is located inside the ISL,
$M_{\rm ISL}/M_{\rm tot}$, vs. 
$\chi_S$, given by
\be
M_{\rm ISL}=
\int_0^{r_t}2\pi rdr\int_0^{\ell_S(r)}\rho(R)dX.
\label{mass_ISL}
\ee
Here $R=\{r^2+[X-(r_t^2-r^2)^{1/2}]^2\}^{1/2}$. In Figure~\ref{ISL}
we also plot $L_S^{-1}=\ell_S(0)/(2r_t)$  (the Str\"omgren length 
for $r=0$ in units of the halo diameter) vs. $\chi_S$, and we indicate the 
corresponding simulation results at the approximate moment of transition of 
the I-front from R-type to D-type, to be discussed in \S~\ref{trapping_sect}. 
At $\chi_S=\chi_{S,crit}$, the ISL includes the entire minihalo, thus
$M_{\rm ISL}/M_{\rm tot}=1$ and $\ell_{S}(0)=2r_t$. 
In Figure~\ref{ISL_shape}, we
plot the shape of the ISL surface for $\chi_S/\chi_{S,0}=0.01$, 0.1, 
1, 10, 100, 
and 250. We see that, even for low values of the parameter $\chi_S$ (equivalent
to low external flux levels) a significant fraction  
of the halo volume becomes ionized quickly due to the relatively low 
density in the halo outskirts. However, from Figure~\ref{ISL} we see that this 
volume still corresponds to a small fraction 
of the total mass (e.g. for $\chi_S/\chi_{S,0}<0.01$,
$M_{\rm ISL}<10\%$). 
As $\chi_S$ approaches $\chi_{S,\rm crit}$, the
neutral gas fraction shielded by the highly concentrated halo core diminishes
until it disappears completely when $\chi_S=\chi_{S,\rm crit}=490\chi_{S,0}$
(for $T_4=2$).

\subsection{Evaporation time}
\label{time_scale_sect}

A rough, order-of-magnitude guide to the time-scale $t_{\rm ev}$ for minihalo 
photoevaporation is the sound-crossing time for the characteristic size of the 
minihalo at the sound speed of the ionized gas, 
$t_{\rm sc}=2r_t/c_s(10^4\rm K)$.
The actual I-front propagation and evaporative wind are 
quite complex, as we shall 
see, and depend upon additional properties like the halo 
mass and density, flux level 
and spectrum, and the time it takes the halo and background universe to evolve.
However, this simple estimate will be a convenient standard 
of reference with which 
to compare the actual simulation results.  

For a minihalo of mass $M$ at high redshift, we obtain
\be
t_{\rm sc}=98\,{\rm Myr}\left({M_{7}}\right)^{1/3}
	\left(\frac{\Omega_0 h^2}{0.15}\right)^{-1/3}
 	\left({1+z}\right)_{10}^{-1},
\label{t_sc}
\ee
using the adiabatic sound speed
$c_s(10^4\rm K)=11.7(T_4/\mu)^{1/2}=15.2\,T_4^{1/2}\rm km\,s^{-1}$,
where $\mu=0.59$ 
is the mean molecular weight for fully ionized gas of H and He if 
the abundance of He is 
$A(\rm He)=0.08$ by number relative to H, and we have assumed $T_4=1$. 
Of course, if $t_{\rm ev}\geq t_H$, the Hubble time at the epoch of photoevaporation, then
the underlying properties of the halo, infall, ionizing source, and 
background universe
will evolve significantly during the process. The Hubble time at high 
redshift in the $\Lambda$CDM universe is
\ba
t_H=\frac{2}{3H(z)}
	\approx 533\,{\rm Myr}
	\left(\frac{\Omega_0h^2}{0.15}\right)^{-1/2}\left({1+z}\right)_{10}^{-3/2}.
\label{hubble_time}
\ea
Thus, the ratio of the two timescales is 
\be
\frac{t_{\rm sc}}{t_H} = 0.184\left({M_{7}}\right)^{1/3}
	\left(\frac{\Omega_0 h^2}{0.15}\right)^{1/6}
 	\left({1+z}\right)_{10}^{1/2}.
\label{time_ratio}
\ee
We see that this ratio is only weakly dependent on the background cosmology, 
and suggests that the photoevaporation of a minihalo over the range of mass and
epochs of interest will be completed in less than a Hubble time, if this estimate
of $t_{\rm ev}$ is correct.

\subsection{Ionizing photon consumption}
\label{photon_consump_sect} 
An important quantity characterizing the effect of minihalos on the 
process of reionization is the consumption of ionizing photons 
which results from 
photoabsorption by minihalo atoms during their photoevaporation. 
We can parametrize
this consumption by the photon-to-atom ratio, $\xi\equiv {N}_{\gamma}/N_a$, 
i.e. how many photons per minihalo atom are required to photoevaporate the 
halo. Collapsed minihalos are significantly denser than the IGM, thus the 
recombination rate inside is significantly higher, 
leading to a much higher photon 
consumption rate per unit time per atom, which could 
in turn have a substantial effect 
on the progress and duration of reionization.  

\begin{figure}
\includegraphics[width=3.6in]{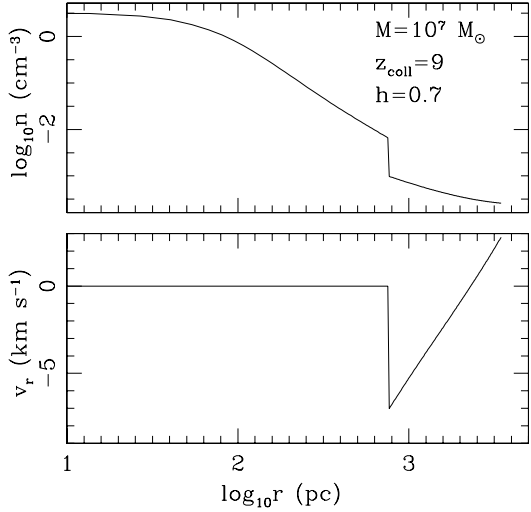}
\vspace{-2cm} 
\caption{Initial conditions for the minihalo photoevaporation 
simulations (TIS halo + self-similar cosmological infall): atomic number 
density (upper panel), and radial velocity (lower panel) profiles vs. 
radius for 
minihalo of mass $M_7=1$ at $(1+z)_{10}=1$.}
\label{init_lam0.7}
\end{figure}

\begin{figure}
\includegraphics[width=3.5in]{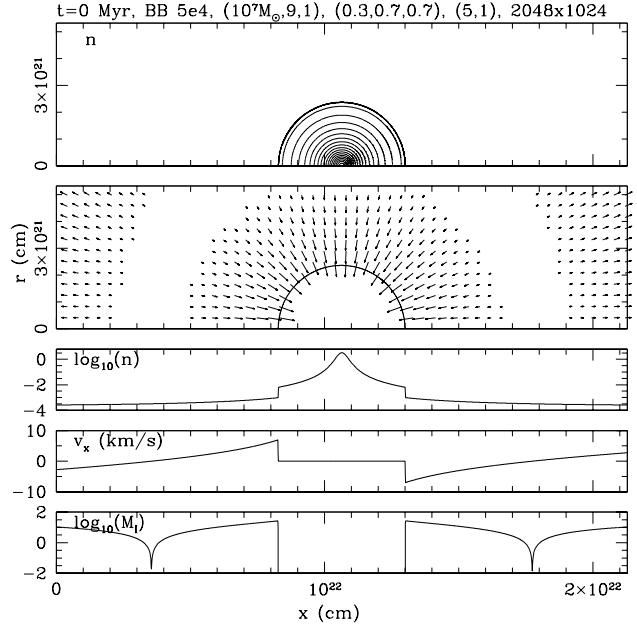}
\caption{Initial time-slice at $t=0$ Myrs ($z=9$ in $\Lambda$CDM universe), 
from top to bottom: (1) isocontours of atomic density, 
logarithmically spaced, in $(r,x)-$plane of cylindrical coordinates; 
(2) flow velocities; arrows are plotted with length proportional to gas velocity.
An arrow of length equal to the spacing between arrows has velocity
$5\, {\rm km\, s^{-1}}$; the minimum velocities plotted are
$\rm 1\,km\,s^{-1}$. Solid line indicates the boundary of the minihalo; 
cuts along the $r=0$ axis of: (3) gas number density $n$,
(4) velocity $v_x$, and (5) Isothermal Mach number $M_I$. Source is 
located outside the box, far to the left along the $x$-axis.}
\label{vel_arrows_new_init_BB5e4}
\end{figure}

We define the effective absorption cross-section at frequency $\nu$ of a minihalo at 
a given time $t$ as
\begin{equation}
\label{eff_cross_sect}
\sigma_{\rm eff,\nu}(t)\equiv\int_0^{\infty}2\pi r
		[1-e^{-\tau_\nu(r,t))}]dr,
\end{equation}
where $\tau_\nu=\tau_{\rm \nu,H}+\tau_{\rm \nu,He\,I}+\tau_{\rm \nu,He\,II}$
is the 
total bound-free opacity at impact parameter $r$ and time $t$, including 
the minihalo gas which is leaving the halo in a wind, where
\be
\tau_i(r,t)=\int_{\nu_{\rm th,i}}^{\infty}\sigma_{\rm bf,i}(\nu)n_idx.
\ee
where $\sigma_{\rm bf,i}$ and $\nu_{\rm th,i}$ are the bound-free cross-section
and ionization threshold frequency and $n_i$ is the number
density for species $i$.
The number of photons absorbed by the minihalo gas per minihalo atom over time 
$t_{\rm ev}$ is then given by
\be
\xi_{\tau}=\frac{1}{N_a}\int_0^{t_{\rm ev}}\int_{\nu_H}^\infty F_\nu(t)\sigma_{\rm eff,\nu}(t)dt,
\label{n_gamma1}
\ee
where $F_\nu(t)$ is the time dependent ionizing photon flux per unit frequency.

Alternatively, $\xi$ can be calculated by the direct count of the number
of recombinations experienced by each atom which was initially in the halo:

\be
\xi_{\rm rec}=1+\frac{1}{N_a}\int_0^{t_{\rm ev}}\!\!\!dt
\int\!dV(\alpha_H^{(2)}n_en_{\rm H\, II}+\alpha_{He}^{(2)}n_en_{\rm He\, II}),
\label{n_gamma2}
\ee
where $n_e$ is the number density of electrons, $n_{\rm H\, II}$ 
and $n_{\rm He\, II}$
are the number densities of H II and He II, and $\alpha_H^{(2)}$ 
and $\alpha_{He}^{(2)}$ are the Case B recombination coefficients for H II and 
He II, respectively, and the volume integral is over all Lagrangian
fluid elements initially inside the minihalo when the
I-front first encountered it. 
We have neglected the recombinations of $\rm He\,III$ 
to $\rm He\,II$ 
because these generally contribute diffuse flux which is absorbed 
on-the-spot by H and He.

A na\"\i ve estimate of $\xi$ is obtained 
if we assume that the minihalo is optically-thin
and instantaneously ionized but remains static at its initial density for a time $t_{\rm sc}$.
Then, ignoring the contribution of He to recombinations, equation~(\ref{n_gamma2}) yields 
\be
\xi=f\frac{C_{\rm int}\langle n_H\rangle\alpha_H^{(2)}}{1+\delta_{\rm TIS}}t_{\rm sc}
\label{xi_ham}
\ee
\citep{HAM01}, where $\langle n_H\rangle$ is the mean H atom 
number density inside a halo, 
$C_{\rm int}\equiv \langle n_{\rm H}^2\rangle/\langle n_H\rangle^2=444^2$ 
is the effective clumping 
factor for the TIS, and $1+\delta_{\rm TIS}=130.6$ is the average overdensity of a TIS halo
with respect to the cosmic mean background density. Since, in reality, such a model is 
oversimplified, we also introduce an ``efficiency'' factor $f$ which we will derive from our 
simulations below. Thus we obtain
\be
\xi=206 fT_4^{-3/4}M_7^{1/3}\left(\frac{\Omega_0 h^2}{0.15}\right)^{-1/3}
 	\left({1+z}\right)_{10}^{-1}.
\label{xi_numeric}
\ee
According to this estimate, if $f\approx1$ as claimed by \citet{HAM01}, 
based on their simulation of a uniformly-ionized minihalo 
with zero optical depth, 
the photoevaporation of minihalos can be an enormous 
sink of ionizing photons during 
the reionization epoch. By contrast, atoms in the IGM at closer to the mean 
baryon density have otherwise been estimated previously to 
consume only $\xi\sim1$ photons per atom during reionization \citep{G00,MHR00}. It is crucial,
therefore, for us to perform the detailed simulations reported here in order to
derive the efficiency factor $f$ in equation~(\ref{xi_numeric}) properly.

\section{The calculation}
\label{calc_sect}
\subsection{Basic equations}
\subsubsection{Gas dynamical conservation equations}
We solve the Eulerian conservation equations of fluid dynamics for 
the vector ${\bf U}$
of the densities of conserved quantities in 2-D, cylindrical symmetry,

\be
{\bf U}=(\rho,\rho v_x,\rho v_r,E,n_{i,z}),
\ee 
mass density $\rho$, momentum densities $\rho v_x$ in the $x$-direction
and $\rho v_r$ in the radial direction, 
energy density $E=\rho(v_x^2+v_r^2)/2+\epsilon$
and number density of chemical species $i$ in ionization stage $z$, $n_{i,z}$. 
Here $v_x$ and $v_r$ are the velocity components, $\epsilon = c_VT$ is the 
specific internal energy, $T$ is the temperature and 
$c_V$ is the isochoric specific heat. The Eulerian conservation 
equations can then be written as
\be
\frac{\pa {\bf U}}{\pa t}+\frac{\pa{\bf F}({\bf U})}{\pa x}
	+\frac{\pa{\bf G}({\bf U})}{\pa r}={\bf S},
\label{euler}
\ee
where ${\bf F}({\bf U})$ and ${\bf G}({\bf U})$ are the fluxes of  
${\bf U}$ in the $x$ and $r$ directions, respectively, given by:
\be
{\bf F}({\bf U})=(\rho v_x,\rho v_x^2+p,\rho v_xv_r,v_x(E+p),n_{i,z}v_x),
\ee
and
\be
{\bf G}({\bf U})=(\rho v_r,\rho v_xv_r,\rho v_r^2+p,v_r(E+p),n_{i,z}v_r),
\ee
and ${\bf S}$ are the source terms:
\ba
{\bf S}=\bigg[-\frac{\rho v_r}{r},-\frac{\rho v_xv_r}{r}-(\nabla\phi)_x,
	-\frac{\rho v_r^2}{r}-(\nabla\phi)_r,\nonumber\\
	\qquad-\frac{v_r(E+p)}{r}-{\bf v}\cdot\nabla\phi-\Lambda+\Gamma,
	-\frac{n_{i,z}v_r}{r}+S_{i,z}\bigg].
\label{source_terms}
\ea
Here $p$ is the pressure, $\phi$ is the gravitational potential, $\Lambda$
is the cooling rate, $\Gamma$ is the heating rate, 
and $S_{i,z}$ is the source
function for the chemical species $i$ in ionization stage $z$, as described in 
\S~\ref{chem_sect}. We use the ideal gas equation of state for monatomic gas
($\gamma = 5/3$)
\be
p=\frac{1}{\gamma-1}\epsilon\rho.
\label{equ_state}
\ee

\subsubsection{Nonequilibrium chemical reaction network}
\label{chem_sect}
We take account of the nonequilibrium ionization balance of H, He, and a possible admixture
of heavy elements C, N, O, Ne, and S, as described in \citet{RML97}, \citet{MRCLBSN98}, and
references therein. [For a description of the details and checks of our microphysics, 
the reader is referred to \citet{M93}]. 

\begin{figure}
\includegraphics[width=3.5in]{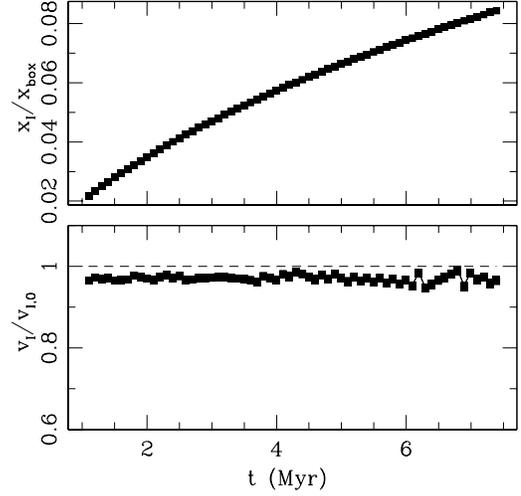}
\vspace{-2cm} 
\caption{R-type I-front speed. 
Position $x_I$ (in units of the box length $x_{\rm box}=5.64$ Mpc; upper 
panel) and velocity $v_I$ (in units of the analytical prediction as 
described in the text; lower panel) of an R-type ionization front 
propagating through the mean IGM at redshift $z=9$.}
\label{I_front_speed_noHe}
\end{figure}

\begin{figure*}
\includegraphics[width=3.4in]{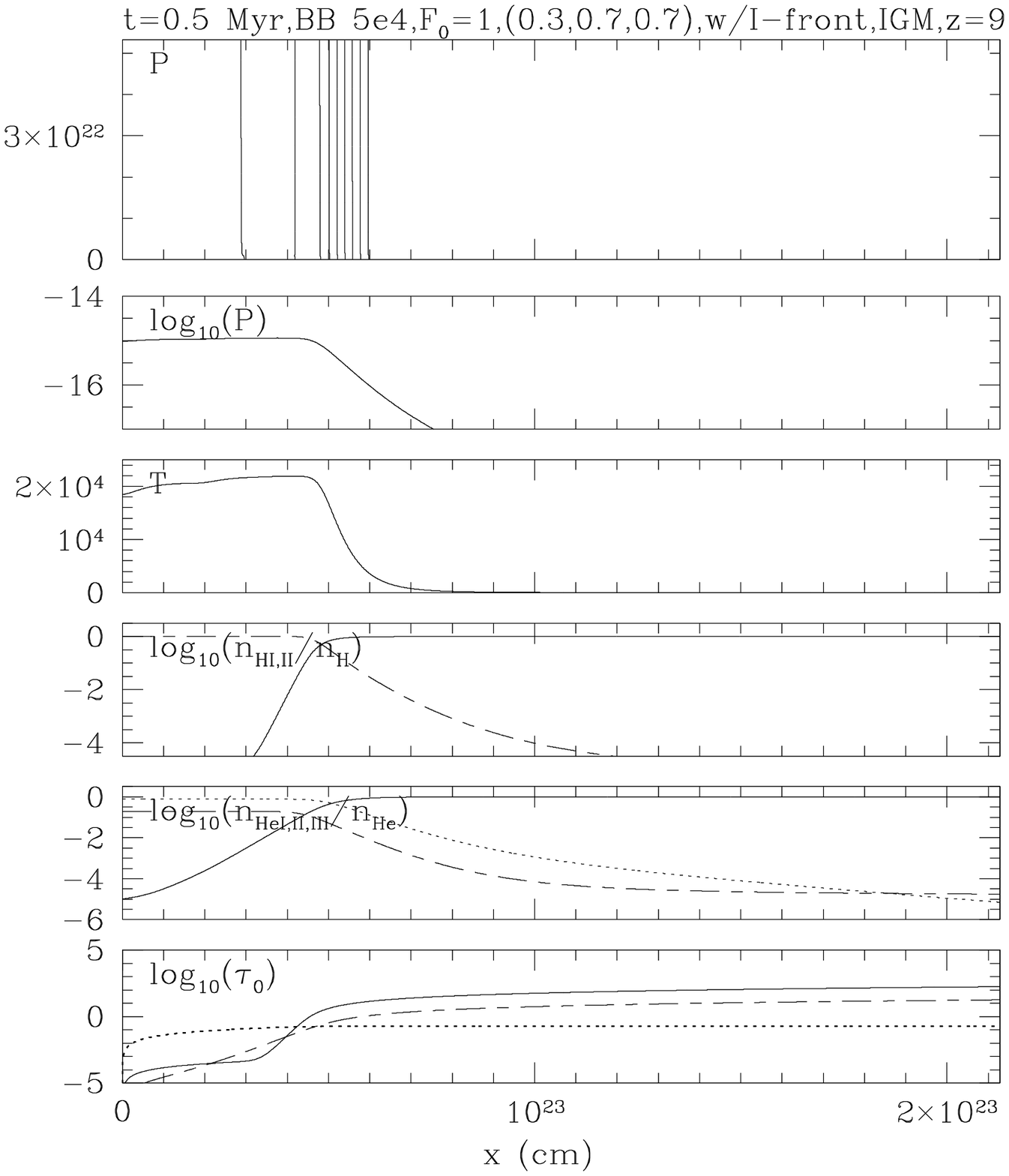}
\includegraphics[width=3.4in]{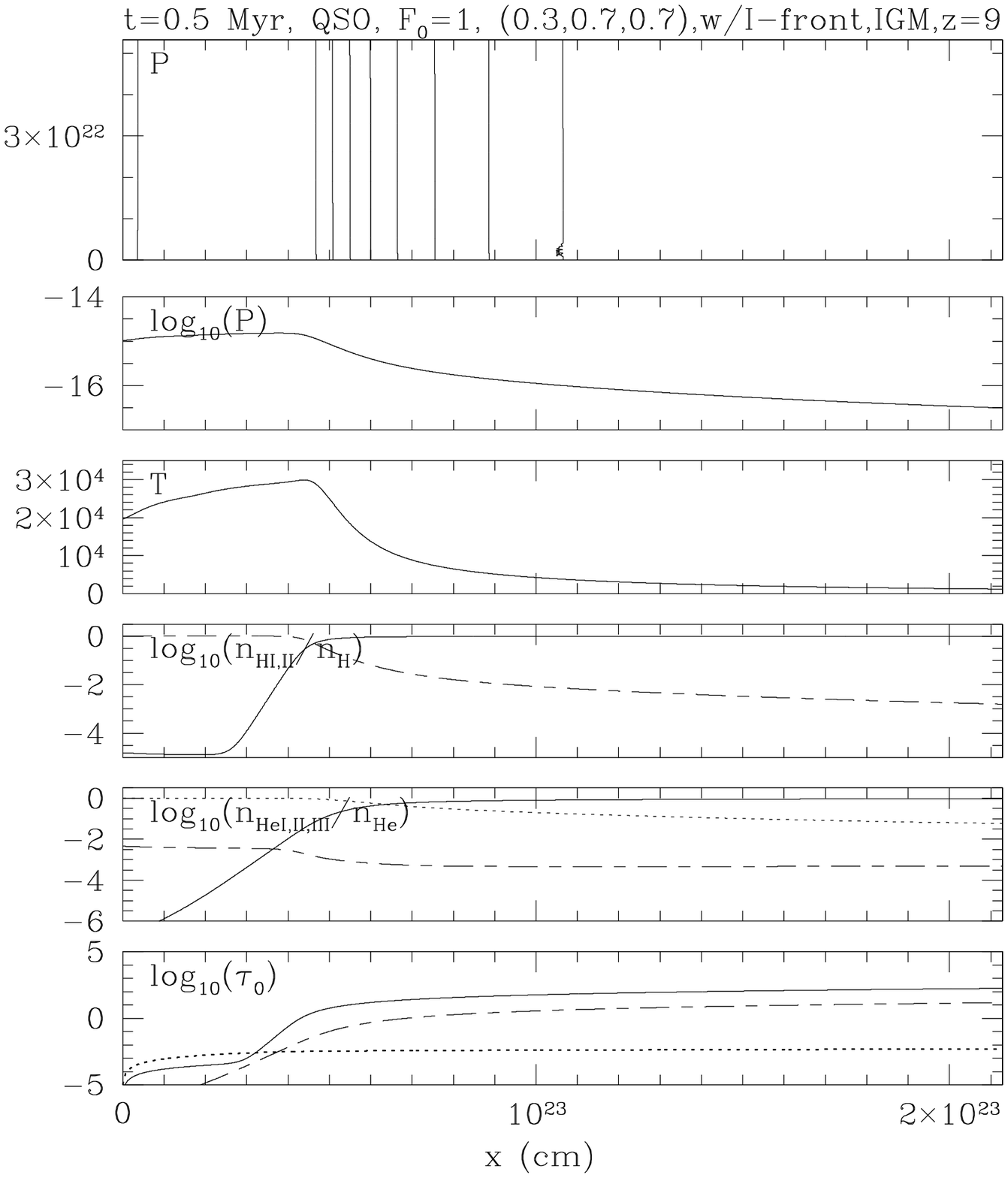}
\caption{Detailed structure of R-type ionization front in the mean IGM at $z=9$ 
at time $0.5$ Myr after the ionizing flux turns on at the 
left-side boundary of the box.
(a) (left) BB 5e4 case and (b) (right) QSO case. 
From top to bottom: (1) isocontours of pressure, logarithmically spaced, in
$(r,x)-$plane of cylindrical coordinates; cuts along the 
$x$-axis (the I-front propagation 
direction) of (2) pressure, (3) temperature $T$, (4) H~I (solid) and 
H~II (dotted) fractions; (5) He~I (solid), He~II (dashed) and He~III (dotted)
 fractions; (6) bound-free optical depths measured from $x=0$ along 
the $x$-axis, for H~I (solid), He~I (dashed), and He~II (dotted) at their 
respective ionization thresholds. 
Source is located 0.5 Mpc, to the left of the box.}  
\label{slices_0.5Myr_QSO_uniform}
\end{figure*}

\begin{figure}
\includegraphics[width=3.5in]{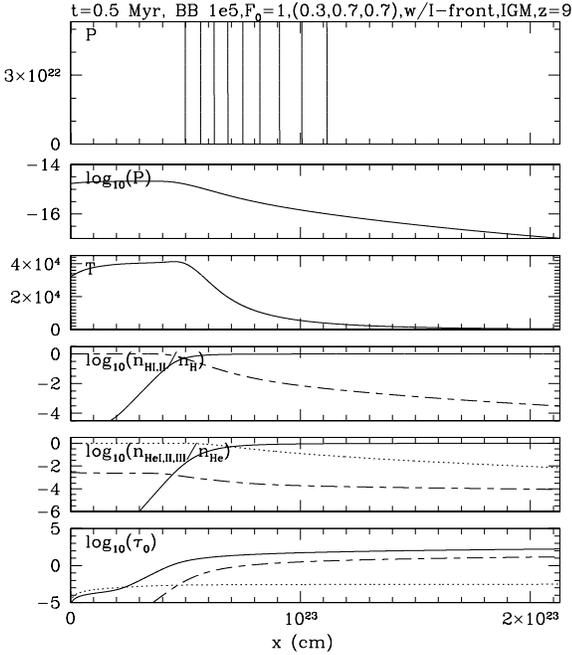}
\caption{Same as in Figure~\ref{slices_0.5Myr_QSO_uniform}, but for BB 1e5 case.}
\label{slices_0.5Myr_BB1e5_uniform}
\end{figure}

The source function for the nonequilibrium chemical reaction network
is given by
\ba
S_i\!\!\!\!&=&\!\!\!\!
n_en_{i,z+1}\alpha_{i,z+1}(T)+n_en_{i,z-1}C_{i,z-1}-n_en_z\alpha_{i,z}(T)
\nonumber\\&&
-n_en_{i,z}C_{i,z}+n_{i,z-1}\phi_{i,z-1}
-n_{i,z}\phi_{i,z},
\label{source_function}
\ea
where, as above, $i$ labels the chemical elements, $z$ labels the ionization 
state, $\alpha_{i,z+1}(T)$ is the recombination rate from stage $z+1$ to $z$,
$C_{i,z-1}$ is the collisional
ionization rate from $z-1$ to $z$, $\phi_{i,z}=\int_{\nu_{\rm th,i}}^{\infty}({F_\nu}\sigma_{\nu,i,z})/({h\nu})d\nu$ 
is the photoionization rate from $z$ to $z+1$, 
and $\nu_{\rm th,i}$ is the photoionization threshold 
for species $i$. Charge exchange reactions are included, although we have
not displayed them in equation~(\ref{source_function}).
Recombination rates include radiative and dielectronic rates.

The electron number density $n_e$ is given by 
$n_e=n_{\rm H\, II}+n_{\rm He\, II}+2n_{\rm He\, III}+n_{\rm C}$.
The contribution of one electron per carbon atom is added to ensure that the 
electron number density is always positive. Since the ionization 
threshold of carbon is below 13.6 eV and the IGM is transparent to these photons
even before reionization, C is always at least singly ionized. The contribution of 
the other metals to the electron density is assumed negligible. 

\subsubsection{Radiative cooling and photo-heating}

Our nonequilibrium cooling function includes collisional
line excitation and ionization, recombination and free-free
cooling due to H, C, N, O, Ne, and S, as described and tested in detail in
\cite{RML97}, to which we have added He cooling and the Compton cooling 
which results 
from inverse Compton scattering of the CMB photons off free electrons.
The Compton cooling rate is given by 
\be
\Lambda_{\rm Comp}=5.65\times10^{-36}n_e(1+z)^4(T-T_{\rm CMB})\, \rm erg \,cm^{-3}s^{-1}
\ee
\citep{SK87}, where $T_{\rm CMB}=2.73(1+z)$ is the CMB temperature at redshift $z$, $n_e$ 
is the local density of free electrons, and $T$ is the local gas temperature.

The photo-heating rate $\Gamma$ (in units of $\rm erg\,cm^{-3}s^{-1}$) is the
sum of the ionization heating terms for each of the species H I, He I, and He II:
\be
\Gamma_{i,z}=n_{i,z}\int_{\nu_{\rm th,i,z}}^{\infty}
	\frac{F_\nu(h\nu-h\nu_{\rm th,i,z})}{h\nu}\sigma_{\rm bf,\nu,i,z}d\nu.
\label{photoheating}
\ee

\subsubsection{Radiative transfer and the ionizing flux}
The energy flux of ionizing photons at a point with coordinates $(r,x)$ for a source on 
the $x$-axis at position $x_0$ is given by
\be
F_\nu(r,x)=\frac{L_\nu}{4\pi\left[r^2+(x-x_0)^2\right]}e^{-\tau_\nu(r,x)}.
\label{ion_flux}
\ee
Here $L_\nu$ is the source luminosity at frequency $\nu$, 
and $\tau_\nu(r,x)$ is the bound-free optical depth from the 
source to the point 
$(r,x)$, given by
\be
\tau_\nu(r,x)=
	\sigma_{\nu,\rm H I}N_{\rm H I}+\sigma_{\nu,\rm He I}N_{\rm He I}
		+\sigma_{\nu,\rm He II}N_{\rm He II},
\label{opt_depth}
\ee
where $N_{i,z}$ is the column density of species $i$ at ionization stage $z$ along the
line from the source to the point $(r,x)$. 

\subsubsection{Gravity force}
We have modified the code to include the gravity field due to the halo of dark matter and gas,
and the infalling matter, as discussed in detail in the Appendix. We neglected the small 
change in this gravity force over time which results from the gas dynamical evolution which 
moves the gas relative to the dark matter. Thus the density field used for the gravity 
calculation is spherically-symmetric at all times and the gravitational
acceleration is given by 
\be
a_{\rm grav}=-\nabla\phi=\frac{GM(\leq R)}{R^2}\,,
\label{gravity}
\ee
where $M(\leq R)$ is the time-varying mass within the spherical
radius $R=(r^2+x^2)^{1/2}$.
This force is included in the Euler equations~(\ref{euler}) through the 
source term ${\bf S}$ as shown in equation~(\ref{source_terms}).

\subsection{Initial conditions}

\subsubsection{Minihalo and cosmological infall}

As discussed above, the halo in our illustrative simulations has radius $r_t=0.76$ kpc, 
and virial temperature $T_{\rm vir}=4000$ K, corresponding to dark-matter velocity 
dispersion $\sigma_V=5.2\,{\rm km\, s^{-1}}$. The initial number density 
and velocity profiles of the halo and the infall are shown in 
Figure~\ref{init_lam0.7}. For comparison with our simulation results at later times, 
we show the initial conditions on the computational grid in 
Figure~\ref{vel_arrows_new_init_BB5e4} 
(top to bottom): (1) the density contours (logarithmically-spaced), (2) the 
initial velocity field, and cuts along the axis of: (3) gas number density $n$, 
(4) velocity $v_x$, and (5) Isothermal Mach number $M_I\equiv v/c_{s,I}$,
where $c_{s,I}$ is the isothermal sound speed. 
The computational box is 
chosen so that the turn-around radius of the cosmological
infall starts well inside the box. 
This ensures that the gas next to the simulation boundaries 
(excluding the axis of symmetry) 
is outflowing throughout the simulation, consistent with our 
transmissive numerical boundary 
conditions (see \S~\ref{num_methods_sect}). 

\subsubsection{Source spectra and evolution}

The source is on the axis of symmetry, far away (thus the rays are close to 
parallel, although the code actually takes the true angles properly into 
account) to the left of the box. 
We consider three possible spectra as described in
\S~\ref{intro}, two stellar cases, black-body spectra with effective temperatures 
$T=50,000$ K (``BB 5e4'') for Pop. II stars and $T=10^5$ K 
(``BB 1e5'') for Pop. III stars, and a power-law QSO-like 
energy spectrum with index 
$\alpha=-1.8$. As the universe expands, the proper distance between the
source and the minihalo grows, as $x_0(t)\propto a(t)$, 
where $a(t)$ is the cosmic 
scale factor, so the incident flux level decreases 
$\propto x_0^{-2}\propto a(t)^{-2}$.

\subsubsection{R-type I-front encounters the simulation volume at $t=0$} 
The I-front has a finite width of $\sim(10-20)$ mean free paths. For the intergalactic 
weak, R-type I-front which encounters the minihalo, this mean free path in the IGM is a
few kpc, which is similar to our simulation box size. 
Therefore, it would be inaccurate to start the simulations by assuming 
a sharp I-front at the left box boundary. Instead, we start 
our simulations shortly
before the I-front arrival at the left boundary, as follows. 
We assume the the gas 
outside the box is uniform, at the mean IGM density, for 
which case there is an exact 
solution for the weak, R-type I-front propagation. 
We then use the frequency-dependent 
optical depths given by the current position of the I-front obtained from the 
analytical solution to attenuate the spectrum of the photoionizing 
source which enters 
the computational box. While not exact (e.g. due to infall which slightly increases 
the local density around the halo) this approach closely imitates the
gradual rise in the photoionizing radiation flux arriving from the 
source as the I-front
approaches and the hardening of the spectrum due to the deeper 
penetration of higher-energy
photons. As an alternative, it is also possible to take a 
much larger 
simulation 
box, which would allow the I-front to relax to its correct structure before 
arriving at the halo. However, such an approach would either degrade the 
resolution unacceptably, or else the simulation would become prohibitively 
(and needlessly) more expensive.

\begin{figure*}
\includegraphics[width=3.4in]{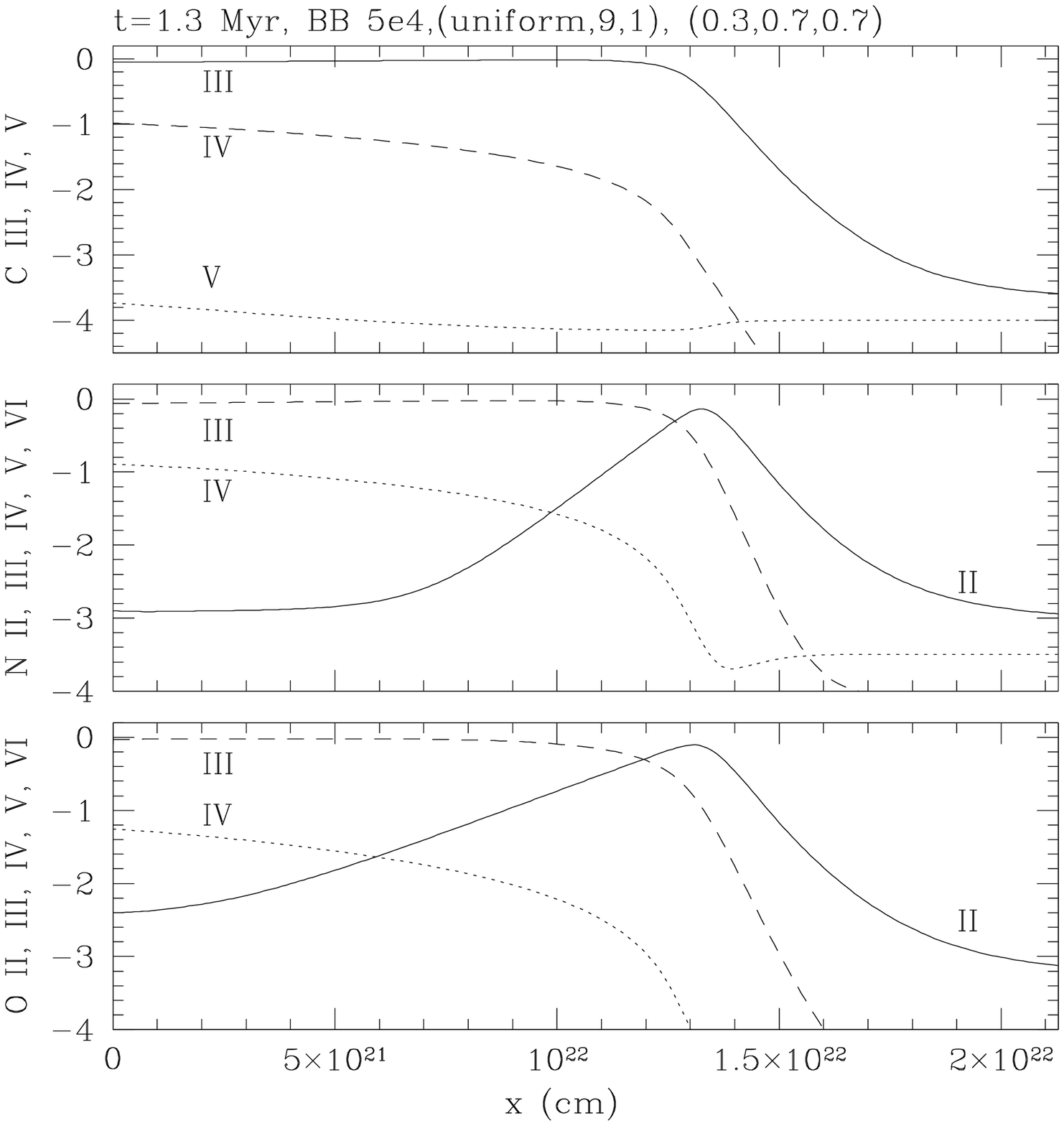}
\includegraphics[width=3.4in]{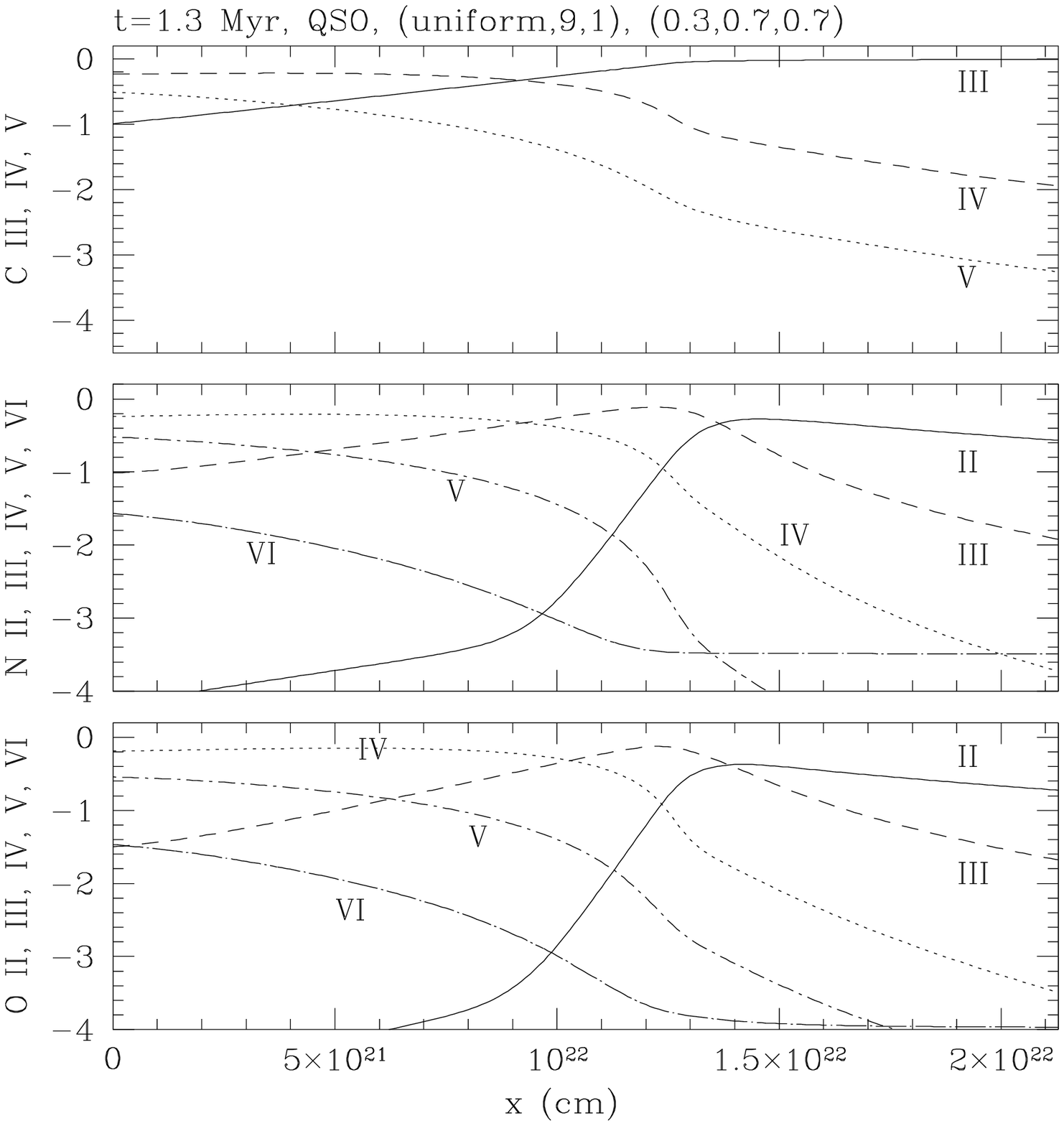}
\caption{Ionization structure of metals in 
R-type I-front propagating in the mean IGM at $z=9$, 1.3 Myr after front
enters the box on the left-hand-side. 
C, N, and O ionic fractions along symmetry
axis: (a) (left) BB 5e4 case; (b) (right) QSO case.}
\label{cno_1.3Myr_QSO}
\end{figure*}

\begin{figure}
\includegraphics[width=3.5in]{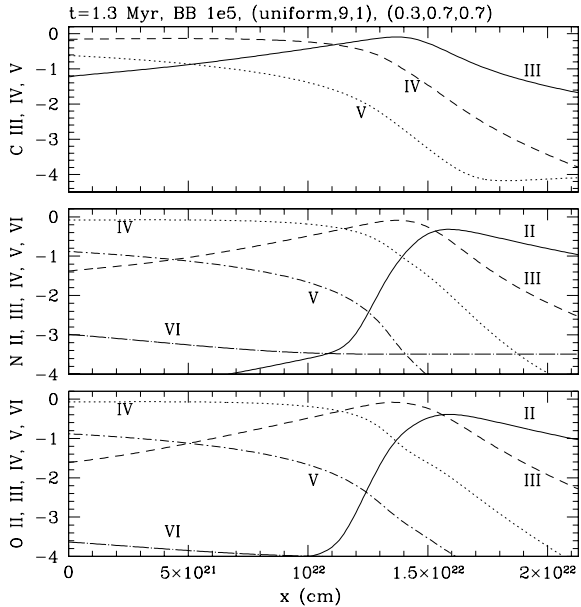}
\caption{Same as in Figure~\ref{cno_1.3Myr_QSO}, but for BB 1e5 case.}
\label{cno_1.3Myr_BB1e5}
\end{figure}

\section{Numerical method and tests}
\label{num_methods_sect}   
\subsection{The method}
\label{method}

We use a 2-D axisymmetric Adaptive Mesh Refinement (AMR) code (called CORAL) described in 
detail in \cite{RTCB95} and \cite{MRCLBSN98}. The code has been extensively
tested and applied to a range of problems including simulations of Herbig-Haro jets 
and the photoevaporation of uniform density interstellar clouds near planetary nebulae. 
Here we will discuss the numerical details of the code only briefly, concentrating on 
the new elements in the code introduced by us, and we refer the reader to these earlier 
papers and references below for a more in-depth discussion. 

The numerical scheme for 
solving the Euler equations is an adaptive grid implementation of the
van Leer Flux Vector Splitting method \citep{vL82}, improved to second order 
accuracy by use of linear gradients within cells as described in \citet{A91}. 
The refinement and de-refinement criteria are based on the gradients of all 
code variables. When the gradient of any variable is larger than a 
pre-defined value the cell is refined, while when the criterion for 
refinement is not met the cell is de-refined.  
We have modified the code to include the gravitational acceleration 
in equation (\ref{gravity}), as discussed in detail
in the Appendix.  
We impose reflective boundary conditions on the axis and transmissive boundary
conditions on the other three boundaries of the box.

The code timestep is set in general 
by the minimum of the timesteps determined by the Courant condition,
the ionization time scale and gravity, 

\begin{equation}
\label{timestep}
\Delta t=\min(\epsilon_{\rm Cour}\Delta t_{\rm Cour},
\epsilon_{\rm ion}\Delta t_{\rm ion}, \epsilon_{\rm grav}\Delta t_{\rm grav})\,.
\end{equation}

\noindent 
The Courant time in equation (\ref{timestep}) is

\begin{equation}
\label{courant}
\Delta t_{\rm Cour}=\Delta x/c_s\,,
\end{equation}

\noindent
where $\Delta x$ is the cell width.  The ionization time scale here is that for
ionizing hydrogen,

\be
\label{t_ion}
\Del t_{\rm ion}=\epsilon_{\rm ion}\frac{n_{\rm H I}}{(dn_{\rm H I}/dt)}
  =\epsilon_{\rm ion}\frac{n_{\rm H I}}{|n_{\rm H I}\phi_{\rm H I}
                -n_{\rm H II}^2\alpha_{\rm H}^{(2)}|},
\ee
where $\phi_{H I}$ is the photoionization rate for hydrogen, $n_{\rm H I}$ and
$n_{\rm H II}$ are the
number densities of neutral and ionized hydrogen, $\alpha^{(2)}_{\rm H}$ is the
Case B recombination coefficient for hydrogen, and $\epsilon_{\rm ion}$ is a
constant to be determined.
The gravity time is just

\begin{equation}
\label{gravitytime}
\Delta t_{\rm grav}=(\Delta x/a_{\rm grav})^{1/2}\,.
\end{equation}

\noindent 
The constants $\epsilon_{\rm i}$ are small numbers whose values are
chosen to ensure accuracy and stability while minimizing the total
number of timesteps per simulation.  In all cases considered here, the
timestep is set by the Courant condition,
except when the fast, R-type I-front propagates through the computational box. 
In the latter case, the ionization time scale is smaller than the 
Courant time, and the value of $\epsilon_{\rm ion}$ is chosen by
experimentation to optimize the accuracy and efficiency of the test
problem described in 
\S~\ref{I_front_sect}.  For the Courant condition,
the van Leer Flux Vector Splitting scheme is stable 
when $\epsilon_{\rm Cour}\leq [2\gamma+M(3-\gamma)]/(\gamma+3)$, where $\gamma=5/3$ is the 
adiabatic index and $M$ is the Mach number. For $M=0$ we obtain 
the condition $\epsilon_{\rm Cour}\leq 5/7$ for the scheme 
to be stable for any Mach number $M$.
For the current simulations we utilized the conservative 
value $\epsilon_{\rm Cour}=0.4$.

The microphysical processes -- chemical reactions, radiative processes, 
transfer of radiation, heating and cooling -- are implemented though the 
standard approach of operator-splitting (i.e. solved each time-step, side-by-side with 
the hydrodynamics and coupled to it through the energy equation).
The energy and chemical rate equations are solved semi-implicitly.
We follow the nonequilibrium evolution of the ionic species of H, He, C II-VI, 
N I-VI, O I-VI, Ne I-VI, and S II-VI. The species C I and S I are assumed fully
ionized since their ionization thresholds are below the ionization
threshold of hydrogen and the gas is largely optically-thin 
to such low energy photons [for details see \cite{MRCLBSN98} and references
therein]. The method for solving the nonequilibrium chemistry equations is from
\citet{SVK87} and is implemented 
as described and tested in \cite{RML97} and \cite{MRCLBSN98}. We use 
Case B recombination coefficients and the corresponding cooling rates, as
is appropriate for the range of densities and temperatures in these simulations. 
\begin{figure}
\includegraphics[width=3.4in]{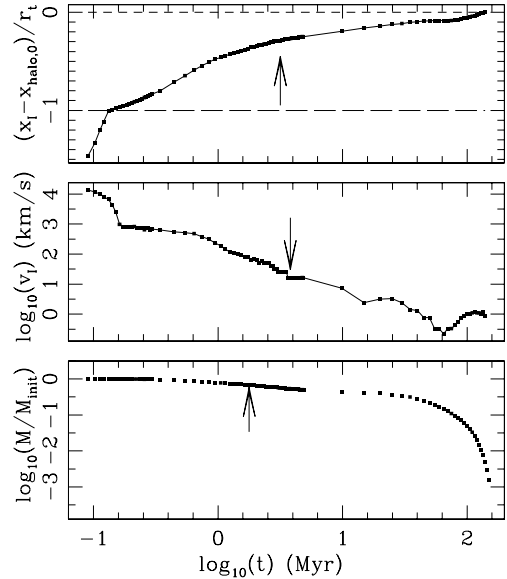}
\caption{Evolution of the I-front from R-type to 
D-type inside the minihalo for BB 5e4 case. (upper panel) 
Position $x_I$ (relative to minihalo center, in units of the minihalo radius $r_{t}$ 
at $t=0$) and (middle panel) velocity 
$v_I$ of the I-front as it travels toward and across the minihalo. 
The positions of the halo boundary (long-dashed line) and centre (short-dashed
line) are also indicated.
(lower panel) Fraction of mass $M_{\rm init}$, the mass which is 
initially inside the minihalo when the intergalactic I-front overtakes it, 
which remains neutral versus time $t$.
The arrows in the top and bottom panels mark 
the analytically-calculated width of the ISL for zero  
impact parameter and the mass fraction inside the ISL, respectively, 
while the arrow on middle panel marks the moment when I-front becomes R-critical, defined by the condition $v_I=2c_{s,I,2}$, the onset of transition
from R-type to D-type, where $c_{s,I}$ is isothermal sound speed of ionized,
post-front gas.}
\label{front_R_to_D_BB5e4}
\end{figure}

\begin{figure*}
\includegraphics[width=3.4in]{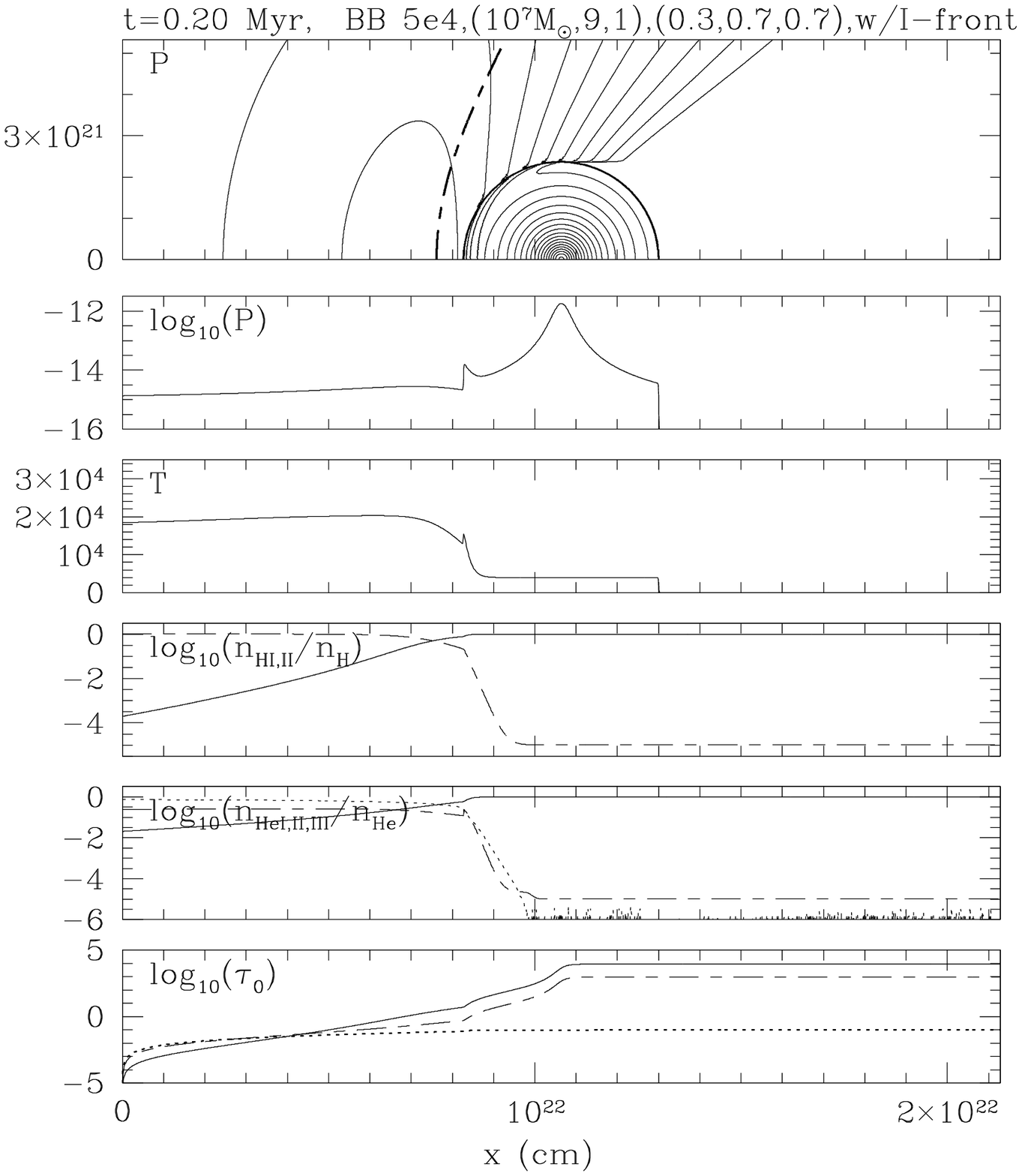}
\includegraphics[width=3.4in]{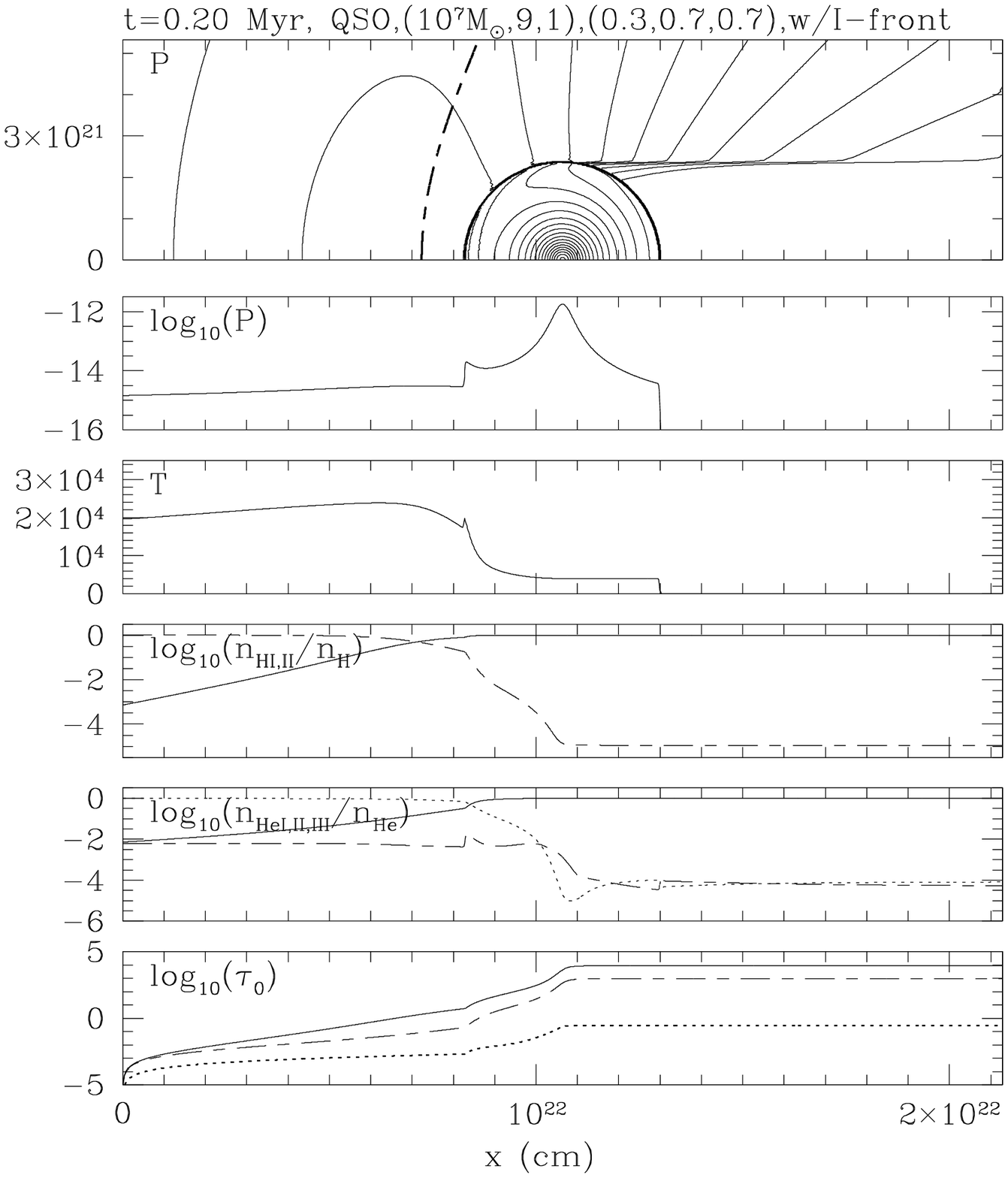}
\caption{Weak, R-type phase of I-front in IGM just about to
overtake a $10^7M_\odot$ minihalo at $z=9$, $t=0.2\,\rm Myr$ after
the I-front entered the box at left-hand side (ionizing source is
located far to the left of
computational box along the $x$-axis): 
(a) (left) BB 5e4 case and (b) (right) QSO case. 
Panels show the same quantities as in Figure~\ref{slices_0.5Myr_QSO_uniform}.
The current position of the I-front (50 \% ionization level of hydrogen) is 
indicated on the uppermost panel with the dashed line.}
\label{slices_0.2Myr_QSO}
\end{figure*}

\begin{figure}
\includegraphics[width=3.5in]{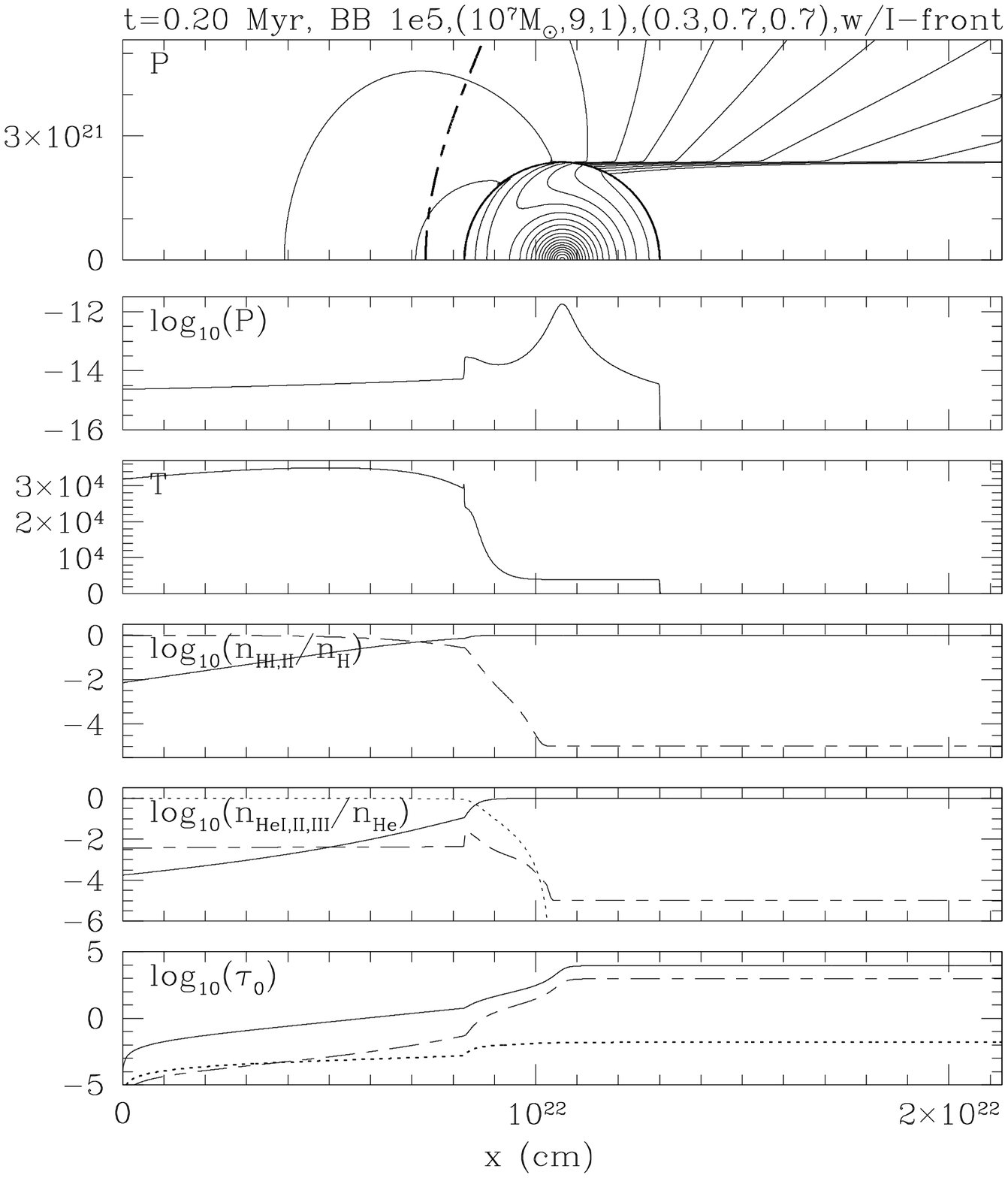}
\caption{Same as Figure~\ref{slices_0.2Myr_QSO},
but for BB 1e5 case.}
\label{slices_0.2Myr_BB1e5}
\end{figure}

Radiative transfer of the ionizing photons is treated explicitly by taking 
into account the bound-free opacity of H and He in the photoionization rates 
and heating rates as explained in detail in \cite{MRCLBSN98}. In order to 
speed up the radiative transfer calculation, the optical depths are approximated as 
described in \citet{TTetal83}, as follows:
\be
\tau_{\nu}=\left\{\begin{array}{lll}
	\tau_{\nu,A}  
		&\textrm{for $\nu_{\rm th, H\, I}\leq\nu\leq\nu_{\rm th, He\, I}$}\\
	\tau_{\nu,B}&\textrm{for $\nu_{\rm th, He\, I}\leq\nu\leq\nu_{\rm th, He\, II}$}\\
	\tau_{\nu,C} &\textrm{for $\nu_{\rm th, He\, II}\leq\nu$},
	\end{array}\right.
\ee
where 
\be
\tau_{\nu,A}=
	\tau_{\nu,\rm H I}\left(\frac{\nu}{\nu_{\rm th,H I}}\right)^{-2.8},
\ee
\ba
\lefteqn{\tau_{\nu,B}=\left[(0.63)^{1.7}\tau_{\nu,\rm H I}+{}\right.}\nonumber\\
	&&{}\left.+(1.81)^{1.7}{\tau_{\nu,\rm He I}}\right]
	\left(\frac{\nu}{\nu_{\rm th,H I}}\right)^{-1.7},
\ea
and
\ba
\lefteqn{\tau_{\nu,C}=\left[{\tau_{\nu,\rm H I}}
	+(2.73)^{2.8}{\tau_{\nu,\rm He I}}+{}\right.}\nonumber\\
	&&{}\left.+4^{2.8}{\tau_{\nu,\rm He II}}\right]
	\left(\frac{\nu}{\nu_{\rm th,H I}}\right)^{-2.8}.
\ea
Here $\tau_{\nu,i,z}=\sigma_{\rm \nu,th,i,z}N_i$ are the
optical depths at the respective 
ionization thresholds of H I, He I and He II. This allows us to precompute
the integrals over frequency involved in calculating
photoionization and photoheating rates for each assumed
source spectrum to make an extensive look-up table for these rates
as functions of the threshold optical depths, for
use in the rapid computation of these rates during each run
by interpolation between entries in the look-up table.
In order to speed up the
calculations further, the optical depths are normally re-calculated
once every 5 time steps. 

\subsection{Simulation parameters}

Our box size in $(r,x)$ dimensions is 3.45 kpc$\times$6.9 kpc, except for the uniform IGM
simulations in \S~\ref{I_front_sect} and \S~\ref{I-front_IGM_sect}, where
we adopted larger boxes, as described there. The resolution at the finest grid 
level (i.e. fully-refined) is 1024$\times$2048 cells, again with the exception of the 
simulations in \S~\ref{I_front_sect} and \S~\ref{I-front_IGM_sect},
where such high resolution was not required, but a longer propagation time was necessary, 
so we used resolutions of 64 $\times$ 8192 and 512$\times$1024, respectively.

The gas consists of hydrogen and helium in primordial abundance (24.2 \% He by 
mass), with a small trace of metals (C, N, O, Ne, and S) at $10^{-3}$ $Z_\odot$.
The initial ionizing flux we adopt is $F_0=1$, except in \S~\ref{I_front_sect}
where we use $F_0=4$. Each photoevaporation simulation took from several days 
to 1 week on a 2 GHz Athlon processor.

\subsection{Test problem: R-type I-front in uniform IGM}

\label{I_front_sect}
In order to  track properly the supersonic, weak, R-type I-fronts 
which sweep into and
around our minihalos, we defined an ``ionization time-step'' 
related to the time-scale 
for ionizing hydrogen, as described above in \S~\ref{method}. 
We tested the scheme and adjusted the value of 
$\epsilon_{\rm ion}$ by performing
a simulation of an ionization front propagating through a uniform density 
IGM in a long and thin (44 kpc $\times$ 5.64 Mpc), quasi-1D simulation box
at resolution 64 $\times$ 8192. In this case, 
the gas is pure hydrogen, so there is a 
well-known exact analytical solution for the I-front propagation \citep{SG87}. 
The density of the gas is equal to the mean cosmic baryon density at $z=9$.
The source is located on the symmetry axis, 0.5 Mpc to the left of the computational box,
with a flux as measured at the left boundary of the box equal to $F_0=4$. 
At time $t=0$, the box is neutral.
We identify the location of the I-front as the position at which 
the ionized fraction is $x=0.5$. The result is plotted in Figure~6.
By experiment, we found that $\epsilon_{\rm ion}=0.05$ 
is sufficient to
obtain the correct velocity for the R-type I-front to better than 5 \%, 
as shown in Figure~\ref{I_front_speed_noHe}. Using larger $\epsilon_{\rm ion}$ 
can lead to an underestimate of the velocity by a significant amount.

\section{Results}
\label{results_sect}

\subsection{Structure of global I-front in the IGM during 
reionization: the weak, 
R-type phase}
\label{I-front_IGM_sect}

To study the detailed temperature and ionization structure of the 
global I-front during reionization in its weak, R-type phase
and the dependence of these on the source spectrum, 
we performed a set of three simulations of an I-front in the uniform
IGM at $z=9$, by ``zooming in'' with a smaller box 
(34.5 kpc $\times$ 69 kpc) and better
length resolution (512$\times$1024, fully-refined) than were used in the test
problem in the previous section. 
The source in each case produced 
a flux $F_0=1$ at the left boundary of the simulation box, with a
different spectrum 
for each simulation. The results are shown in 
Figures~\ref{slices_0.5Myr_QSO_uniform}--\ref{cno_1.3Myr_BB1e5}. 
As expected, for the cases with 
harder spectra (i.e. QSO, BB 1e5), the I-front is 
broader than in the case BB 5e4, due to the more deeply 
penetrating hard photons of the former cases. 
There is significant pre-heating and pre-ionization in the former cases,
ahead of the point inside the I-front at which the H
neutral fraction is 0.5. In all cases, the ionization of He is 
predominantly from He I on the neutral side to He III 
on the ionized side, with only 
a small fraction of He II. The position of the hydrogen I-front 
($x_{\rm H}=0.5$) is nearly
coincident with the corresponding point for helium ($x_{\rm He}=0.5$) 
for the softer 
BB 5e4 spectrum, while for the harder QSO and BB 1e5 spectra the 
He I-front is more 
advanced by $\sim7$ kpc. Another important difference to note is 
that the post-front 
temperature is $\sim 2\times10^4 K$ in the BB 5e4 case, but 
$\sim 3\times10^4 K$ in 
the QSO case, and even higher, $\sim4\times10^4 K$ in the BB 1e5 case. 
The ionization states of the trace of 
heavier elements also differ significantly for different source spectra, 
as shown in 
Figures~\ref{cno_1.3Myr_QSO} and \ref{cno_1.3Myr_BB1e5}. The BB 5e4 spectrum
ionizes C, N, and O mostly to C~III, N~III, and O~III, respectively, with 
only small fractions, of order 10 \%, of C~IV, N~IV, and O~IV, while the BB 1e5
spectrum ionizes these species mostly up to ionization stage IV, with 
a notable fraction ($>10\%$) of ionization stage V, and tiny fractions of
O~VI and N~VI. Finally, the hard QSO spectrum is able also to produce 
fractions as high as a few per cent for O~VI and N~VI, while the 
bulk of the metals reside in the ionization stages IV and V.

\subsection{I-front encounters the minihalo: weak, R-type phase inside minihalo}
\label{trapping_sect}

As the weak, R-type I-front propagates towards the minihalo, 
it encounters a steeply 
rising density, first the infall region and 
then the virialized region of the minihalo, 
itself, and its velocity
drops precipitously. Figure~\ref{front_R_to_D_BB5e4} shows the evolution 
of the I-front position $x_I$ (upper panel), its velocity $v_I$ 
(middle panel) and the mass fraction $M/M_{\rm init}$ of the gas 
initially within the original hydrostatic sphere (bottom panel) which
remains neutral over time in the BB 5e4 case. 
The initial speed of the I-front as it 
enters the box is about 10,000 $\rm km\, s^{-1}$, dropping to 
$\sim 1000\, \rm km\, s^{-1}$ by the time the front crosses 
the virial radius of the halo. 
As a result, the I-front is initially a weak, R-type front even after
it enters the minihalo.
It takes about 5 Myr for the front to slow to the 
R-critical front speed $v_R\cong2c_{s,2}\approx 20\,\rm km\, s^{-1}$ 
(where the isothermal, post-front sound speed 
$c_{s,2}\cong11\,\rm km\, s^{-1}$) 
(marked on the plot by an arrow) after which a compressive shock 
must form to 
lead the I-front and further decelerate it, so as to transform
it to a D-type front. On the top panel, we have 
indicated (by another arrow) the position of the
inverse Str\"omgren surface at zero impact parameter, as calculated 
in \S~\ref{strom_sect}. The two arrows almost coincide, confirming our expectations 
of approximate correspondence between the ISL (in the static analytical 
calculation) and the surface on which the I-front is R-critical,
close to the moment of conversion of the I-front to D-type (also 
indicated in Fig.~\ref{ISL}). This correspondence is not perfect, 
however, due to 
the significant approximations made in the ISL calculation. For example, due to 
the centrally-concentrated halo density profile the I-front transition to 
D-type occurs in fact at different times for each impact parameter, leading 
to a continuously evolving I-front shape different from the shapes shown in 
Figure~\ref{ISL_shape}.

Figures~\ref{slices_0.2Myr_QSO}~and~\ref{slices_0.2Myr_BB1e5} 
show the time-slice at 0.2 Myr, shortly before the weak, R-type
I-front enters the minihalo. The instantaneous position of the I-front 
is indicated on the top panel which shows the contours of pressure. 
The I-front moves faster further away from the halo due to the lower 
density in the infall profile
and has already significantly slowed down closer to the axis. 
The harder photons 
clearly penetrate significantly deeper in the QSO and BB 1e5 cases
and the halo shadow is already clearly outlined at that time by
the pressure contours, even though the 
front itself has not passed the halo yet. These same hard photons also 
start to heat the halo on the source side, while deeper into the halo and 
in the shadow region behind it virtually no photons can penetrate due to the 
extremely high optical depth, $\tau\sim10^4$ (bottom panels).
While the shadow is mostly neutral in all cases, the QSO case produces a
$10^{-4}$ fraction of He II and He III there.

\begin{figure*}
\includegraphics[width=3.4in]{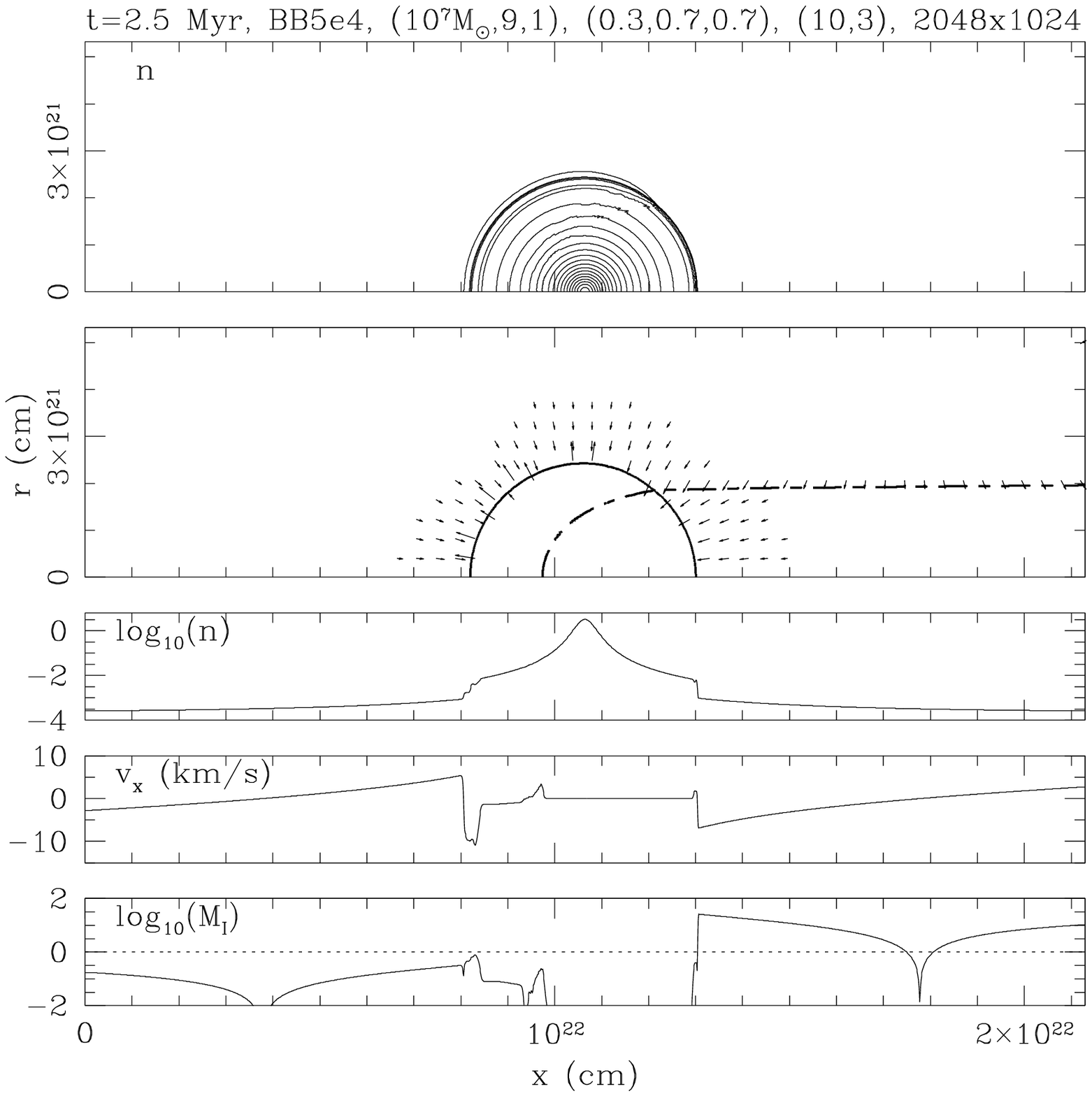}
\includegraphics[width=3.4in]{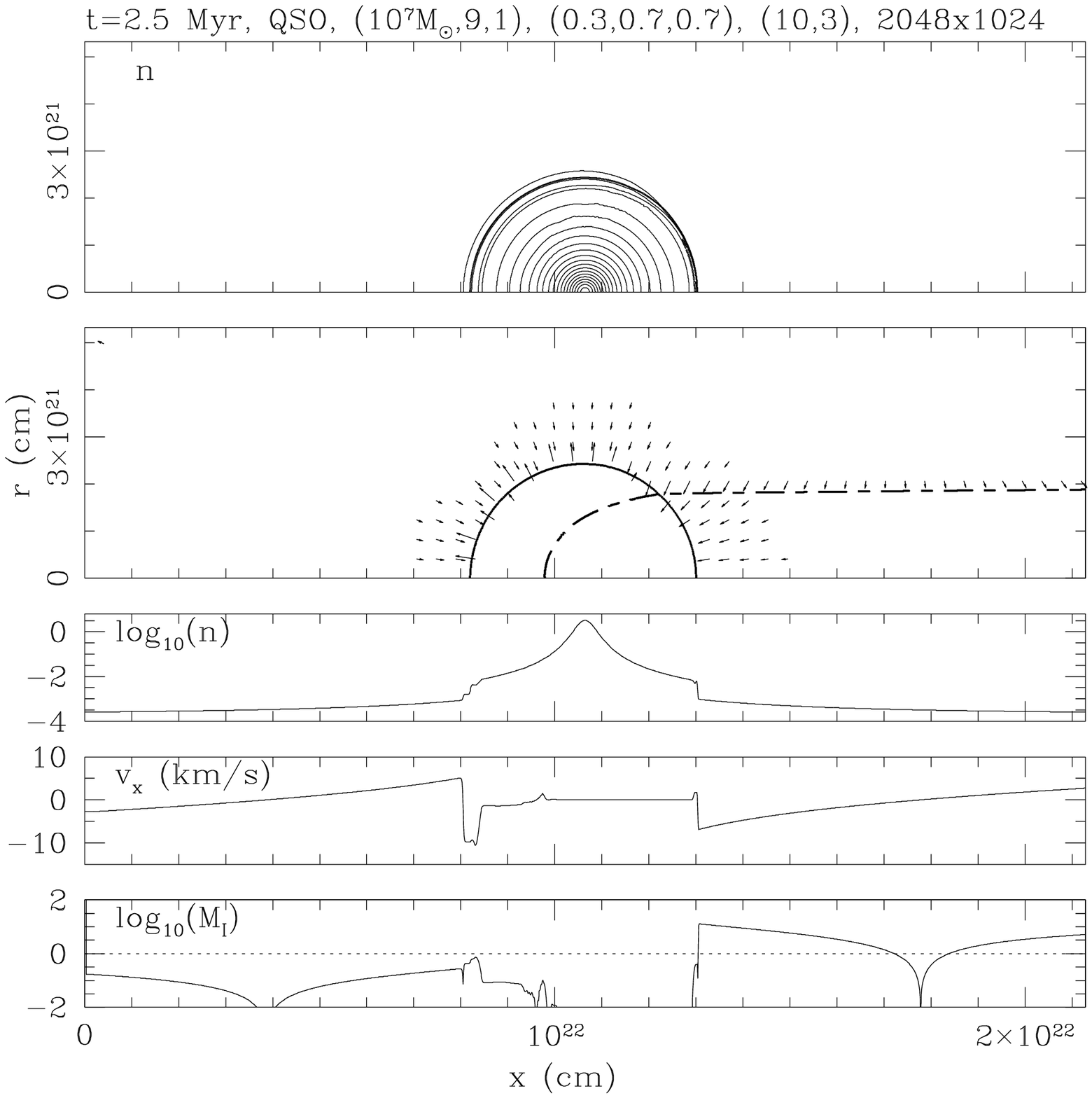}
\caption{Weak, R-type phase of I-front inside the minihalo.
Same as Figure~5, but for later 
time-slice, $t=2.5$ Myr after I-front enters the box:
(a) (left) BB 5e4 case and (b) (right) QSO case.
A velocity arrow of length equal to the spacing between arrows has velocity
$10\, {\rm km\, s^{-1}}$; minimum velocities plotted are
$\rm 3\,km\,s^{-1}$. Solid line shows current extent of gas
initially inside minihalo at $z=9$.
Dashed line indicates the current position of the I-front 
(50\% H-ionization contour).}
\label{vel_arrows_2.5Myr_QSO}
\end{figure*}

\begin{figure}
\includegraphics[width=3.4in]{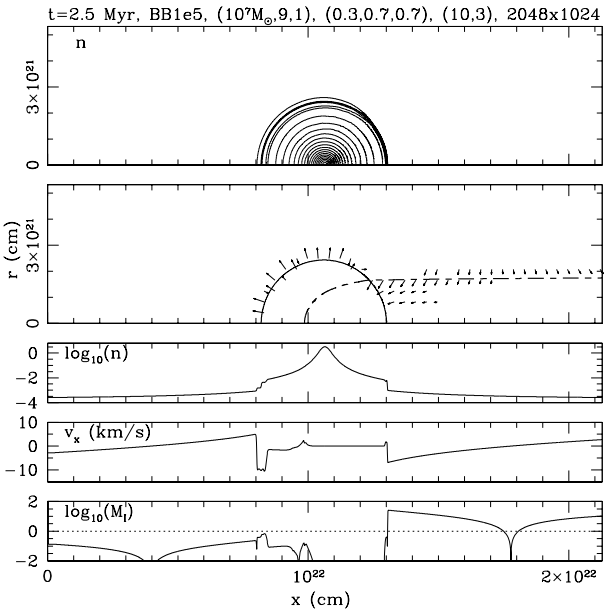}
\caption{Same as in Figure~\ref{vel_arrows_2.5Myr_QSO}, but for BB 1e5 case.}
\label{vel_arrows_2.5Myr_BB1e5}
\end{figure}

\begin{figure*}
\includegraphics[width=3.4in]{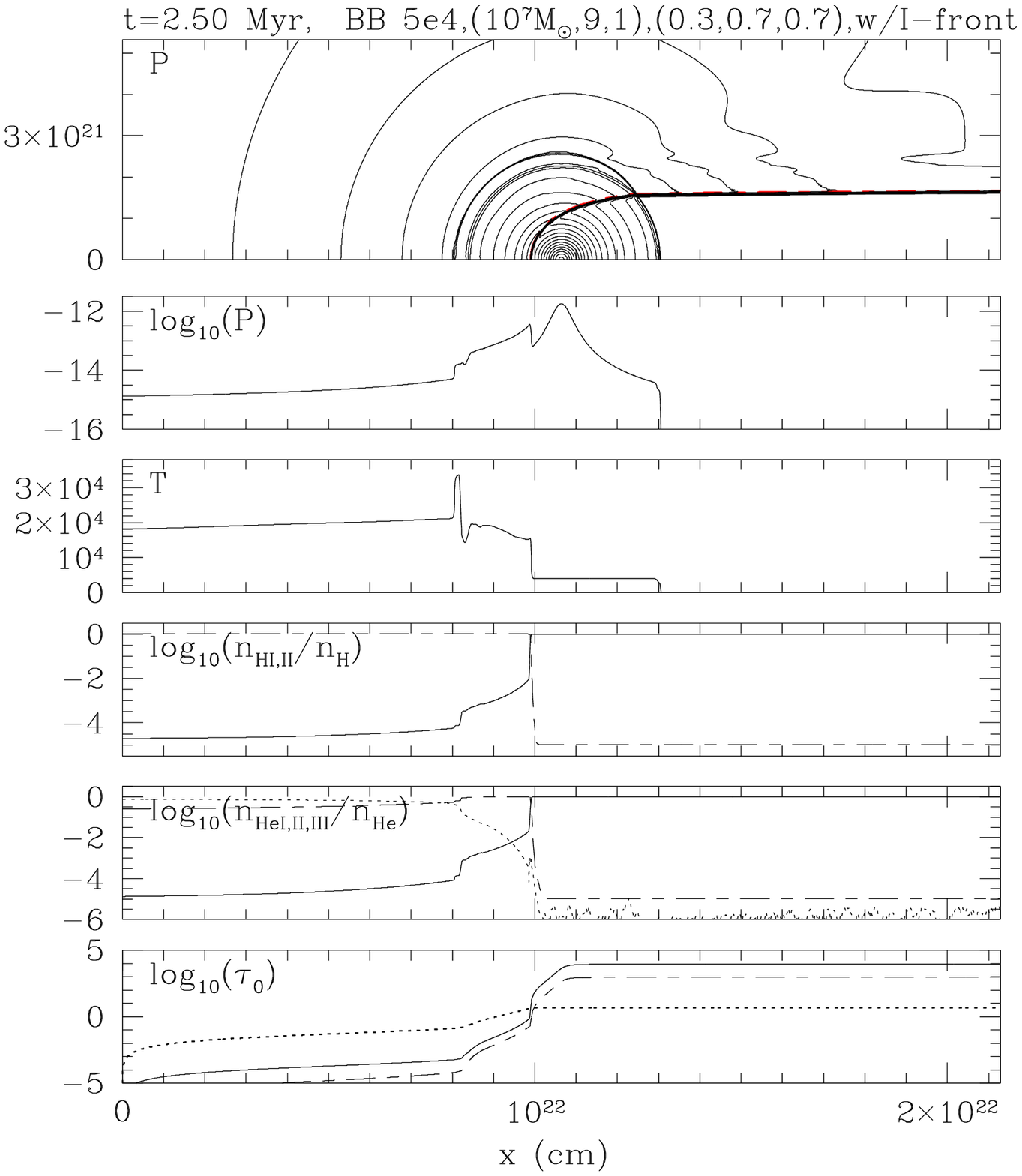}
\includegraphics[width=3.4in]{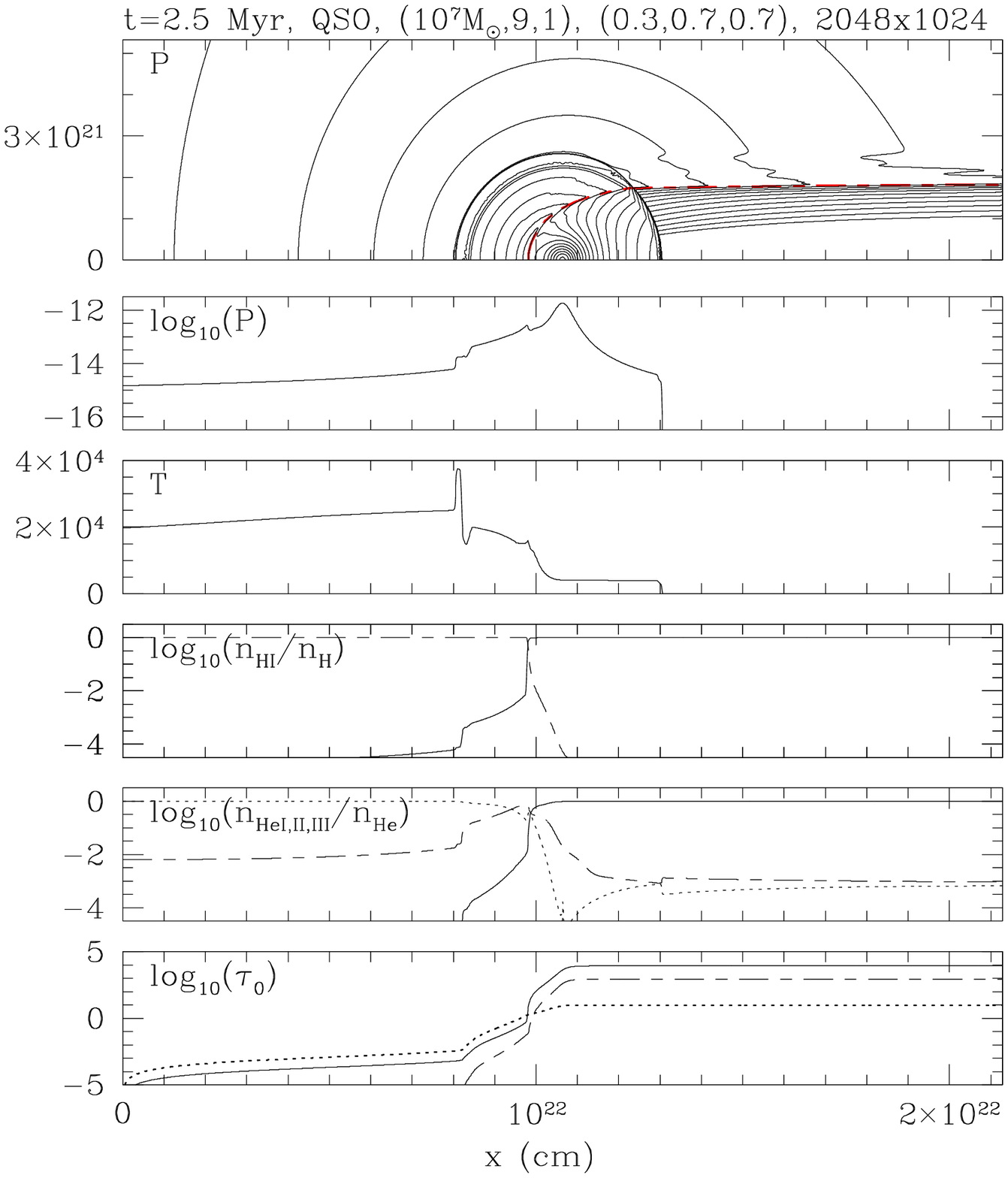}
\caption{Weak, R-type phase of I-front inside the minihalo.
Same as Figure~\ref{slices_0.2Myr_QSO}, but for $t=2.5\,\rm Myr$.}  
\label{slices_2.5Myr_BB5e4}
\end{figure*}

\begin{figure}
\includegraphics[width=3.4in]{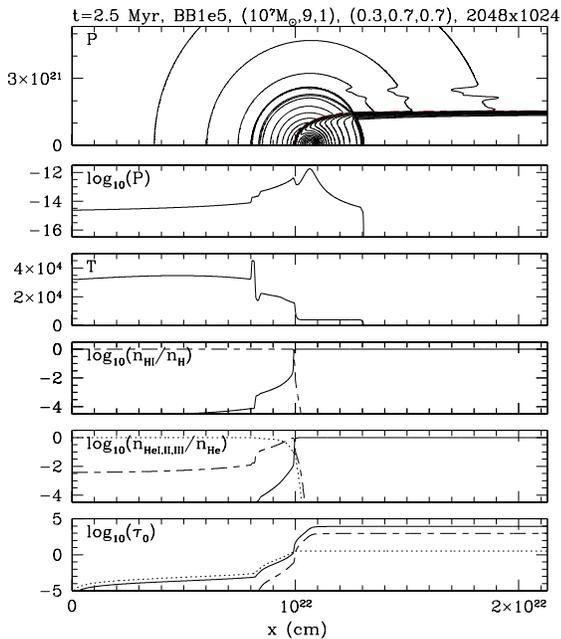}
\caption{Same as in Figure~\ref{slices_2.5Myr_BB5e4}, but for BB 1e5 case.}
\label{slices_2.5Myr_BB1e5}
\end{figure}

In Figures~\ref{vel_arrows_2.5Myr_QSO}-\ref{slices_2.5Myr_BB1e5} we show the 
structure of the flow at $t=2.5$ Myr, when the I-front is inside the
minihalo but still a weak, R-type front,
before much hydrodynamical back-reaction has begun.
Figure~\ref{vel_arrows_2.5Myr_QSO} and
\ref{vel_arrows_2.5Myr_BB1e5} show that the gas which was initially
infalling on the source side next to 
the halo is already reversing its velocity, however,
and starting to form a shock which will sweep the 
IGM outward. On the other hand, on the shadow side of the halo, the gas infall 
continues uninterrupted. Also evident is some squeezing of the 
gas in the shadow 
due to the much higher pressure of the ionized gas outside it 
after the passage of 
the fast R-type I-front there.

\subsection{I-front trapping: R-critical phase and the transition from
R-type to D-type}

\begin{figure*}
\includegraphics[width=3.4in]{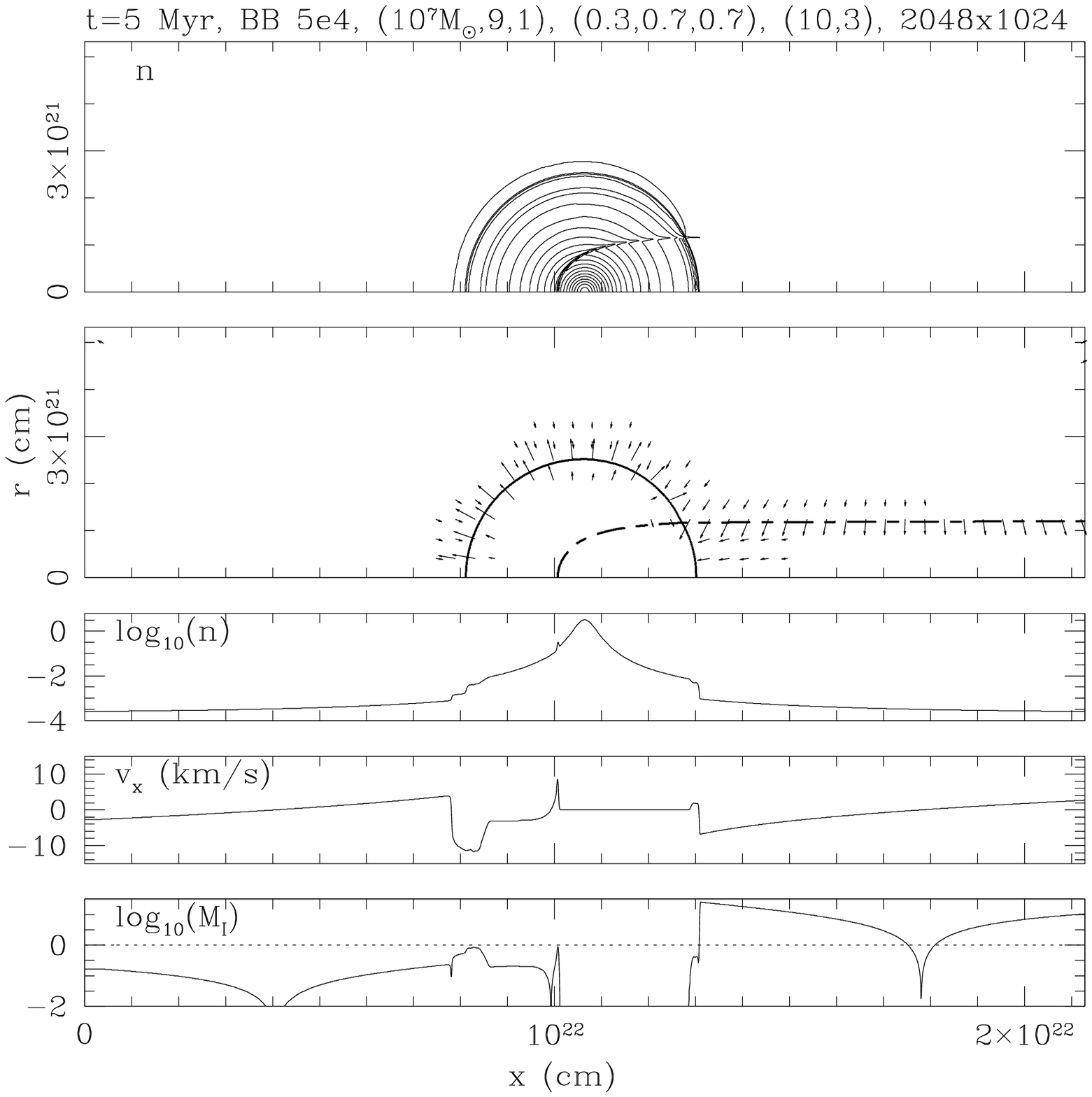}
\includegraphics[width=3.4in]{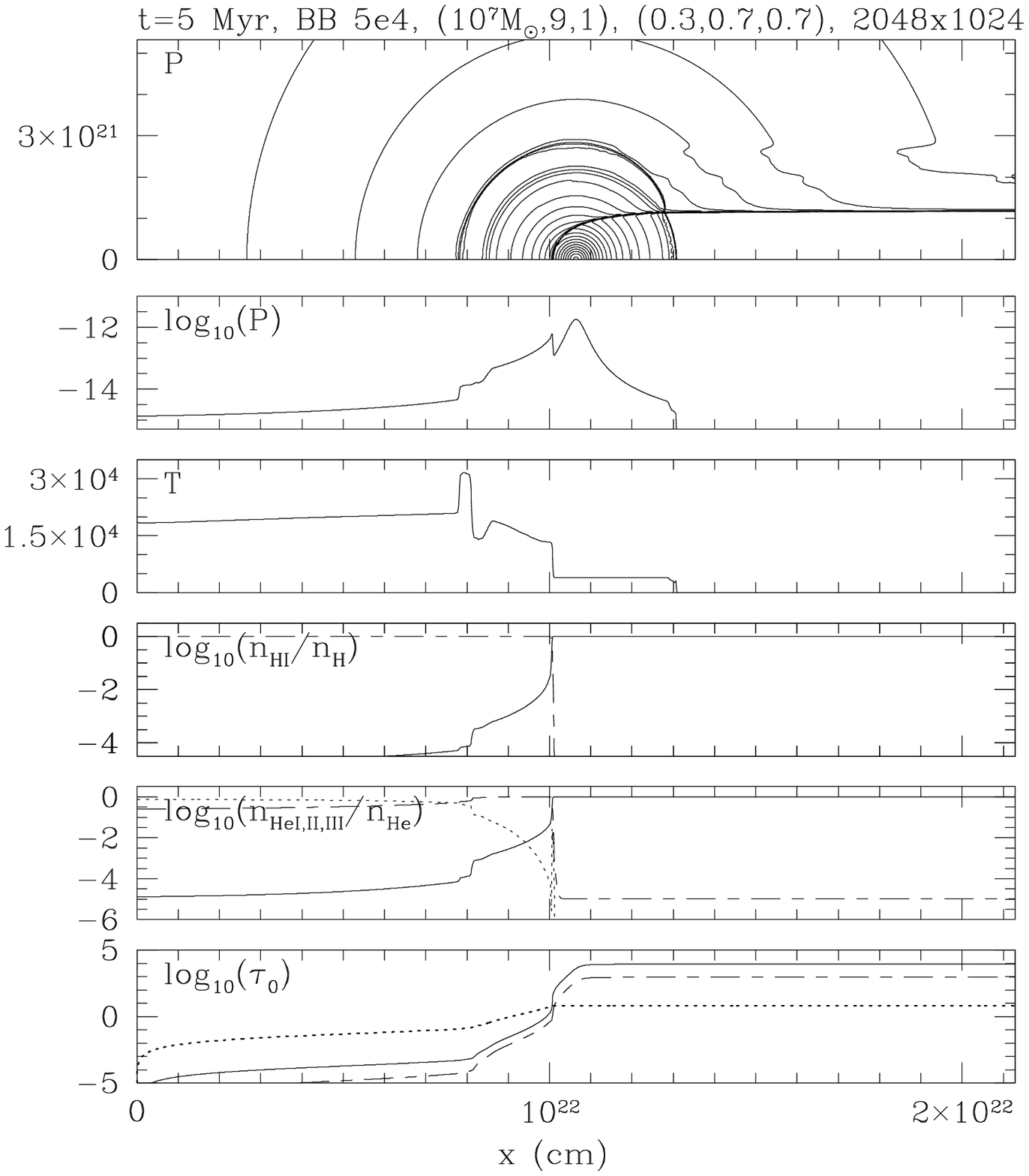}
\caption{R-critical phase of I-front just before transition from R-type
to D-type. (a) (left) Panels show same quantities as
in Fig.~\ref{vel_arrows_2.5Myr_QSO}, but for $t=5\,\rm Myr$ and
BB5e4 case; (b) (right) Same quantities as in Fig.~\ref{slices_0.2Myr_QSO},
but for the $t=5\,\rm Myr$ and BB5e4 case.}
\label{5Myr_BB5e4}
\end{figure*}

\begin{figure*}
\includegraphics[width=3.4in]{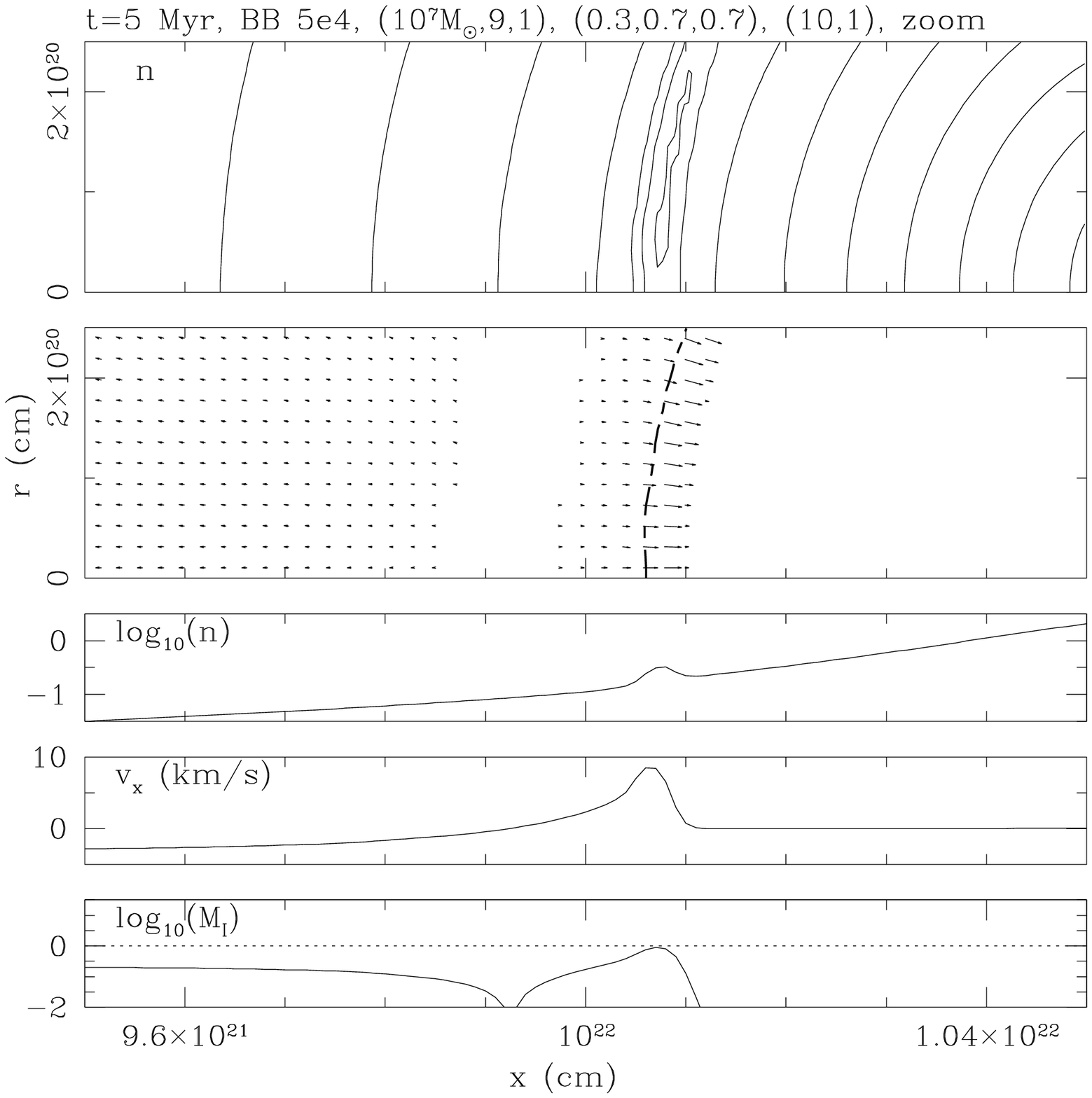}
\includegraphics[width=3.4in]{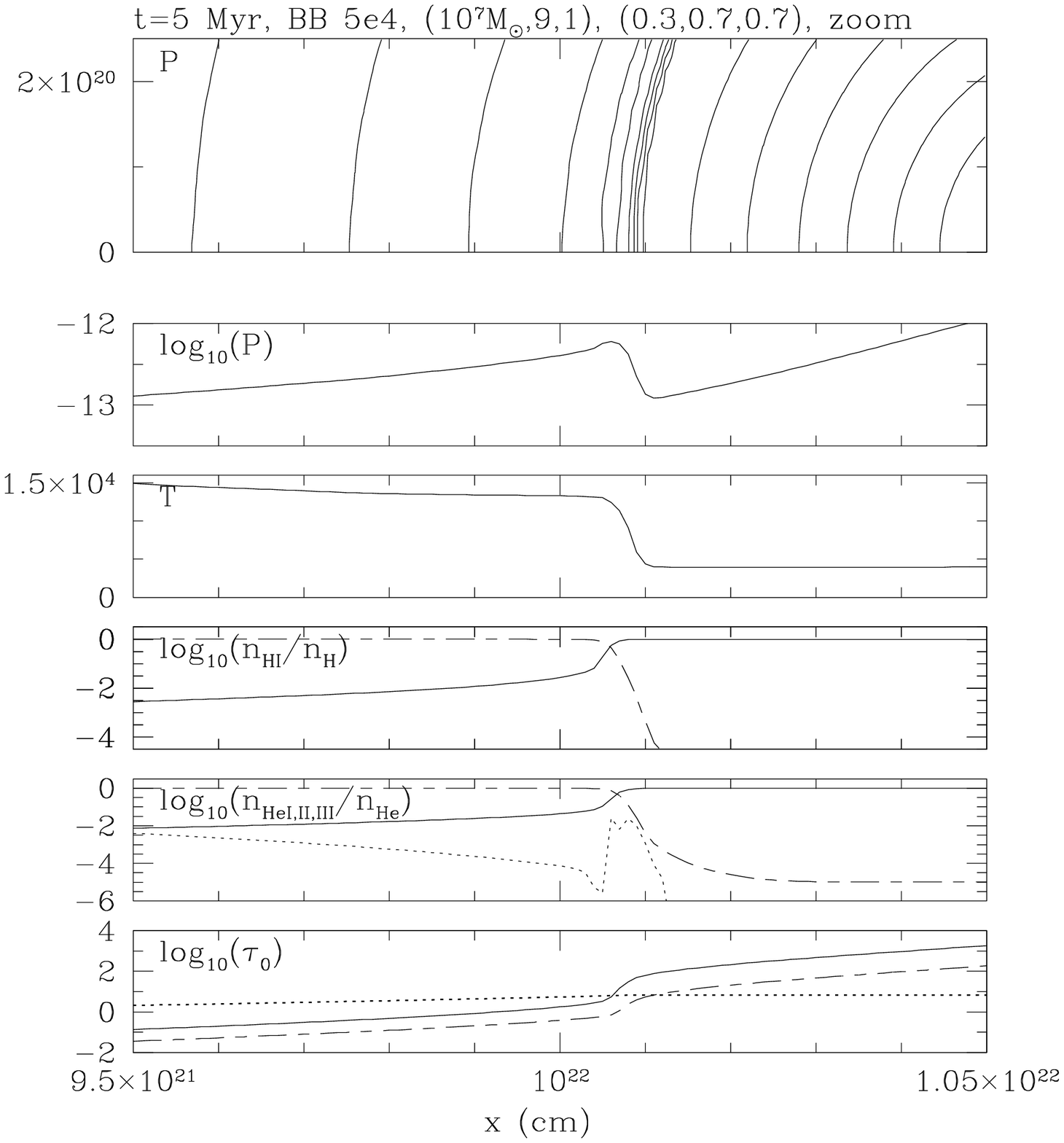}
\caption{Same as Fig.~\ref{5Myr_BB5e4}, but ``zoom-in'' to
enlarge the view of R-critical I-front along the $x$-axis.}
\label{5Myr_zoom_BB5e4}
\end{figure*}

\begin{figure*}
\includegraphics[width=3.4in]{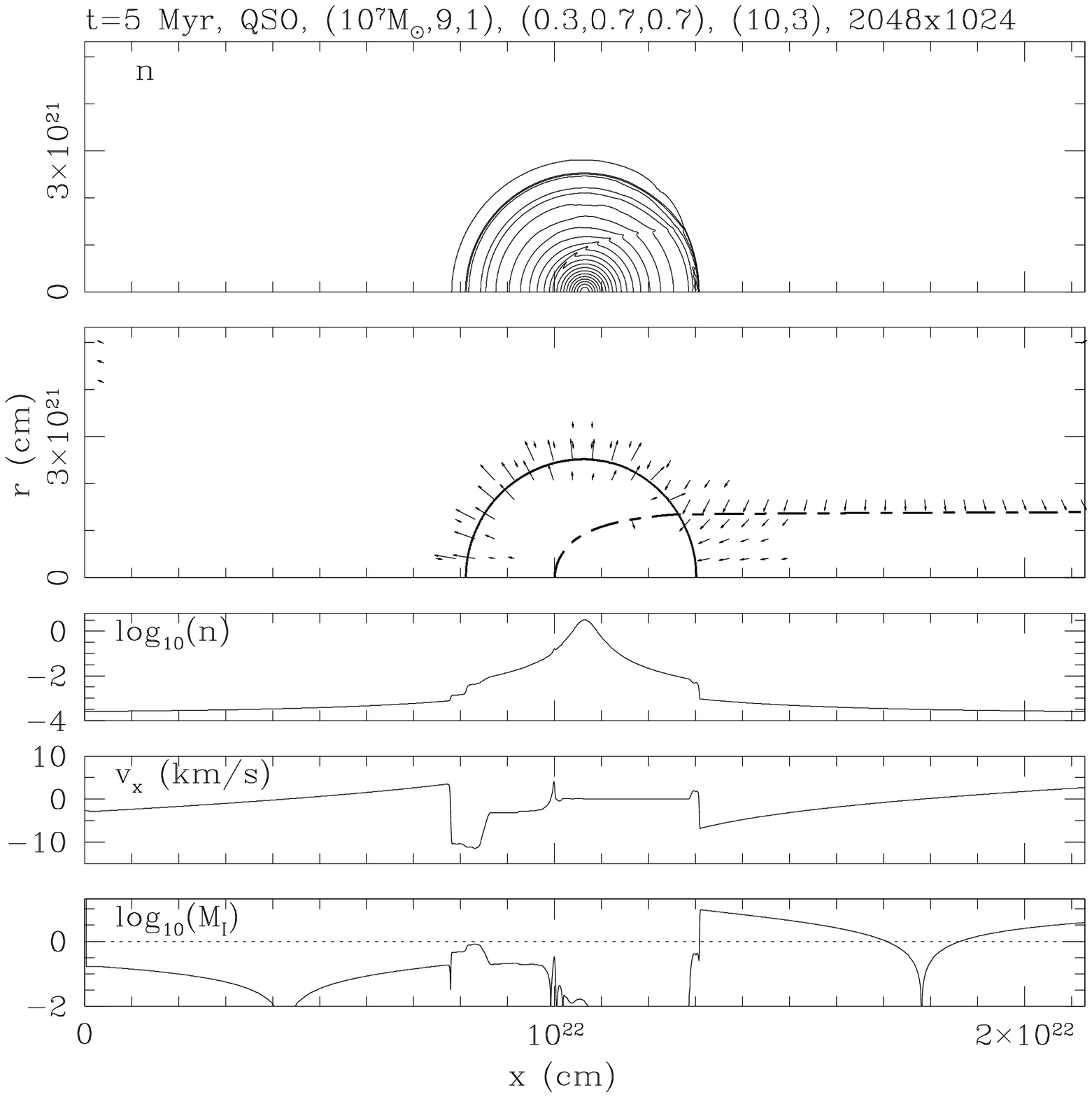}
\includegraphics[width=3.4in]{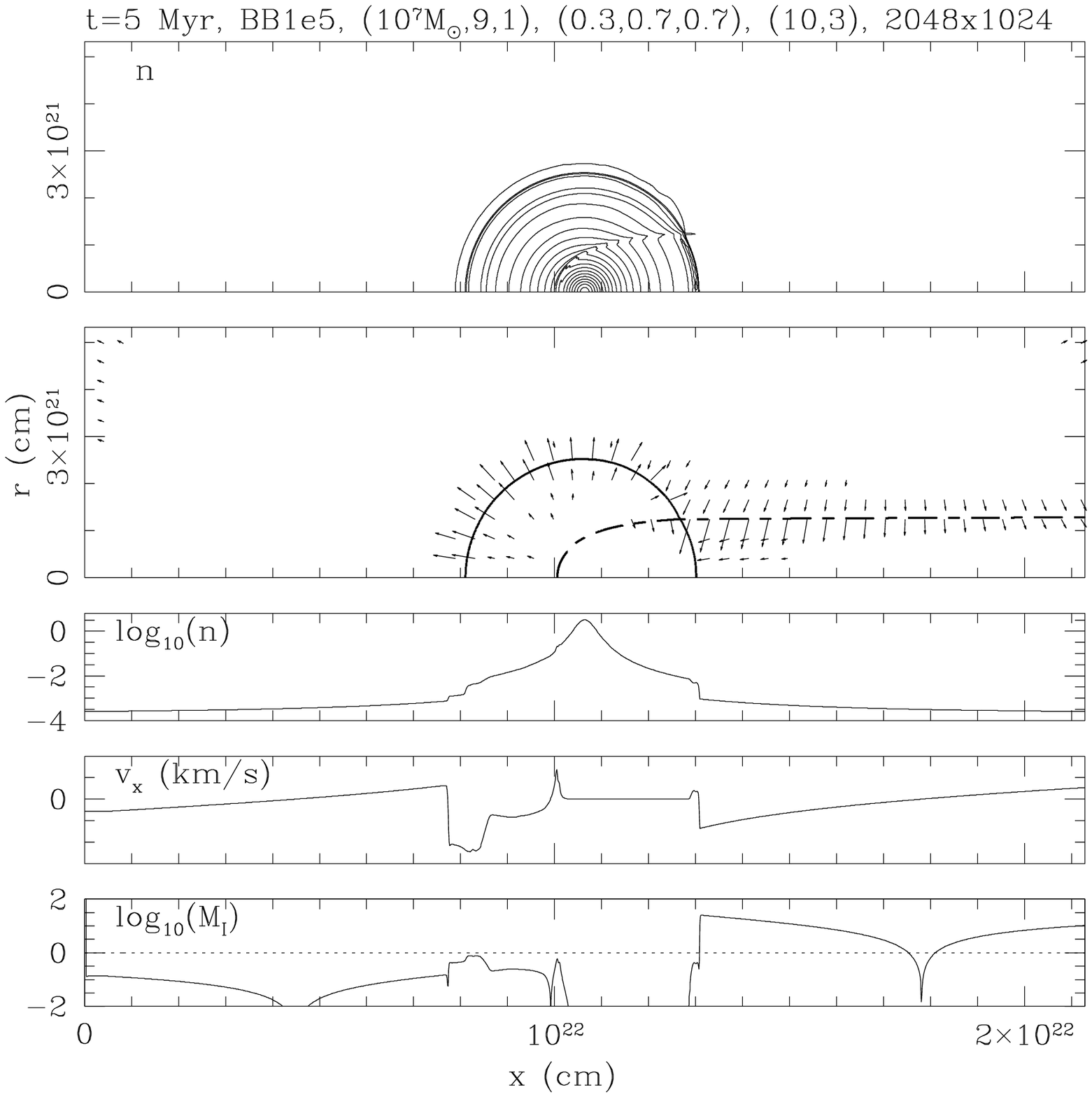}
\caption{R-critical phase of I-front just before transition from R-type
to D-type. Panels show same quantities as in Fig.~\ref{vel_arrows_2.5Myr_QSO},
but for $5\,\rm Myr$: (a) (left) QSO case; (b) BB1e5 case.}
\label{vel_arrows_5Myr}
\end{figure*}

\begin{figure*}
\includegraphics[width=3.4in]{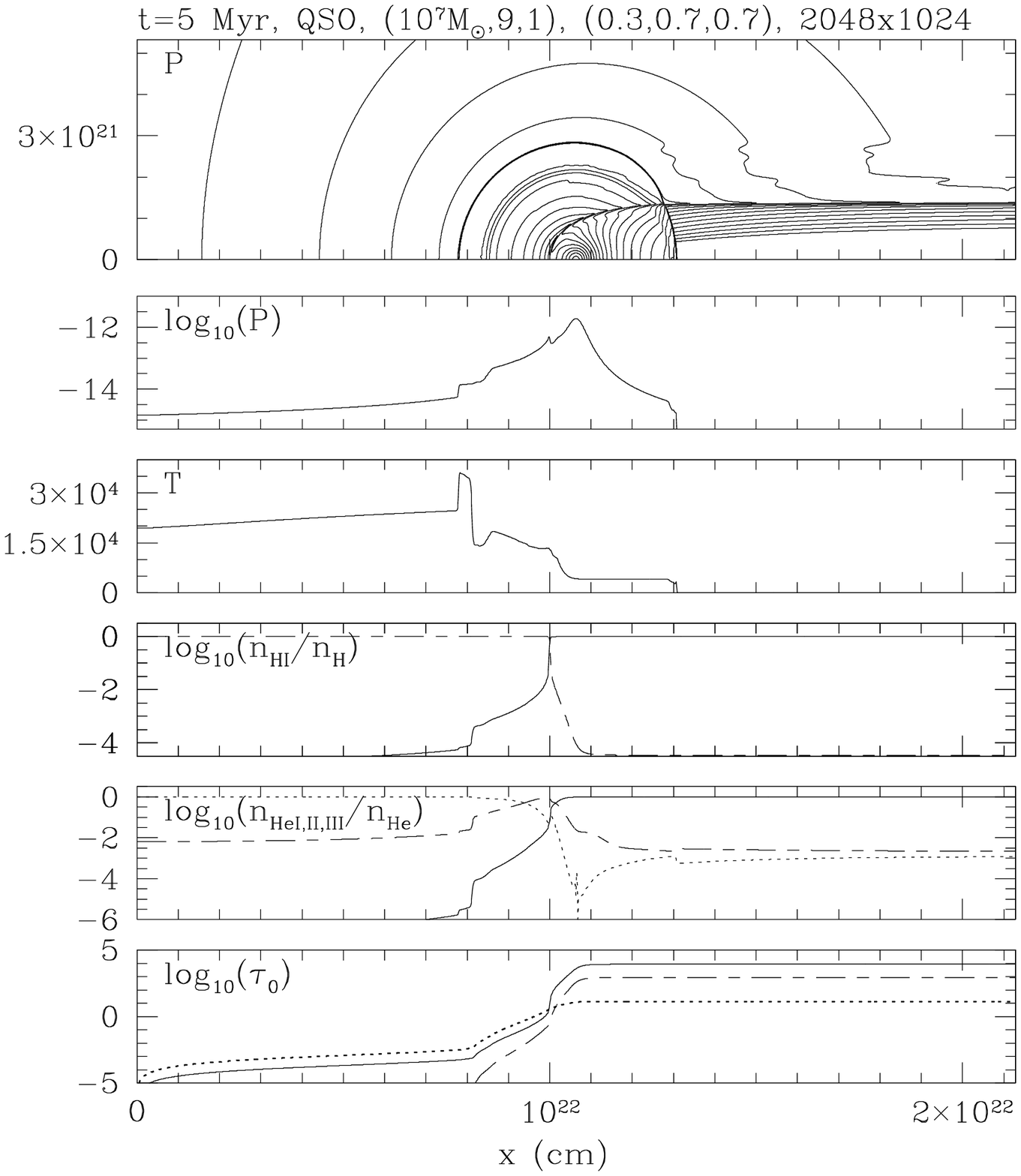}
\includegraphics[width=3.4in]{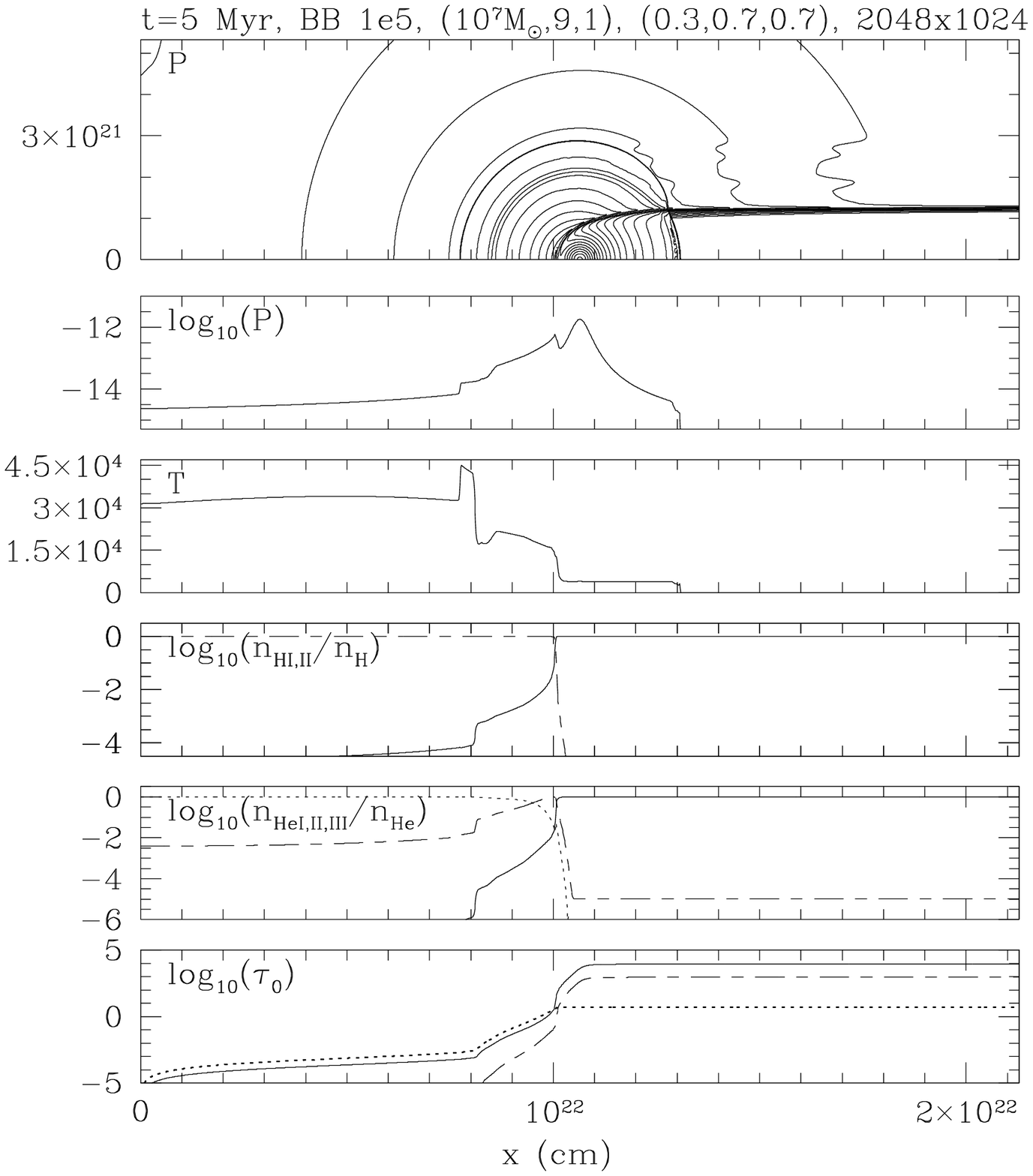}
\caption{R-critical phase of I-front just before transition from R-type
to D-type. Panels show same quantities as in Fig.~\ref{slices_0.2Myr_QSO},
except for $t=5\,\rm Myr$: (a) (left) QSO case; (b) BB1e5 case.}
\label{slices_5Myr}
\end{figure*}

\begin{figure*}
\includegraphics[width=3.4in]{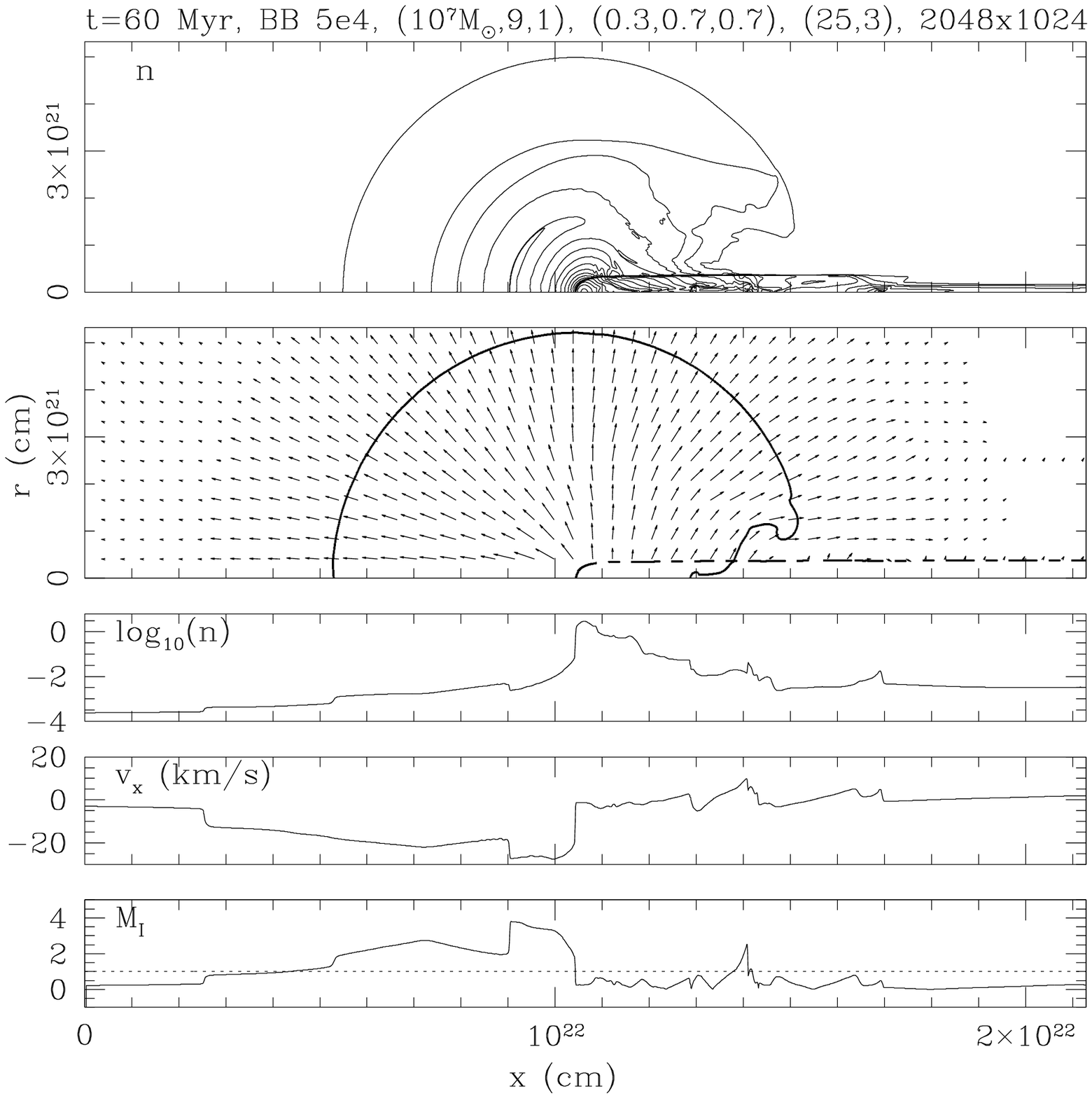}
\includegraphics[width=3.4in]{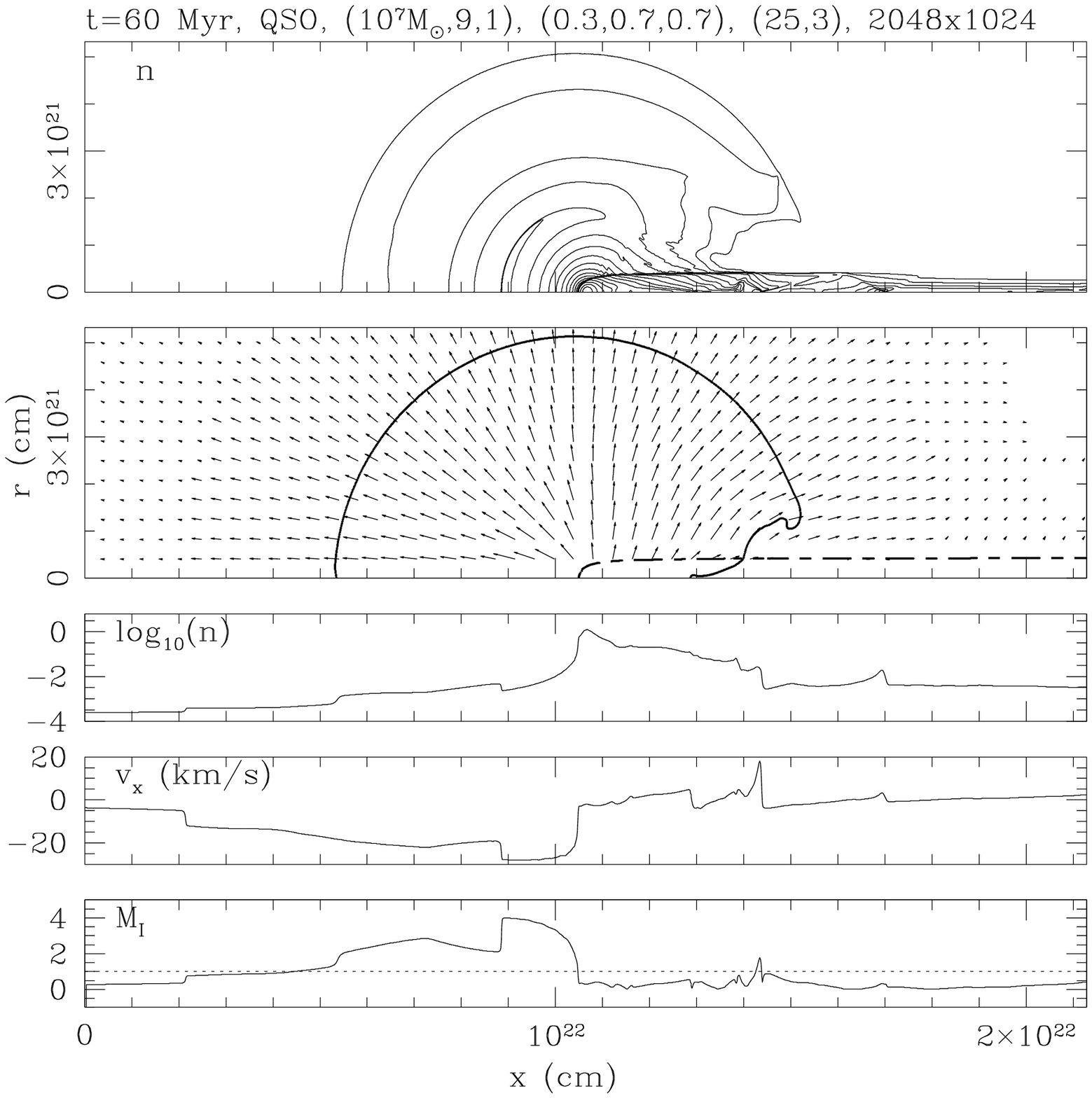}
\caption{D-type phase of I-front: Photoevaporative Flow
60 Myr after I-front
overtakes the minihalo:
(a) (left) BB 5e4 case and (b) (right) QSO case.
Same quantities plotted as in Fig.~\ref{vel_arrows_2.5Myr_QSO},
except $t=60\,\rm Myr$.}
\label{vel_arrows_60Myr_BB5e4}
\end{figure*}

The I-front finally slows to become an R-critical
front by 5 Myr. We show the results for this time-slice for all three
source spectra in 
Figures~\ref{5Myr_BB5e4}--\ref{vel_arrows_60Myr_BB5e4}.
To illustrate clearly that this is the R-critical phase, we focus
on the Pop~II stellar case BB5e4 in 
Figure~\ref{5Myr_BB5e4}
and ``zoom in'' on the region around the I-front along the $x$-axis in
Figure~\ref{5Myr_zoom_BB5e4}.
All of the important properties of an R-critical I-front are evident
in Figure~\ref{5Myr_zoom_BB5e4}.
The I-front velocity at this epoch is $v_I\approx16\,\rm km\,s^{-1}$ which is
not far from the estimated R-critical
value of $v_R\approx20\,\rm km\,s^{-1}$, calculated
based upon the immediate post-front temperature of $T_2\approx10,000\,\rm K$,
the pre-front $T_1=4,000\,\rm K$ (i.e. the neutral side is undisturbed
minihalo gas at this virial $T$),
$\mu_1=1.11$ (i.e. 90\% neutral), and $\mu_2=0.63$ (i.e. 90\% ionized H),
which yield the post-front and pre-front isothermal sound speeds
$c_{s,I,2}=11\,\rm km\,s^{-1}$ and
$c_{s,I,1}=5.3\,\rm km\,s^{-1}$, respectively.
The numerical results show the predicted jump in density
by a factor of close to two across the R-critical I-front, 
required by the I-front jump conditions. 
In addition, in
the lab frame in which the neutral minihalo
gas on the pre-front side is at rest (i.e. the frame in which
we have plotted velocity and isothermal Mach number in 
Figures~\ref{5Myr_BB5e4} and \ref{5Myr_zoom_BB5e4}),
the ionized gas just behind the front should be moving toward
the neutral side at the isothermal
sound speed of the ionized post-front gas, or
about $11\,\rm km\,s^{-1}$, which our plots confirm (i.e. note that the
isothermal Mach number $M_I=1$ at that point in lower panel of
Fig.~\ref{5Myr_zoom_BB5e4}). It is interesting to note
that, while the gas on the ionized side of the I-front
is generally optically thin during most of the evolution, the optical
depth is not negligible at $t=5\,\rm Myr$. According to 
Figure~\ref{5Myr_zoom_BB5e4}, the H optical depth across the ionized, 
post-front layer is of order
a few, so the speed of the I-front
is lower than would be calculated using the unattenuated flux in the
I-front continuity jump condition.

This R-critical phase is the beginning of the transition to D-type,
which should occur shortly, thereafter, when a shock wave forms to
compress the gas ahead of the front, to enable the front velocity
to drop to the D-crit value. In this case, using the sound speeds at 
5 Myr reported above, the D-critical velocity is 
$v_D\approx c_{s,I,1}^2/(2c_{s,I,2})
\cong1.35\,{\rm km\,s^{-1}}\ll v_R$. As we can see from the plot of $v_I$
in Figure~\ref{front_R_to_D_BB5e4}, however, the transition from R-type to
D-type is quite extended in time, lasting 10's of Myr.
During that same time, the I-front advances into denser and denser neutral
minihalo gas, which also slows the front quite apart from any dynamical
compression which might lead the front.
In the meantime, from the R-critical phase onward, the 
hydrodynamical response to the I-front becomes more and more dramatic.

\begin{figure*}
\includegraphics[width=3.4in]{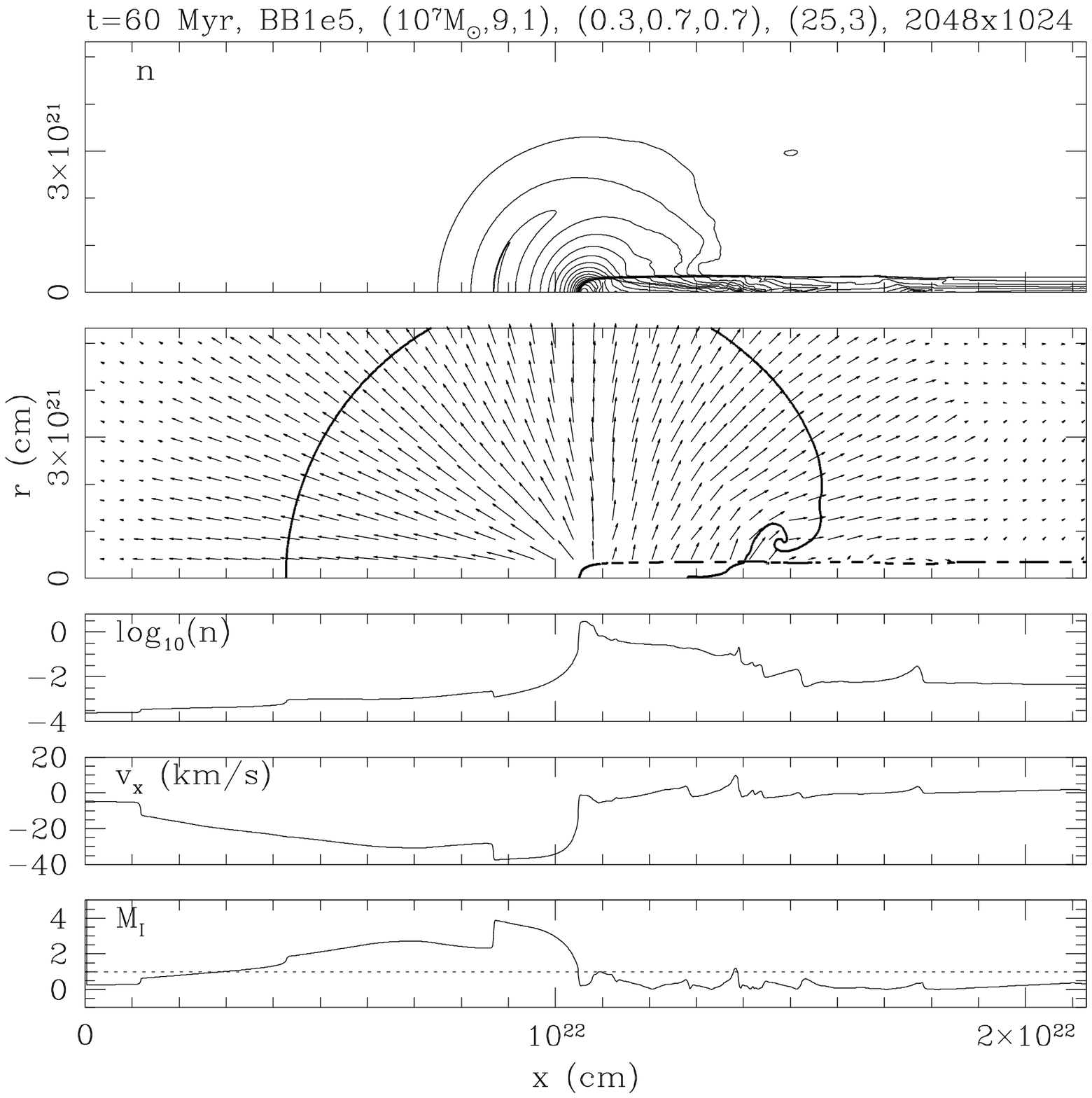}
\includegraphics[width=3.2in]{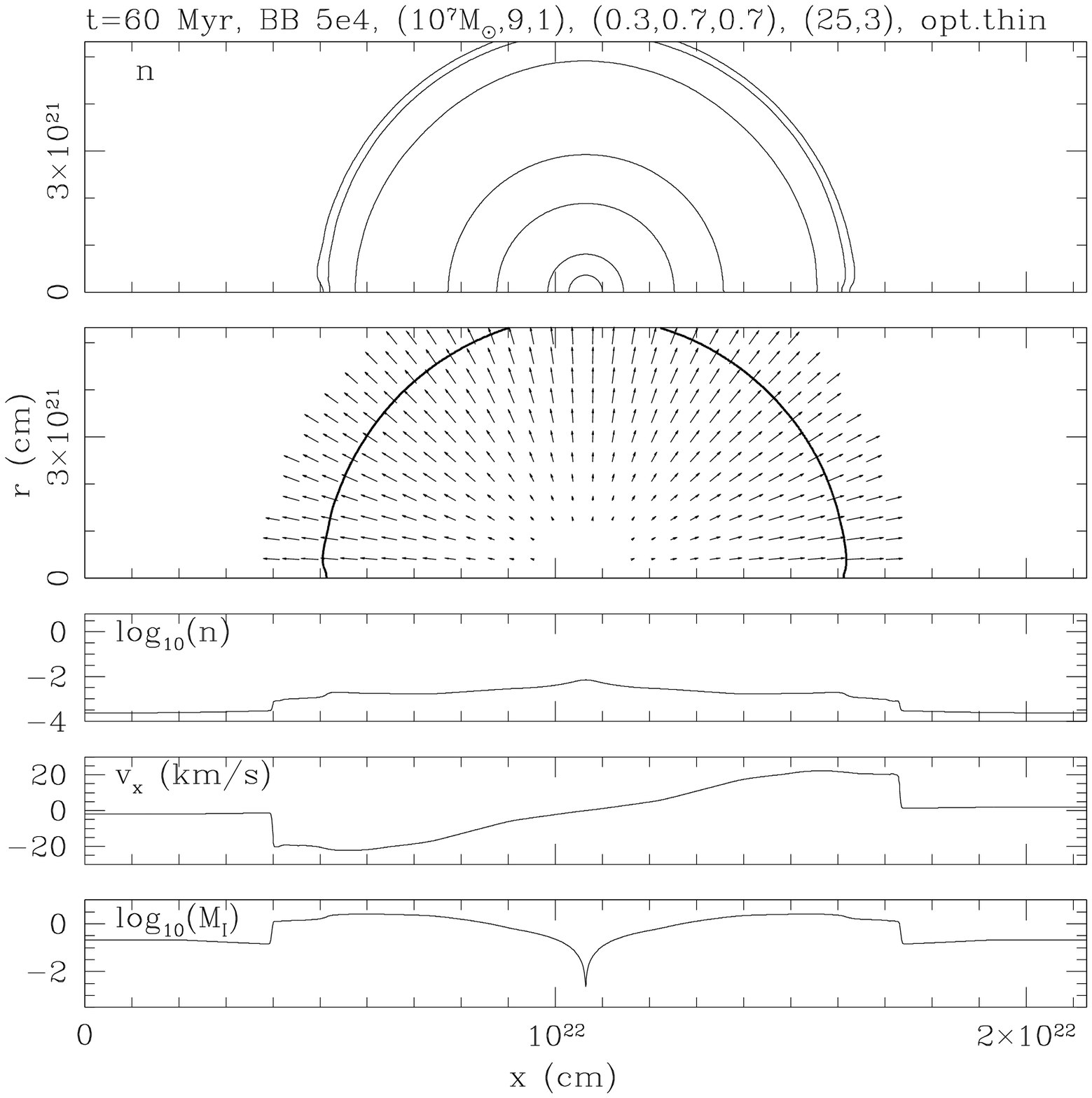}
\caption{(a) (left) Same as Figure~\ref{vel_arrows_60Myr_BB5e4},
 but for BB 1e5 case. (b) (right) Same time-slice
as Figure~\ref{vel_arrows_60Myr_BB5e4},
 but for optically-thin BB 5e4 case.}
\label{vel_arrows_60Myr_BB1e5}
\end{figure*}

\begin{figure*}
\includegraphics[width=3.4in]{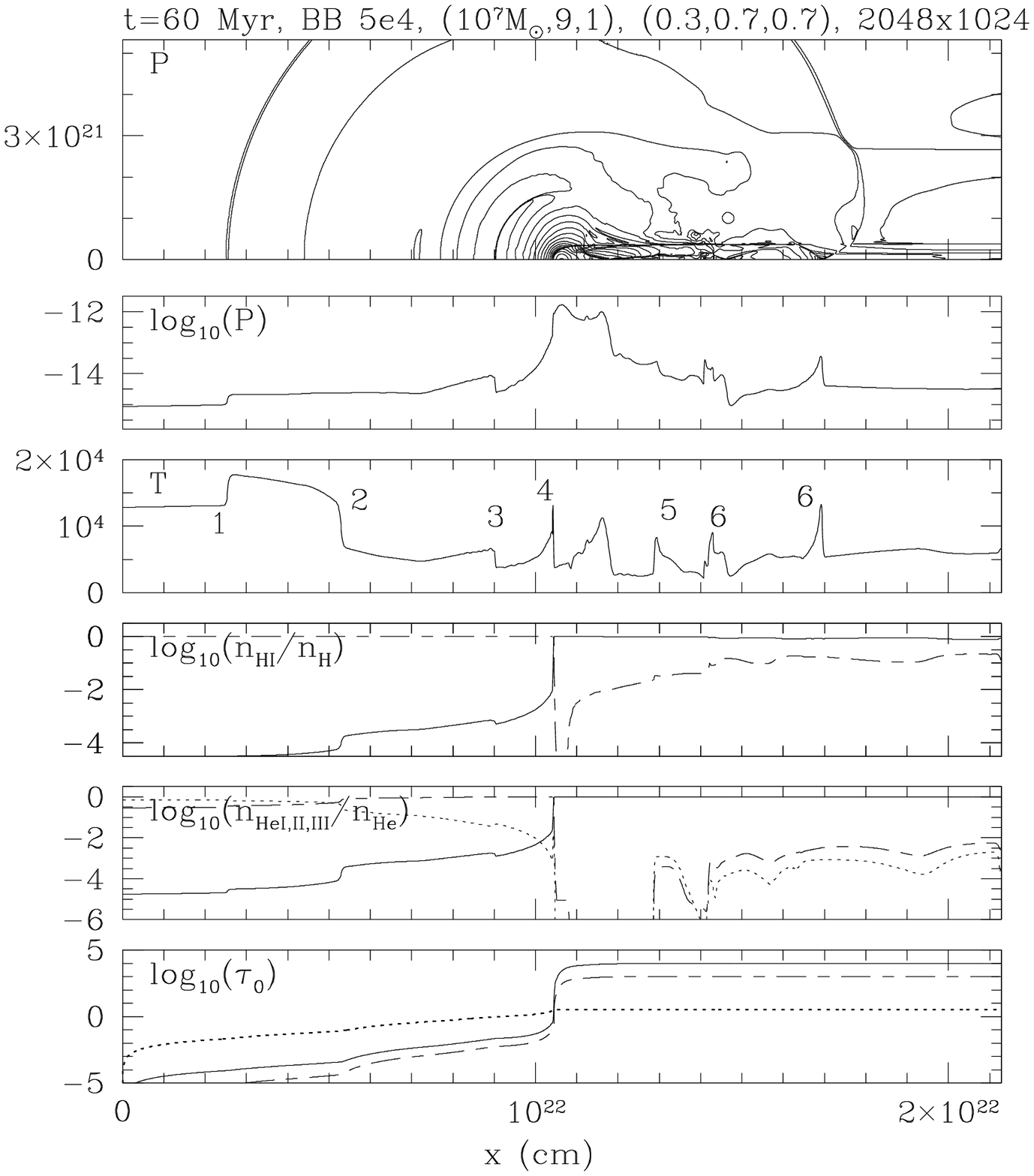}
\includegraphics[width=3.4in]{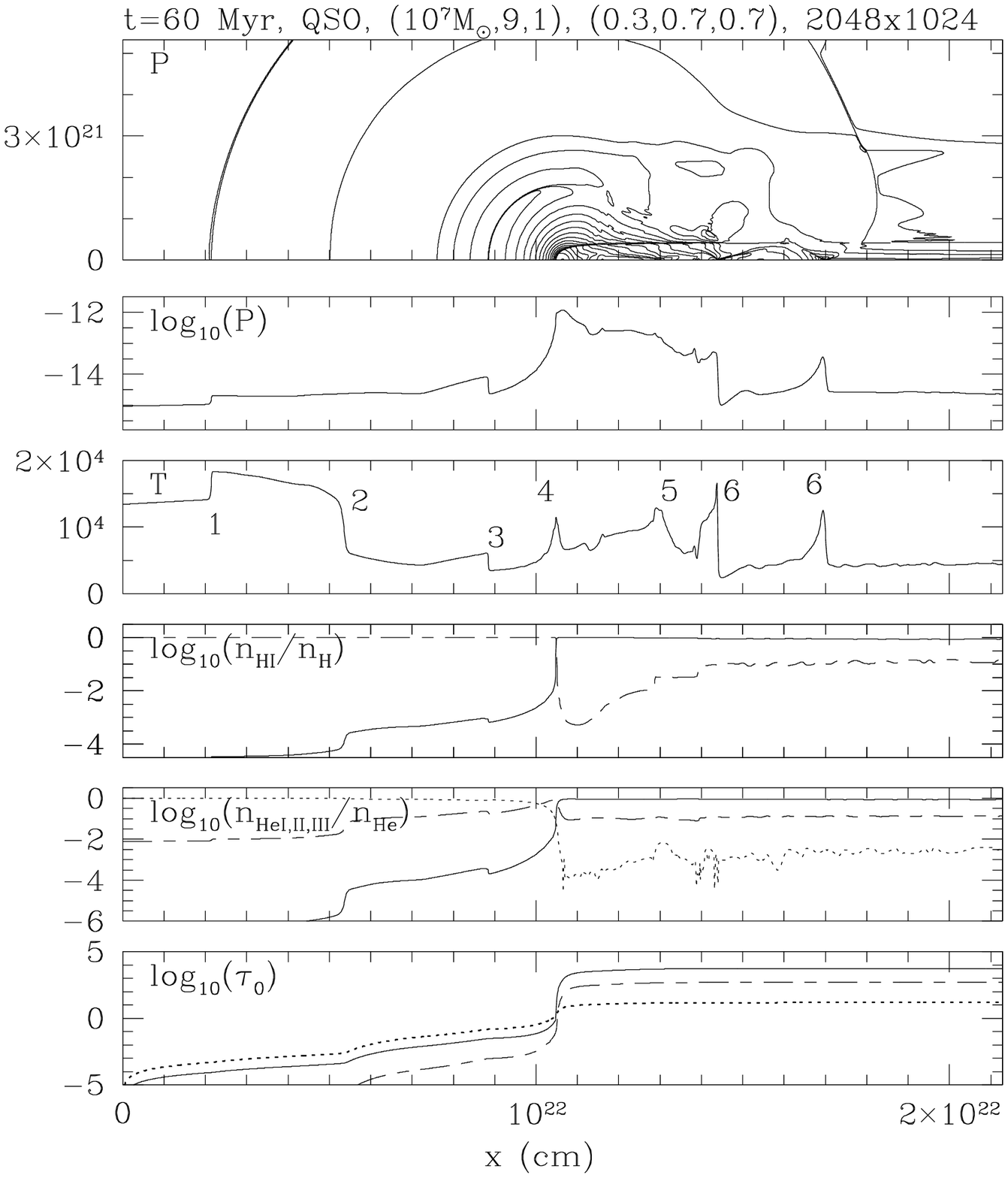}
\caption{D-type phase of I-front: Photoevaporative Flow
$60\,\rm Myr$ after I-front overtakes the minihalo: 
(a) (left)  BB 5e4 case and (b) (right) QSO case. Panels show the same 
quantities as in Figure~\ref{slices_0.2Myr_QSO}.
Key features of the flow are indicated by the numbers which label them on 
the temperature plots: \hbox{1 = IGM} shock; \hbox{2 = contact}
discontinuity which separates
shocked halo wind (between 2 and 3) from swept-up IGM (between 1 and 2); 
\hbox{3 = wind}
shock; between 3 and \hbox{4 = supersonic} wind; \hbox{4 = I-front}; 
\hbox{5 = boundary} of gas initially inside minihalo at $z=9$; 
6 = shock in shadow region caused by compression of shadow gas 
by shock-heated gas outside shadow.}
\label{slices_60Myr_BB5e4}
\end{figure*}

 \begin{figure*}
  \includegraphics[width=3.4in]{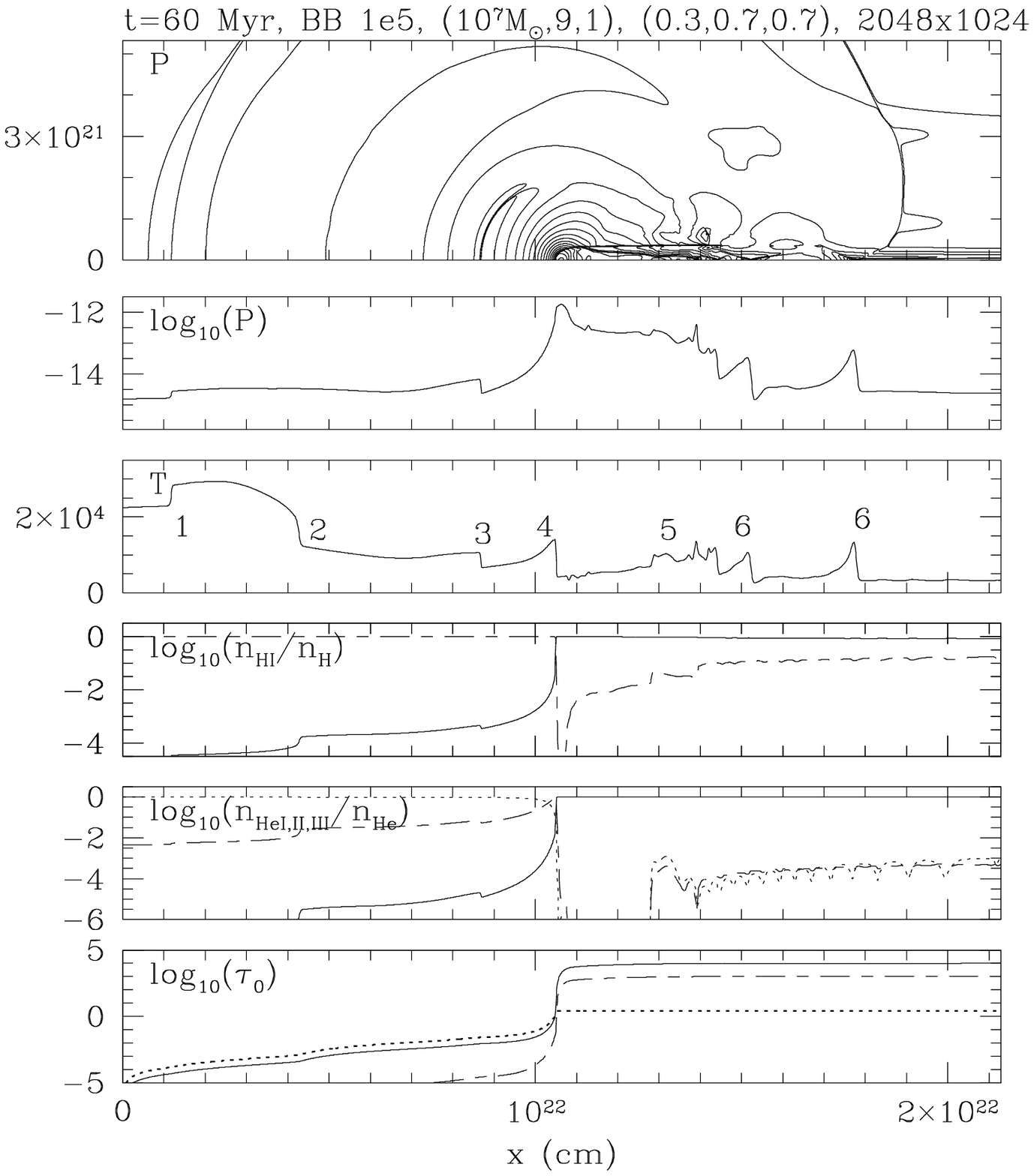}
  \includegraphics[width=3.2in]{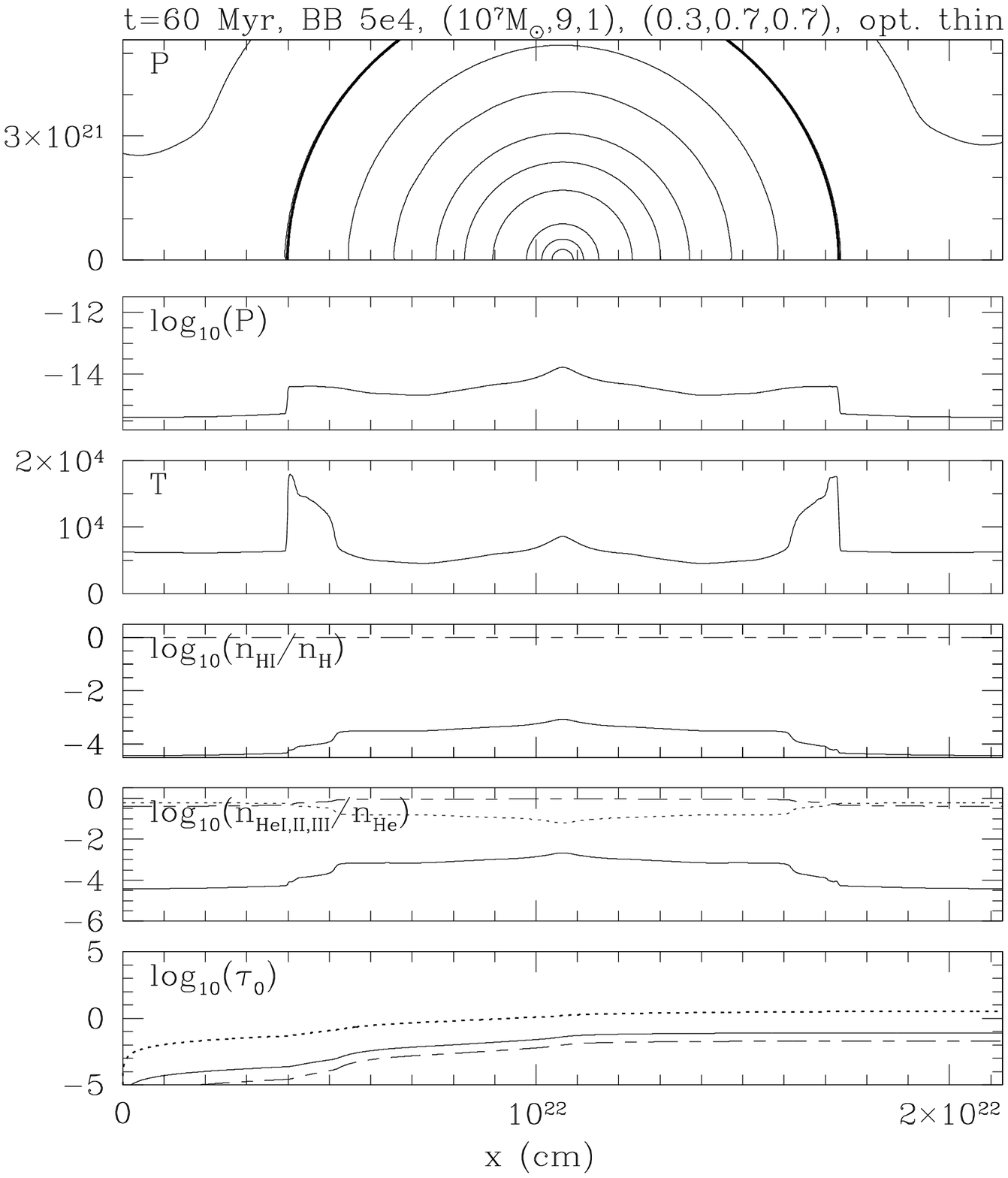}
 \caption{(a) (left) Same as Figure~\ref{slices_60Myr_BB5e4},
 but for BB 1e5 case. (b) (right) Same as Figure~\ref{slices_60Myr_BB5e4},
 but for optically-thin BB 5e4 case.}
 \label{slices_60Myr_BB1e5}
 \end{figure*}

\begin{figure*}
 \includegraphics[width=3.4in]{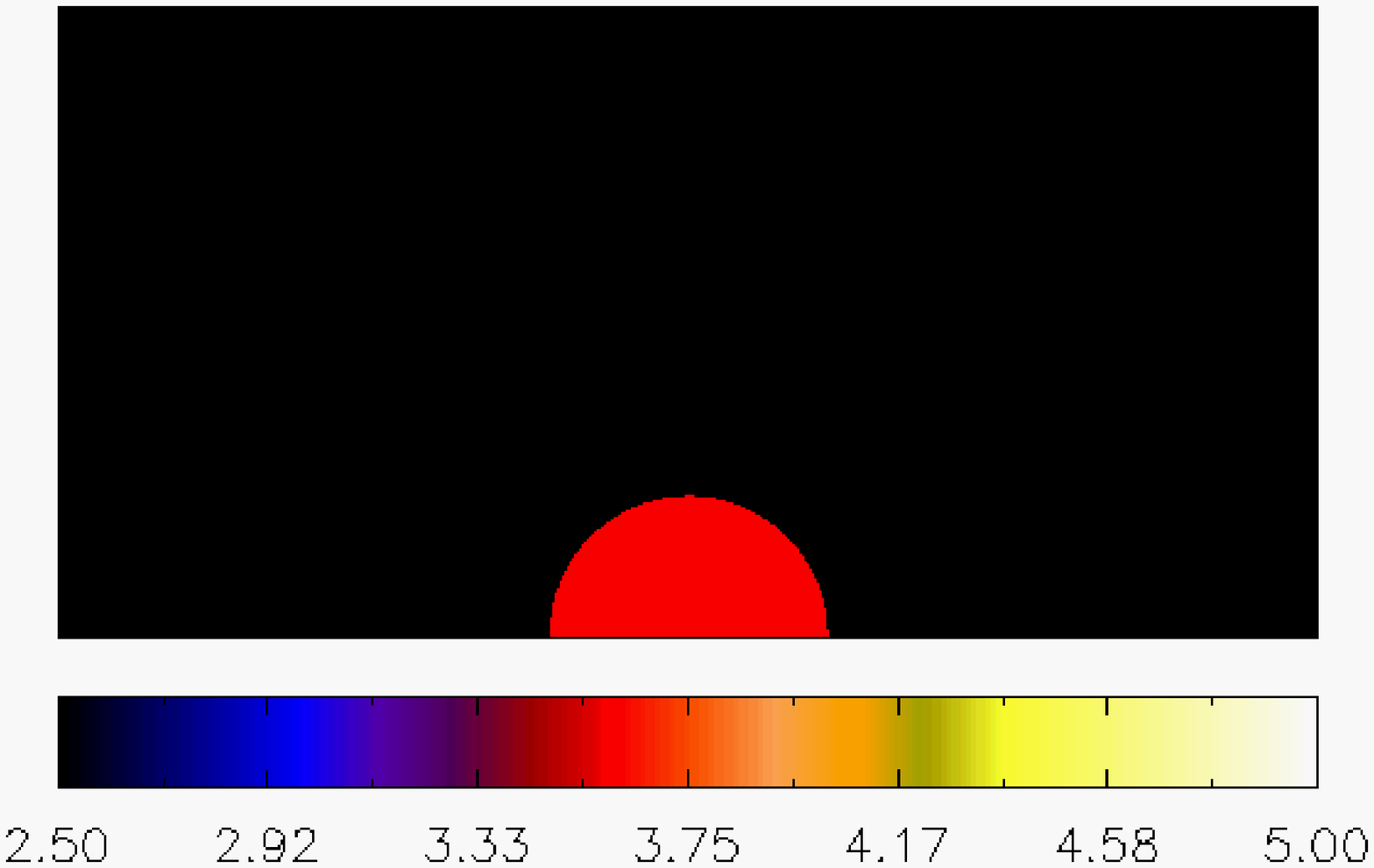}
 \includegraphics[width=3.4in]{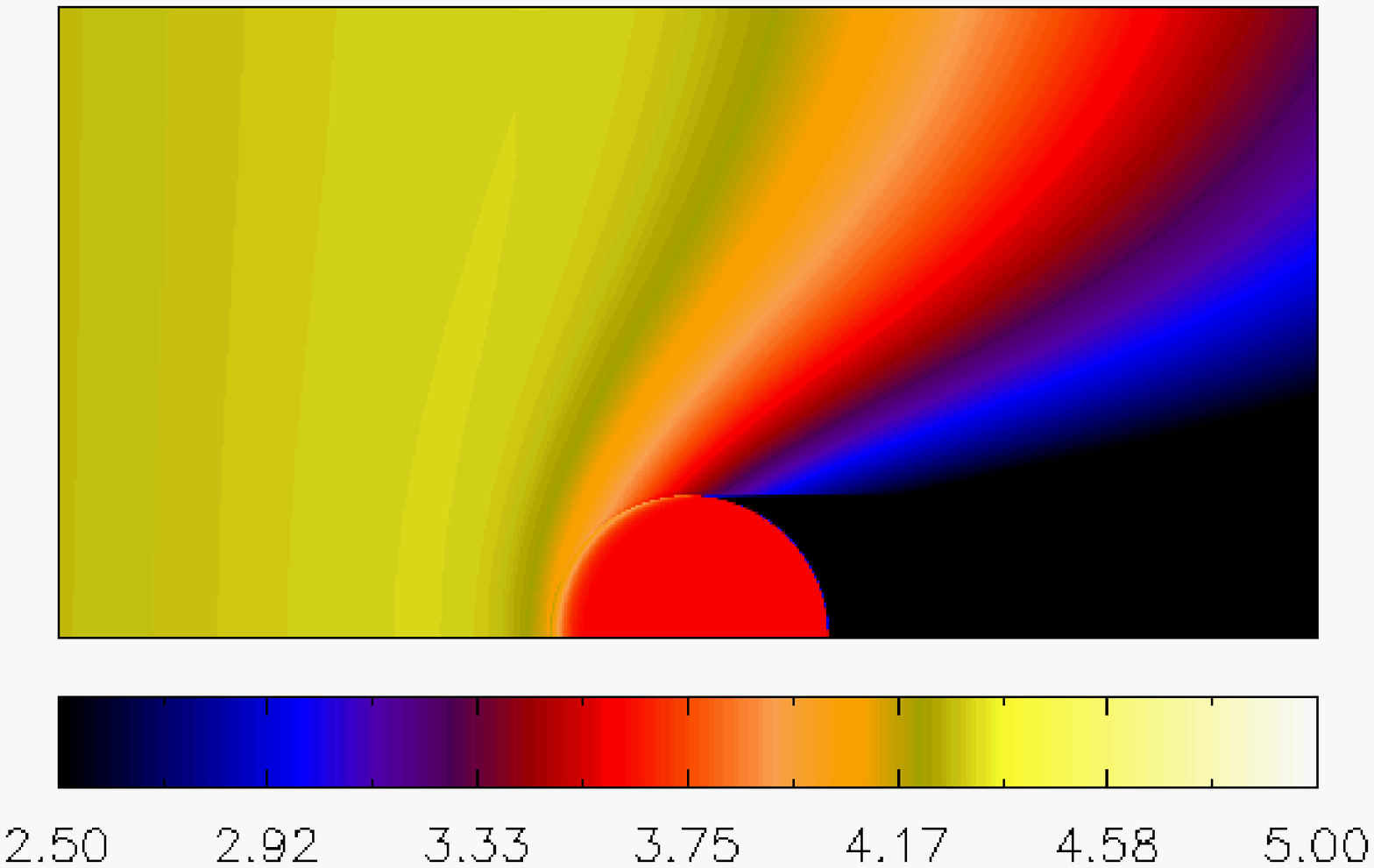}
 \includegraphics[width=3.4in]{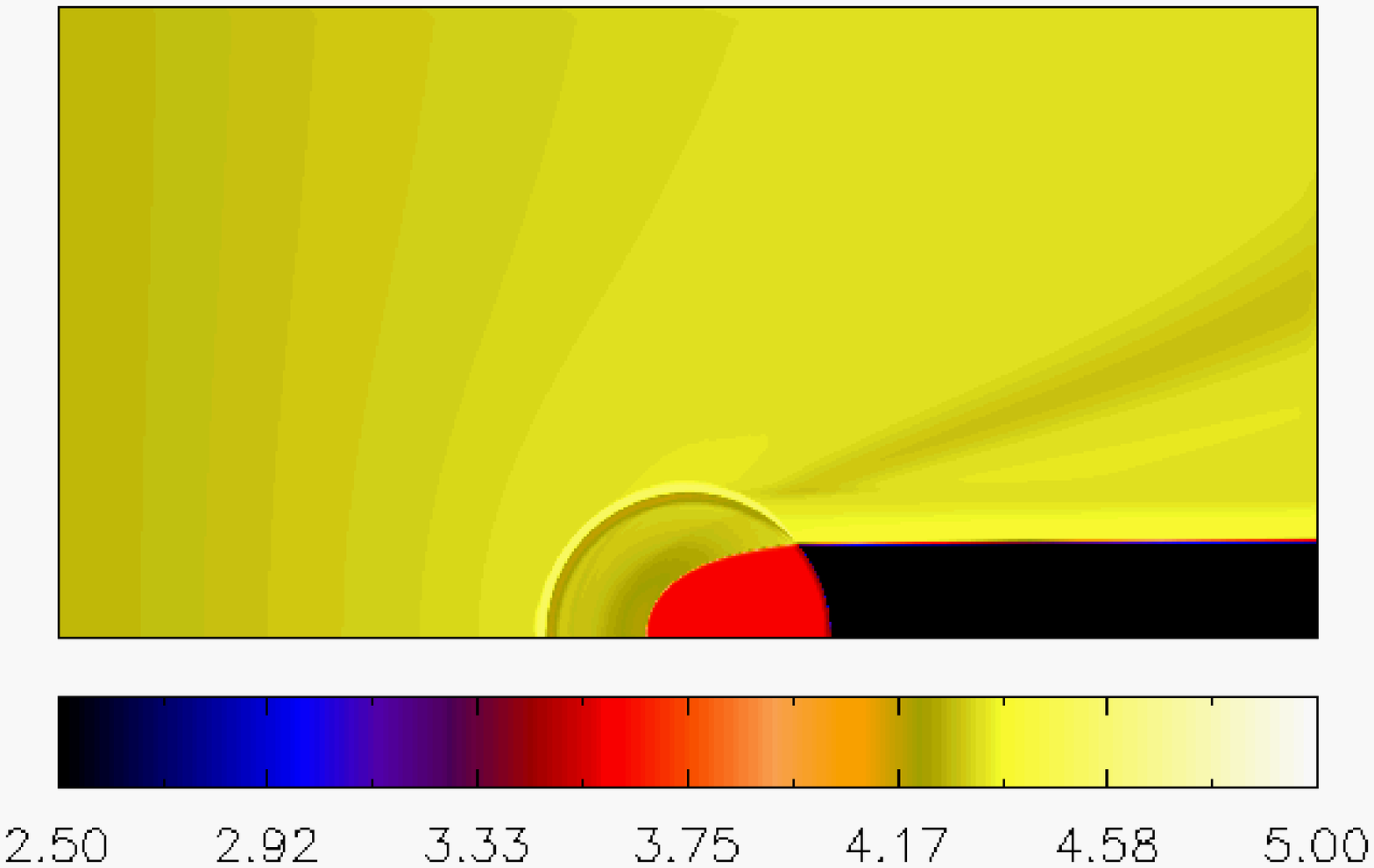}
 \includegraphics[width=3.4in]{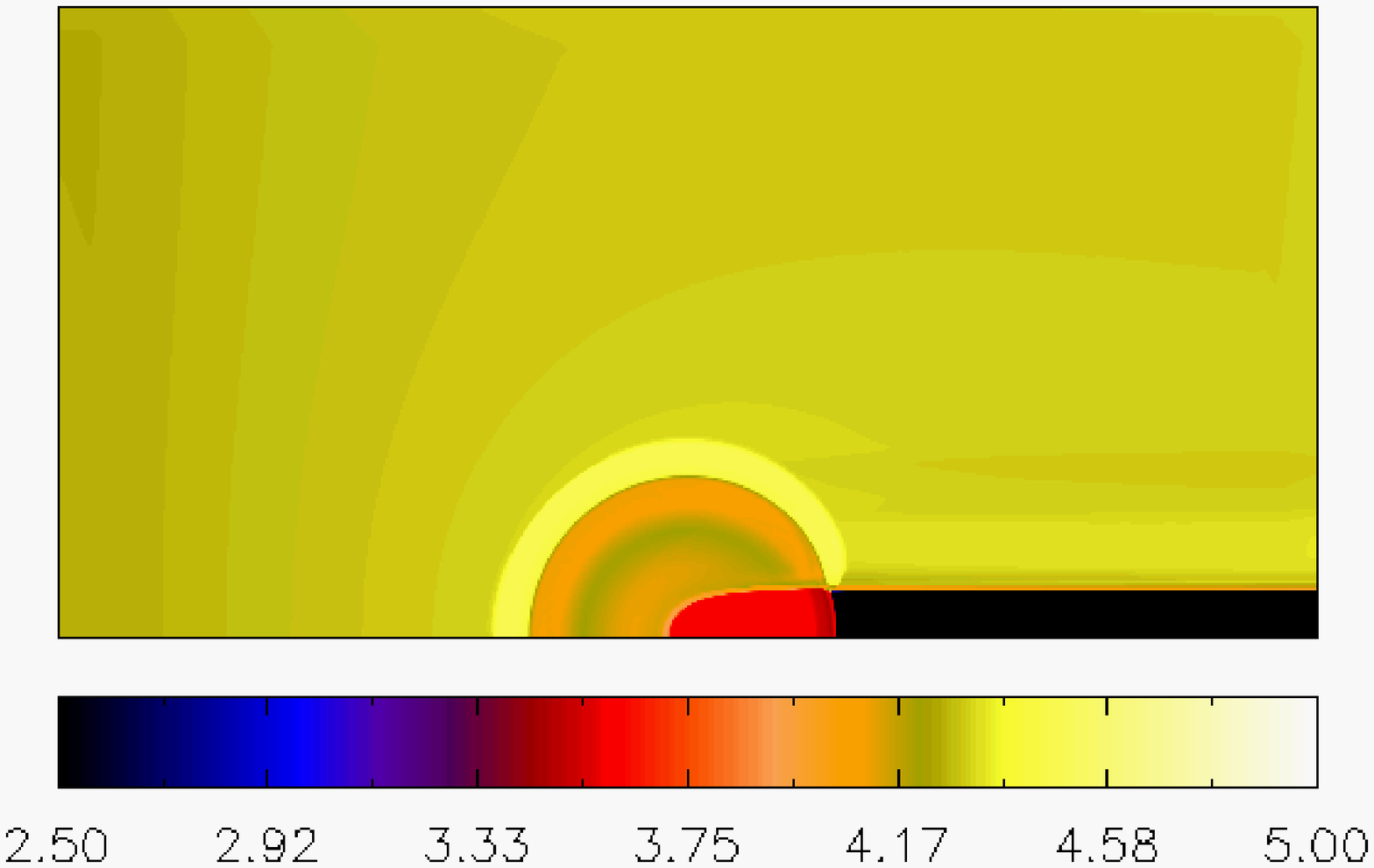}
 \includegraphics[width=3.4in]{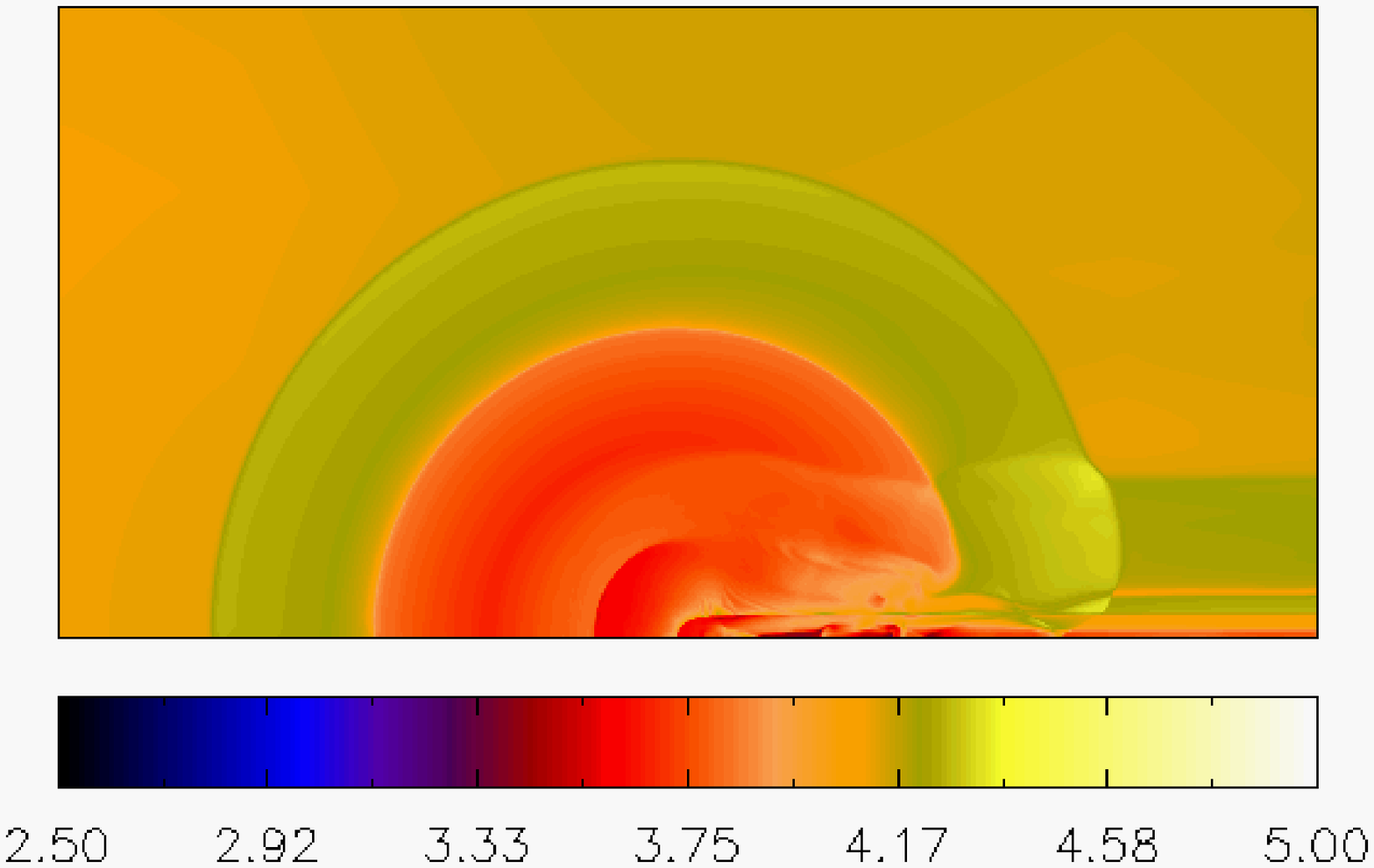}
 \includegraphics[width=3.4in]{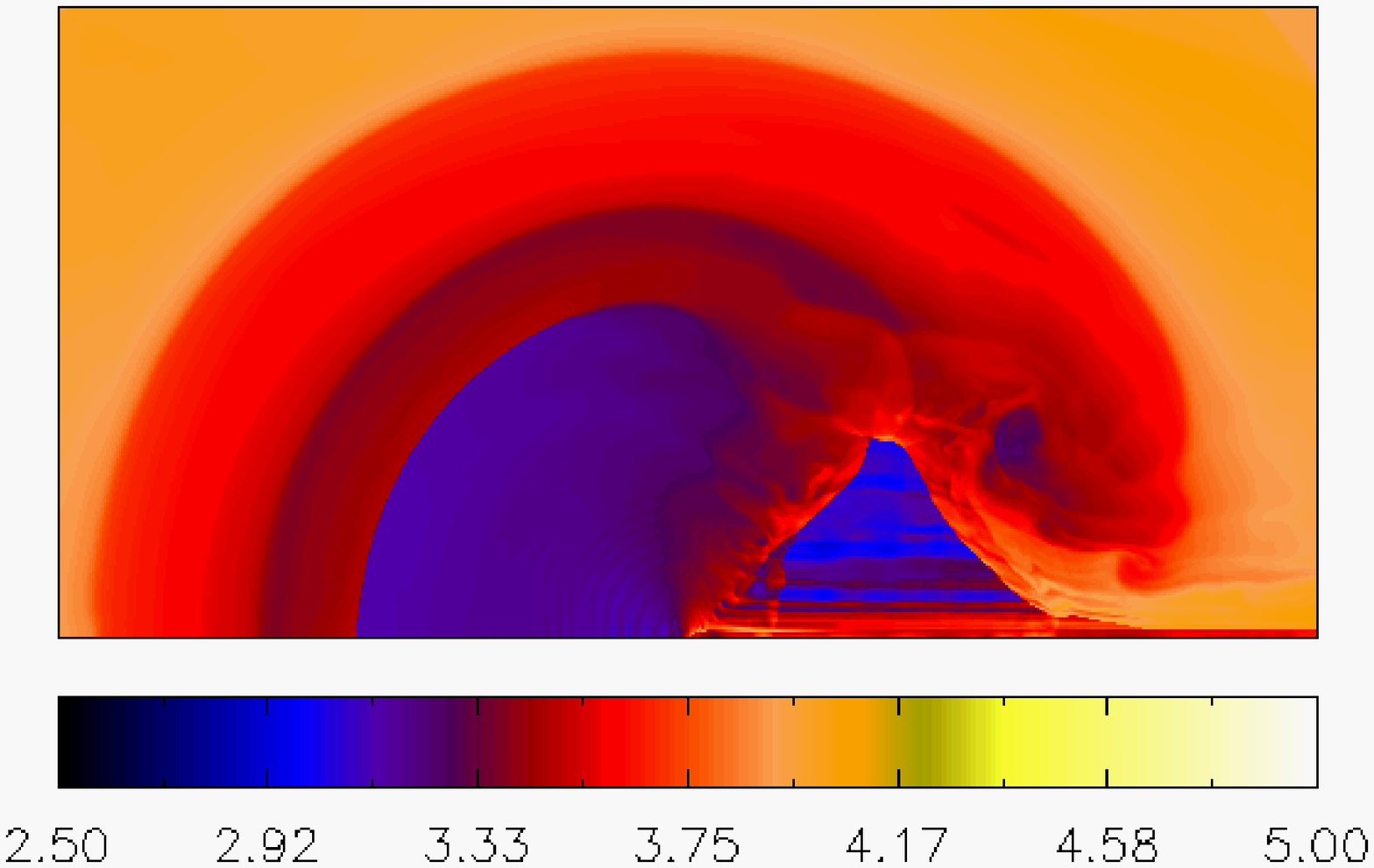}
 \caption{Snapshots of the temperature at times $t=0$ (top left), 
0.2 (top right), 2.5 (middle left), 10 (middle right), 60 (bottom left), and 
150 (bottom right) Myr 
in the BB5e4 case. Shaded isocontours are logarithmic in temperature.
Colors indicate the values of $\log_{10}(T)$,
as labelled on the color bar. }
\label{color_images}
 \end{figure*}

 \begin{figure*}
  \includegraphics[width=3.4in]{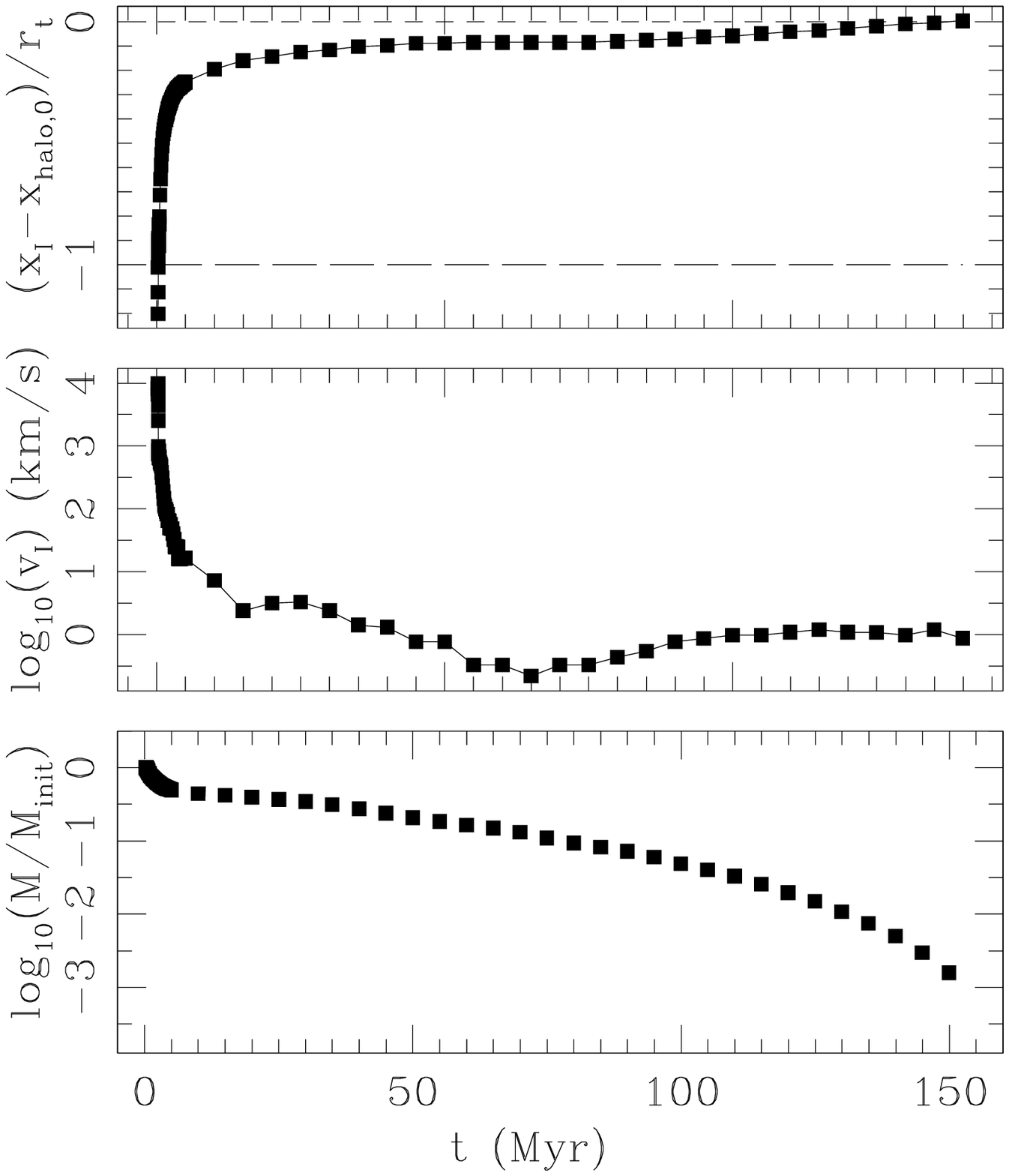}
  \includegraphics[width=3.4in]{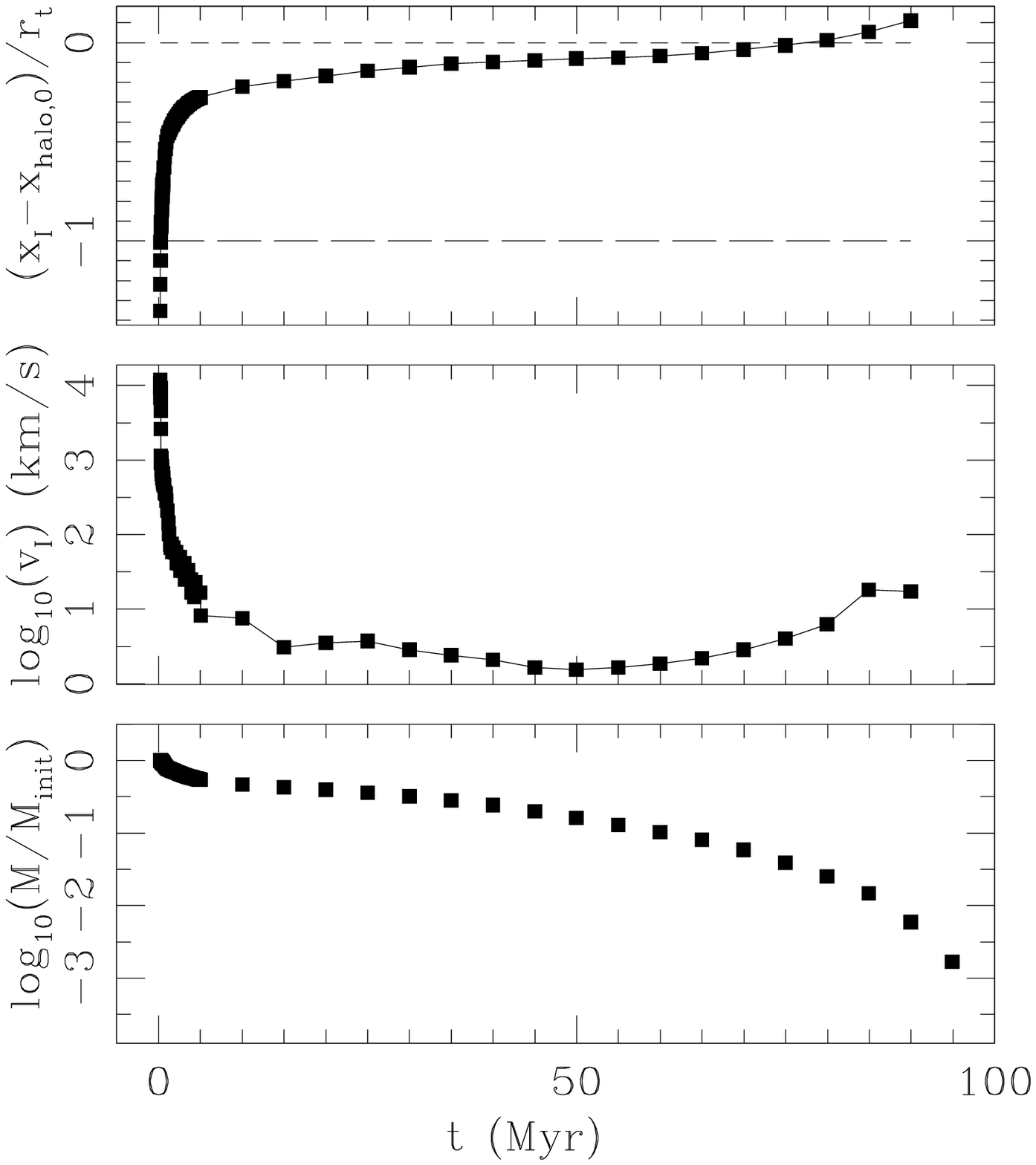}
 \caption{Evolution of I-front position and velocity and 
 of the of neutral gas content of photoevaporating
 minihalo: (a) (left) BB 5e4 case and (b) (right) QSO case. (upper panels) 
 Position $x_I$ (in units of the minihalo radius $r_{t}$ at $t=0$) and 
 (middle panels) velocity 
 $v_I$ of an ionization front propagating toward and through the minihalo. 
 The positions of
 the boundary (long-dashed line) and centre of the halo (short-dashed
 line) are also indicated.
 (lower panels) Fraction of mass $M_{\rm init}$, the mass which is 
 initially inside the minihalo when the intergalactic I-front overtakes it, 
 which remains neutral versus time $t$.}
 \label{I_front_mass_BB5e4}
 \end{figure*}

\subsection{The structure of the photoevaporative flow
during the D-type I-front phase}
\label{flow_sect}

When the I-front slows to become a D-type front, the hydrodynamical
response of the gas catches up with it and leads to the complete
photoevaporation of the minihalo.
This photoevaporative flow exhibits
many generic features which are independent of the
spectrum of the ionizing source. The side facing the source
expels a supersonic wind backwards towards the source.
There is a wind shock which thermalizes this wind outflow and
separates the unshocked supersonic wind close to the ionized side of
the I-front from the shell of shocked wind further away from the I-front
on this side. The shocked wind gas acts as a piston which sweeps up
the photoionized IGM outside the minihalo
and drives a shock into the IGM, reversing its infall velocity.
The wind grows more isotropic with time as
the remaining neutral halo material is photoevaporated. Since this gas 
was initially bound to a dark halo with $\sigma_V<\rm 10\,km\,s^{-1}$, 
photoevaporation proceeds unimpeded by gravity.
In Figures~\ref{vel_arrows_60Myr_BB5e4}--\ref{color_images}, we show the 
structure of the flow at a time
60 Myr after the global I-front first overtakes the minihalo.
In Figure~\ref{vel_arrows_60Myr_BB5e4} and the left panels of 
Figure~\ref{vel_arrows_60Myr_BB1e5} we
show the density contours of the gas, the current position of the
ionization front,
the velocity field, the current extent of the original halo 
material, and profiles of gas number density, velocity
and Mach number along the $x$-axis. 
In all three cases, the typical speed of the photoionized
outflow is supersonic and roughly proportional to the sound 
speed of the ionized 
gas. Hence, the outflow is fastest in the BB 1e5 case, due to the higher 
photoionization temperature reached in this case. 

For comparison, we have also performed a simulation of this problem
in the optically-thin approximation (i.e. zero optical depth),
to demonstrate the inadequacy of such an approximation.
In Figure~\ref{vel_arrows_60Myr_BB1e5} 
(right panels), we plot the results for this optically-thin
photoevaporation simulation 
for exactly the same halo and source parameters as the BB 5e4 spectrum
case shown in Figure~\ref{vel_arrows_60Myr_BB5e4} (left panels),
for the same time-slice. In the optically-thin case, the 
halo is suddenly and uniformly ionized and photoheated
to $T\sim20,000-35,000$ K, well 
above its virial temperature. Thereafter, the large pressure gradient
causes the gas to expand isotropically in all 
directions unimpeded by gravity, producing a completely different flow
structure.

For the more realistic simulations which included radiative transfer,
Figures~\ref{slices_60Myr_BB5e4} and \ref{slices_60Myr_BB1e5}
show the pressure contours, and the profiles along the symmetry
axis for pressure, temperature, H I-II fractions,  He I-III
fractions, and optical depths at the ionization thresholds of H I, 
He I, and He II, for the photoevaporative flow 60~Myrs after the 
global I-front first overtakes the minihalo. Key features of
the flow are indicated by the labels on the temperature panels. For
comparison, we also show the very different results for these
quantities in the optically-thin simulations
(Figure~\ref{slices_60Myr_BB1e5}, right panels).

\subsection{Evolution of the temperature structure}

The different stages of the photoevaporation process are illustrated
in Figure~\ref{color_images} by shaded isocontours
of temperature for a sequence of time-slices of the BB 5e4 case. 
The panels show: (a) (time $t=0$ Myr), the initial 
condition of the isothermal, virialized halo surrounded by a cold,
infalling, IGM; (b) ($t=0.2$ Myr), the fast, 
R-type I-front sweeping through the computational box is just 
about to encounter the minihalo along the $x$-axis, and is 
starting to photoheat 
the halo on the source side; (c) ($t=2.5$ Myr), when the weak, R-type I-front is
inside the minihalo before it slows to R-critical and begins to
transform to D-type, a distinct shadow is apparent behind the 
neutral portion of the halo,
while the rest of the gas is heated to $T>10^4$ K, 
and the photoevaporation process 
begins; (d) ($t=10$ Myr), the R-type to D-type transition phase, when
the I-front slowly advances along the axis, the shadow 
region behind the shielded, neutral
minihalo core is reduced, and the IGM shock
and the shocked
wind are clearly visible; (e) ($t=60$ Myr), strong D-type phase, when
the shadow is being ablated and the 
remaining gas heated, by shocks reflected off the $x$-axis
behind the halo. The key features of 
the flow indicated in Figure~\ref{slices_60Myr_BB5e4} are all clearly seen; 
(f) ($t=t_{\rm ev}=150$ Myr, the evaporation time), when the gas is almost
completely evaporated from the halo and resides in a much larger, low-density 
region which is cooling adiabatically as it expands, 
leaving behind a 
dark-matter halo almost completely devoid of gas. Movies of these simulation 
results, including the time evolution of
additional quantities like pressure, 
density, H I and He I fractions, and the results for the cases
with different source spectra are available at {http://galileo.as.utexas.edu}. 

 \begin{figure}
 \includegraphics[width=3.4in]{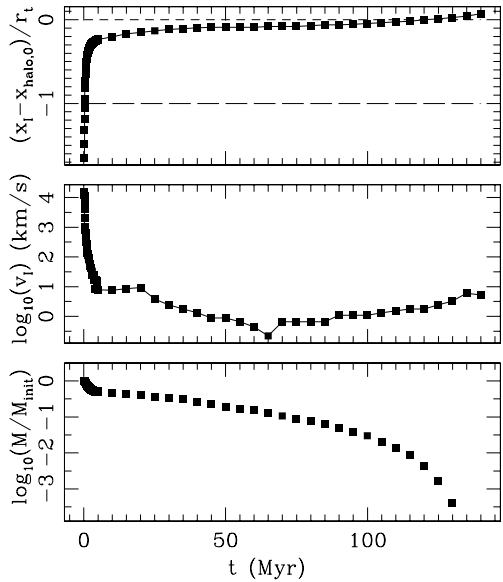}
 \caption{Same as Figure~\ref{I_front_mass_BB5e4},
but for BB 1e5 case. }
 \label{I_front_mass_BB1e5}
 \end{figure}

\subsection{I-front evolution}

The evolution of the position $x_I$ and velocity $v_I$
of the I-front, and of the remaining neutral mass fraction $M/M_{\rm init}$
are shown for all three source spectra cases
in Figures~\ref{I_front_mass_BB5e4} and \ref{I_front_mass_BB1e5},
for comparison.
The weak, R-type I-front slows down to the R-critical velocity
about 5 Myr after it reaches the minihalo, after
which a shock must form ahead of the I-front to compress the
gas and slow it down to D-critical velocity or below,
as discussed in detail in \S~\ref{trapping_sect}. The further 
evolution is slower and more gradual. Eventually, the I-front speed
drops below the D-critical front speed 
$v_D\approx1\rm km\,s^{-1}$. For the rest of the evaporation 
process, the velocity of the I-front is $\la1\,\rm km\,s^{-1}$, 
i.e. sub-critical.
At 60 Myr, for example, the I-front velocity has dropped to 
about $v_I\sim0.3\,\rm km\,s^{-1}$, while the D-critical velocity
is still $v_I\approx1\,\rm km\,s^{-1}$.
This behaviour is different from the usual approximation made in 
analytical calculations of photoevaporation, in which the I-front is 
assumed to be D-critical. 

What kind of D-type
front is it during the D-type phase? Weak, D-type fronts
move subsonically with respect to both the neutral and ionized gas. Strong
D-type fronts move subsonically into the neutral gas
but supersonically with respect to the ionized gas. In the lab frame (i.e.
rest frame of the neutral gas before ionization), 
as the D-type front advances into the neutral side
(i.e. away from the source), the
ionized gas behind it always moves {\it toward} the
source (i.e. opposite to direction of I-front).
This ionized gas motion can be either subsonic (for weak fronts),
or almost sonic (for D-critical fronts), but it is
supersonic only for {\it strong} D-type fronts. If we apply this
description to our 60 Myr time-slice for the BB5e4 case and
look both at the I-front velocity evolution plot in
Figure~\ref{I_front_mass_BB5e4} and the cuts along the $x$-axis in
Figure~\ref{vel_arrows_60Myr_BB5e4}, we see that the I-front is
subcritical at that point. From the Mach number plot at
points just behind the
I-front on the immediate post-front (ionized) side,
the lab frame velocity is supersonic toward the source, which indicates
that the front is a strong, D-type. This is consistent with the density cut
along the $x$-axis which shows that density {\it drops} by a very
large amount from the neutral to the ionized side, as it should for
both weak and strong, D-type fronts, but the strong type
has an even bigger drop in density, as required to explain the
numerical results. In short, the I-front at
60 Myr is a subcritical, strong, D-type.

\subsection{Evaporation times}
\label{evap_time_sect}

The neutral mass fraction of the original minihalo gas
quickly drops to $\sim60$ per cent during the 
initial slowing-down phase 
of the I-front evolution, and subsequently declines much more gradually
as the minihalo photoevaporates. We define the 
photoevaporation time $t_{\rm ev}$ 
as the time when only 0.1 per cent of the mass remains neutral (i.e. when 
$M/M_{\rm init}=10^{-3}$). We see from Figures~\ref{I_front_mass_BB5e4} and 
\ref{I_front_mass_BB1e5} that, once the neutral mass fraction reaches $\sim 1$ 
per cent, it drops fairly precipitously, indicating that the value of
$t_{\rm ev}$ is not sensitive to our particular choice of the
final value for $M/M_{\rm init}$ used to define this
evaporation time. We obtain 
$t_{\rm ev}\approx150$, 125 and 100~Myrs 
for the BB 5e4, BB 1e5 and QSO cases, respectively. Compared to
the BB 5e4 case, the evaporation time in the BB 1e5 case is shorter due to the 
higher outflow velocity in this case (see Fig.~\ref{vel_arrows_60Myr_BB1e5}),
caused by the higher temperature in the photoionized region. In the QSO case,
on the other hand, the outflow velocity is similar to the one in the BB 5e4 
case, but the photoevaporation process in this case is 
accelerated by the significant
pre-heating due to the hard photons present in this spectrum, which ultimately
leads to an even shorter $t_{\rm ev}$. These values of $t_{\rm ev}$ are
all within a factor of two of the sound crossing time for an ionized gas
sphere of the same diameter as our minihalo reported in
\S2.4. However, we caution that this result is not a proof that $t_{\rm sc}$
is a good estimator of $t_{\rm ev}$. In fact, results for a wide range
of halo and source parameters which we shall present in a
companion paper demonstrate that $t_{\rm ev}$ departs significantly from
$t_{\rm sc}$, in general.

\subsection{Ionizing photon consumption}
\label{phot_consump_sect}

\subsubsection{Minihalo effective cross-section}

\begin{figure*}
 \includegraphics[width=3.4in]{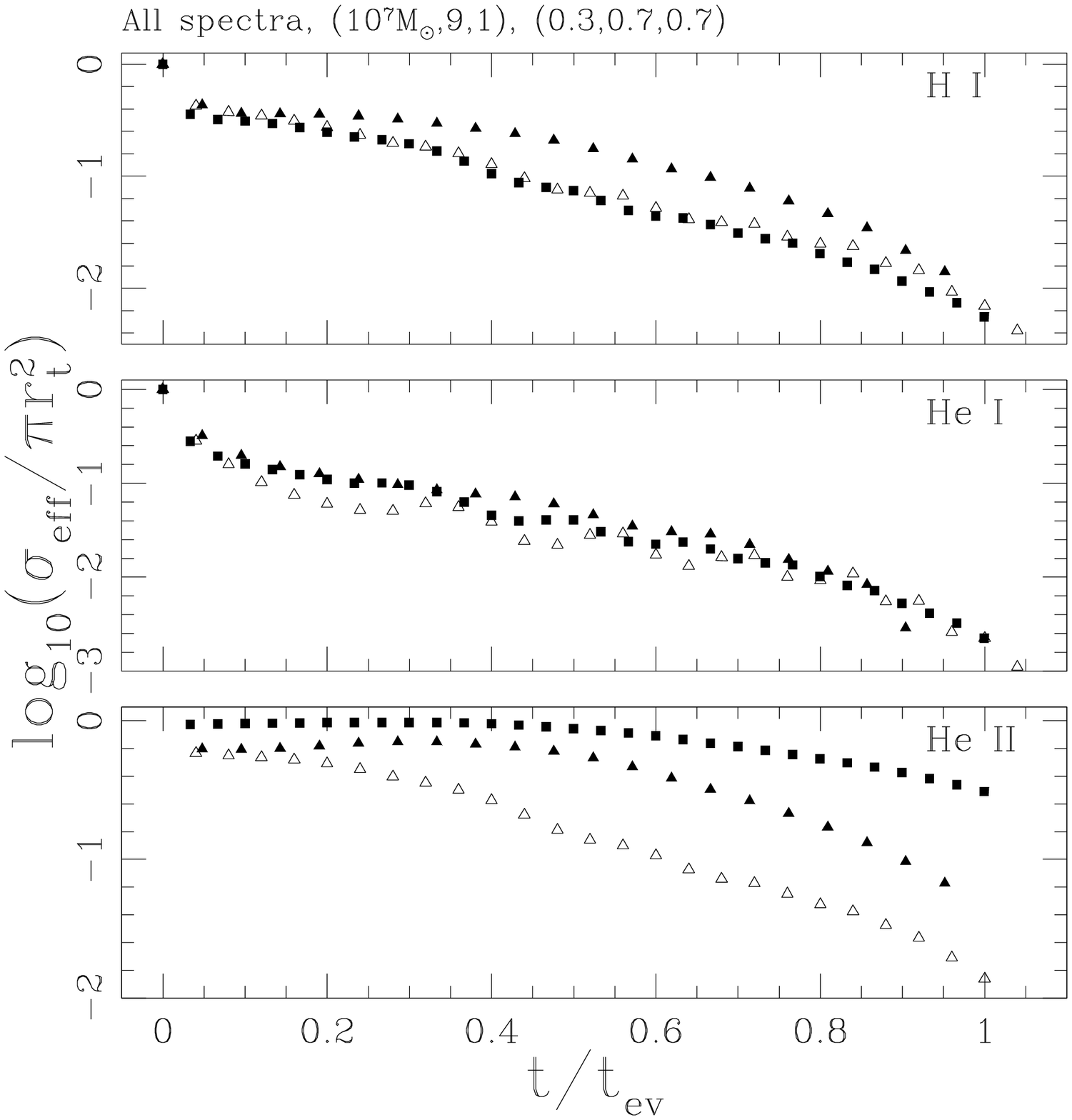}
 \includegraphics[width=3.4in]{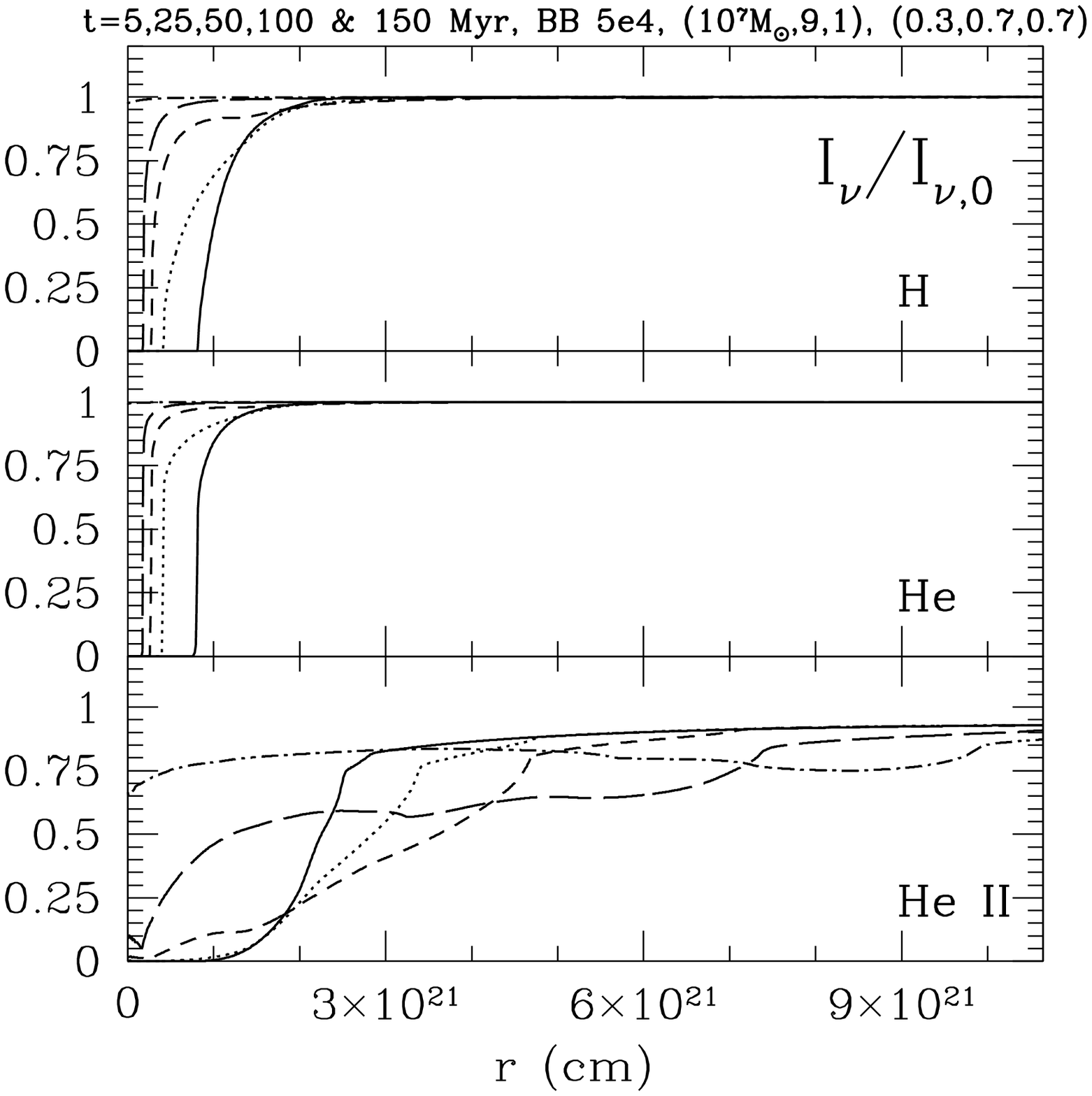}
 \caption{(a) (left) Fraction of minihalo initial
geometric cross section $\pi r_t^2$ 
 ($r_t=0.76$ kpc) which is opaque to source photons that can ionize
 H~I (top panel), He~I (middle panel), or He~II (bottom panel) versus
 time (in units of $t_{\rm ev}$), for the BB 5e4 (filled squares), QSO (filled 
triangles) and BB 1e5 (open triangles) cases. 
($t_{\rm ev}=150,100, \,\rm and \,125\, {\rm Myr}$, respectively).
 (b) (right) Transmitted flux $I_\nu/I_{\nu,0}$ at the
ionization thresholds of H (top),
 He I (middle), and He II (bottom) vs. impact parameter
(or radius in axisymmetry) $r$ at
 times $t=5$ (solid), 25 (dotted), 50 (short-dash), 100 (long-dash), and 150 
(dot-dash) Myr for the BB 5e4 case.} 
 \label{cross_BB5e4}
 \end{figure*}

The gradual decay of the opaque cross section of the minihalo as seen by the 
source [as defined in eq.~(\ref{eff_cross_sect})]
is illustrated by Figure~\ref{cross_BB5e4} (left panels) for 
photons at the ionization thresholds of H I, He I, and He II. 
For the BB5e4 case,
the cross section at the H I threshold drops to 10\% of its original value
after $0.4t_{ev}=60$ Myr, while the cross section at the He I threshold 
drops somewhat faster. The cross section of the minihalo at the
He II threshold, however, initially rises 
as the partially ionized gas expands, and only later,
around $0.4t_{ev}=60$ Myr, 
does it start to decline slowly, as He II starts to become ionized to He III.
The evolution of the H I effective cross-section is similar for the two stellar 
spectra, but decreases somewhat more slowly in the QSO case, while the evolution
of the He II cross-section is similar for all three spectra. Finally, the 
evolution of the He I cross section is markedly different in all three cases,
due to the different photon energy distribution of the three spectra.

In Figure~\ref{cross_BB5e4}, we show the transmitted fluxes
at the ionization thresholds of H I, He I, and He II vs. impact parameter $r$  
at times $t=5$, 25, 50, 100, and 150 
Myr for the BB 5e4 case. We see that the
simulation box is optically-thin
to the photons with energies below the He II ionization threshold along most
lines of sight along the $x$-direction outside the minihalo;
only the minihalo is 
significantly opaque, once the global I-front has swept past the minihalo. 
The minihalo's cross-section declines quickly
with time once the I-front enters it and is 
somewhat smaller for He I ionizing photons than for H I 
ionizing photons. The simulation
box is never completely optically-thin to He II ionizing
photons and has rather complex time behaviour as the wind
expands and more He II is photoionized to He III.

\subsubsection{Number of ionizing photons per atom consumed
during photoevaporation}

Based on our simulation data, we have calculated the number of ionizing photons
per atom, $\xi$, required to evaporate a minihalo both by using the effective 
cross-section method of equation~(\ref{n_gamma1}) and by counting the total 
number of recombinations [as in eq.~(\ref{n_gamma2})]. Using the first
method, we obtain $\xi=(4.51,5.41,3.43)$ in the BB 5e4, QSO, and BB 1e5 cases, 
respectively, while the second method
yields $\xi=(5.1,5.0,3.3)$. Since the two methods give similar results,
we can, henceforth, rely upon either one. For the 
``efficiency'' factor $f$ in equation~(\ref{xi_numeric}) (assuming $T_4=1$),
these results imply
$f=(0.029,0.034,0.022)$ and $f=(0.032,0.032,0.021)$, respectively.

In the optically-thin approximation case,
the effective cross-section is, by definition, 
zero at all times, and thus only the second method for calculation of $\xi$ is 
applicable. By contrast with our results which took proper account of
the optical depth in bound-free opacity, for which $f\ll1$, our
optically-thin results for $\xi$ by the second approach yield 
$f\approx 1$ [consistent with the result obtained in this optically-thin
approximation by \cite{HAM01}]. This enormous overestimate of $\xi$ by the
optically-thin approximation
indicates that the optically-thin approximation used by previous authors 
is completely inadequate for determining $\xi$. 

\begin{figure}
 \includegraphics[width=3.2in]{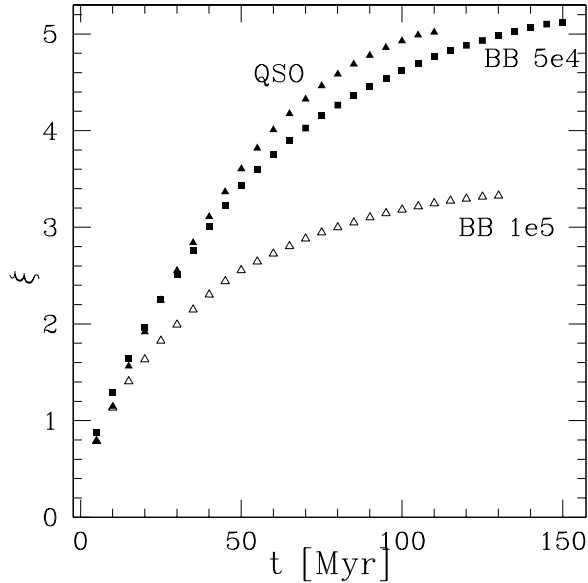}
 \caption{Evolution of the consumption of ionizing photons as given by 
equation~(\ref{n_gamma2}) for all three spectra, as labelled.}
 \label{consump_time}
\end{figure}

The reason for this
significant discrepancy 
is easily understood if we consider where and when most of the 
recombinations take place. In Figure~\ref{consump_time}, we show the evolution 
of $\xi$ for our illustrative simulations. During the initial R-type phase, 
the I-front deceleration phase, 
a significant fraction of the mass ($\sim 40\%$ in our sample simulations)
becomes ionized almost instantly, just as does the entire minihalo
in the optically-thin approximation. This
gas is in the halo ourskirts, however, where the density is low and
that density 
drops even further as the gas expands in the evaporative
outflow, into the IGM, making recombination in
this gas inefficient within $10-20$ Myr. Most of the minihalo gas is initially 
shielded, however, 
and remains neutral for a much longer time. From this point on, the 
additional ionization of gas
occurs only where the photoevaporation process is removing it by
expelling it in a supersonic wind. For this gas,
the density becomes too low too soon
for significant recombinations to 
occur there. Therefore, during 
most of the evolution, the bulk of the recombinations occur in a 
thin dense layer 
in the immediate vicinity of the I-front. Each gas atom spends 
only a short time 
in this layer, however, as it rushes supersonically away
from the neutral region into the  
low-density expanding wind, hence experiencing many fewer 
recombinations than the 
optically-thin approximation would predict. 

As Figure~\ref{consump_time} shows, 
$\xi$ increases gradually throughout the evaporation time, and there
are noticeable differences among the three cases. For the hard QSO spectrum,
the transition layer is thicker and penetrates deeper into the denser 
and colder parts
of the halo, thus increasing the number of recombinations per atom
per evaporation time. However, this same 
pre-heating effect shortens the evaporation time in this case, ultimately
leading to a rough cancellation of the two effects and the same 
total $\xi$ as in
the BB 5e4 case. The BB 1e5 spectrum is not as hard as in the QSO case, 
which decreases
the number of recombinations in the I-front layer. The evaporation time 
in this case
is also shorter than in the BB 5e4 case,  due to a higher photoionization 
temperature and
thus to faster outflow speed. Additionally, the higher temperatures 
in the wind for the BB 1e5 case
($\sim10^4$ K, as opposed to $\sim5000$ K in the other two cases, see 
Figs.~\ref{slices_60Myr_BB5e4} and \ref{slices_60Myr_BB1e5}) also 
leads to lower rates 
of recombination in the wind and again lower $\xi$. Thus, overall,
the Pop. III sources appear significantly more efficient than
Pop. II or QSO sources
in terms of the total
number of ionizing photons needed to complete the photoevaporation process. 

In principle,
the optically-thin approximation might become correct in the limit of 
$\chi_S\gg\chi_{S,\rm crit}$ (see \S~\ref{strom_sect}), in which case 
the I-front is not trapped, the minihalo is ionized almost instantly
compared to the evaporation time,
and the radiative transfer effects are less significant. 
However, for most minihalos this is not a relevant limit, since it requires
excessively high ionizing flux of $F_0\simgreat100-1000$ (depending on the halo 
mass and redshift of collapse) (see \S~\ref{ion_flux_sect}). 

\subsection{Observational diagnostics}
\label{observ_sect}
\subsubsection{Absorption lines}

\begin{figure*}
 \includegraphics[width=3.4in]{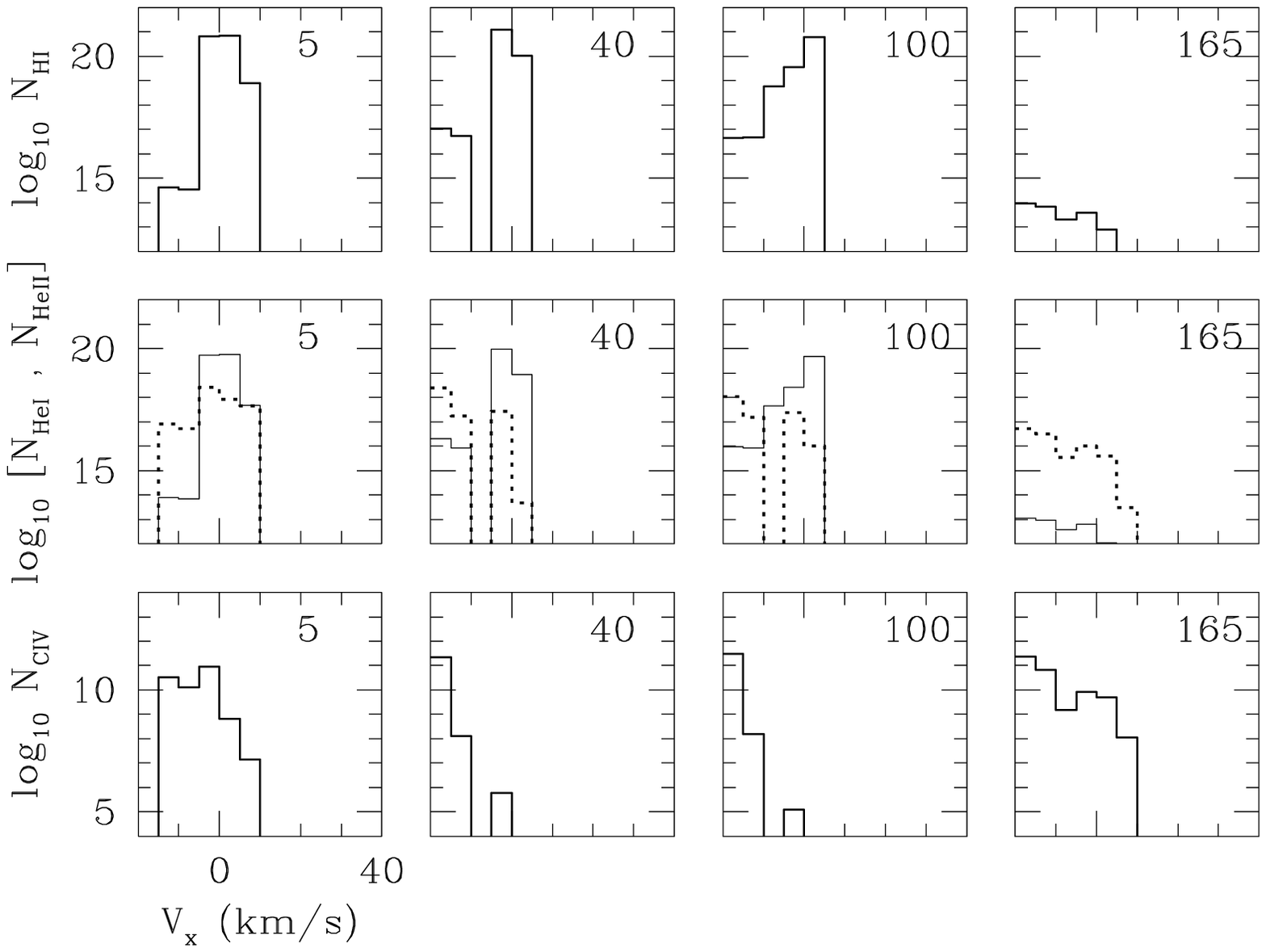}
 \includegraphics[width=3.4in]{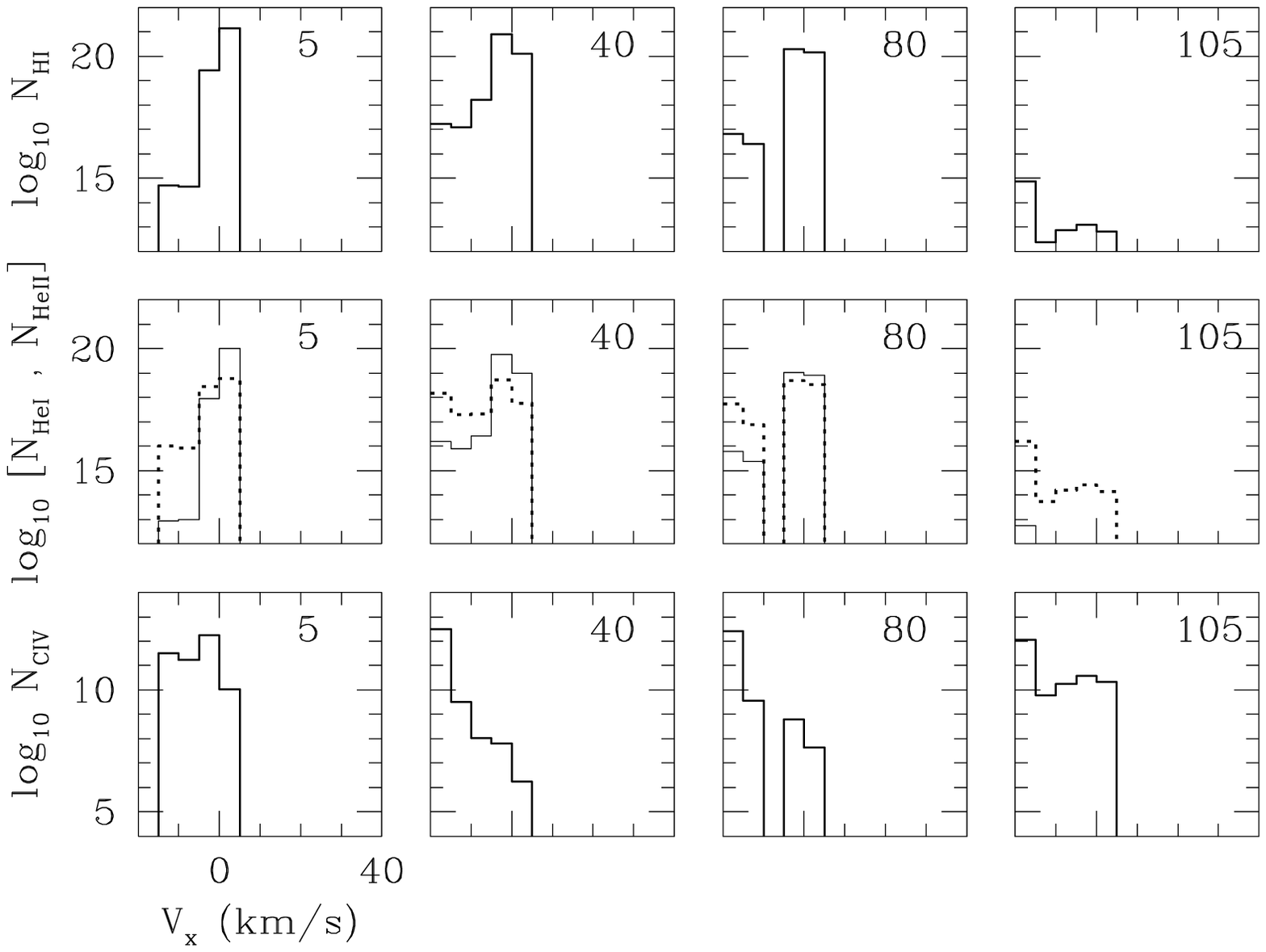}
\vspace{-2.5cm} 
 \caption{Observational diagnostics I. Absorption lines.
Minihalo column densities ($\rm cm^{-2}$) along symmetry axis for gas at
different velocities, for photoevaporating minihalo. 
(a) (left) BB 5e4 case; (b) (right) QSO case.
(Top) H~I; (Middle) He~I (solid) and
He~II (dotted); (Bottom) C~IV (if $\rm[C]/[C]_\odot=\eta\times10^{-3}$,
then plotted values are $N_{\rm CIV}/\eta$).
Each box labeled with time (in Myr) since  arrival of intergalactic
I-front.} 
\label{column_dens_QSO}
 \end{figure*}

\begin{figure}
 \includegraphics[width=3.4in]{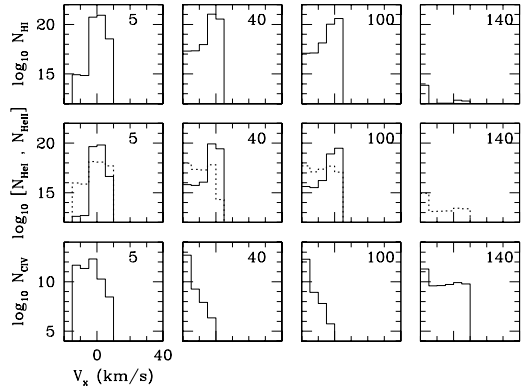}
\vspace{-2.5cm} 
 \caption{Same as Figure~\ref{column_dens_QSO}
but for BB 1e5 case.}
\label{column_dens_BB1e5}
 \end{figure}

Some observational signatures of the photoevaporation  process are shown in 
Figures~\ref{column_dens_QSO}--\ref{cno_60Myr_BB1e5}.
In Figures~\ref{column_dens_QSO} and \ref{column_dens_BB1e5}, we show
histograms of the column 
densities of H~I, He~I and II, and C~IV for minihalo gas at
different radial velocities 
as seen along the symmetry axis at different times, to
illustrate the kind of absorption lines which a photoevaporating
minihalo might produce.
At early times, the minihalo gas resembles a weak
damped Ly$\alpha$ (``DLA'') absorber with small velocity width 
($\simgreat10\rm\,km\,s^{-1}$) and $N_{\rm H\,I}\simgreat10^{20}\rm cm^{-2}$,
with a Ly$\alpha$-Forest (``LF'')-like red wing 
($\hbox{velocity width}\,\simgreat10\,\rm km\,s^{-1}$)
with $N_{\rm H\,I}\sim10^{15}\rm cm^{-2}$ on the side moving toward
the source. As photoevaporation proceeds, this red wing increases in
H I column density to $N_{\rm H\,I}\sim10^{17}\rm cm^{-2}$.
The He~I profile mimics that of H~I but with 
$N_{\rm He\,I}/N_{\rm H\, I}\sim {\rm [He]/[H]}$, and there is a weak C~IV 
feature with $N_{\rm C\,IV}\sim10^{10}\,(10^{12})\eta\,\rm cm^{-2}$ 
for the BB 5e4 (QSO, BB 1e5) cases, respectively, displaced in
this same asymmetric way to the red side of the velocity of peak H~I column
density, where $\eta\equiv[{\rm C}]/[{\rm C}]_\odot\times10^3$.
For He~II at early times, the QSO and BB 5e4 cases have 
$N_{\rm He\,II}\approx10^{18}\rm cm^{-2}$ at velocities close to those of
the H~I peak, and a red wing shifted by $\sim10\,\rm km/sec$ to the red,
with $N_{\rm He\,II}\sim10^{17}\rm cm^{-2}$ which
increases over time to $10^{18}\rm cm^{-2}$. He~II
qualitatively follows the 
H~I and He~I profiles in these cases. For the BB 1e5 case, however, 
$N_{\rm He\,II}\approx10^{18}\rm cm^{-2}$ in the HI peak, at early times, decreasing to
$N_{\rm He\,II}\approx10^{17}\rm cm^{-2}$ over time, with a red wing with
$N_{\rm He\,II}\approx10^{17}\rm cm^{-2}$ which {\it increases} to
$N_{\rm He\,II}\approx10^{18}\rm cm^{-2}$. In this case,
the He~II profile notably
diverges from the H~I profile.
After about an evaporation time, both H~I and He~I column densities 
decrease to LF-like values of
$10^{13}-10^{14}\,\rm cm^{-2}$, and shortly thereafter 
the minihalo is
completely evaporated.
Finally, the C~IV column density is greatest  in the red wing formed by the
evaporative wind, except at the earliest times.


\subsubsection{Ionization structure of metals}

\begin{figure*}
 \includegraphics[width=3.4in]{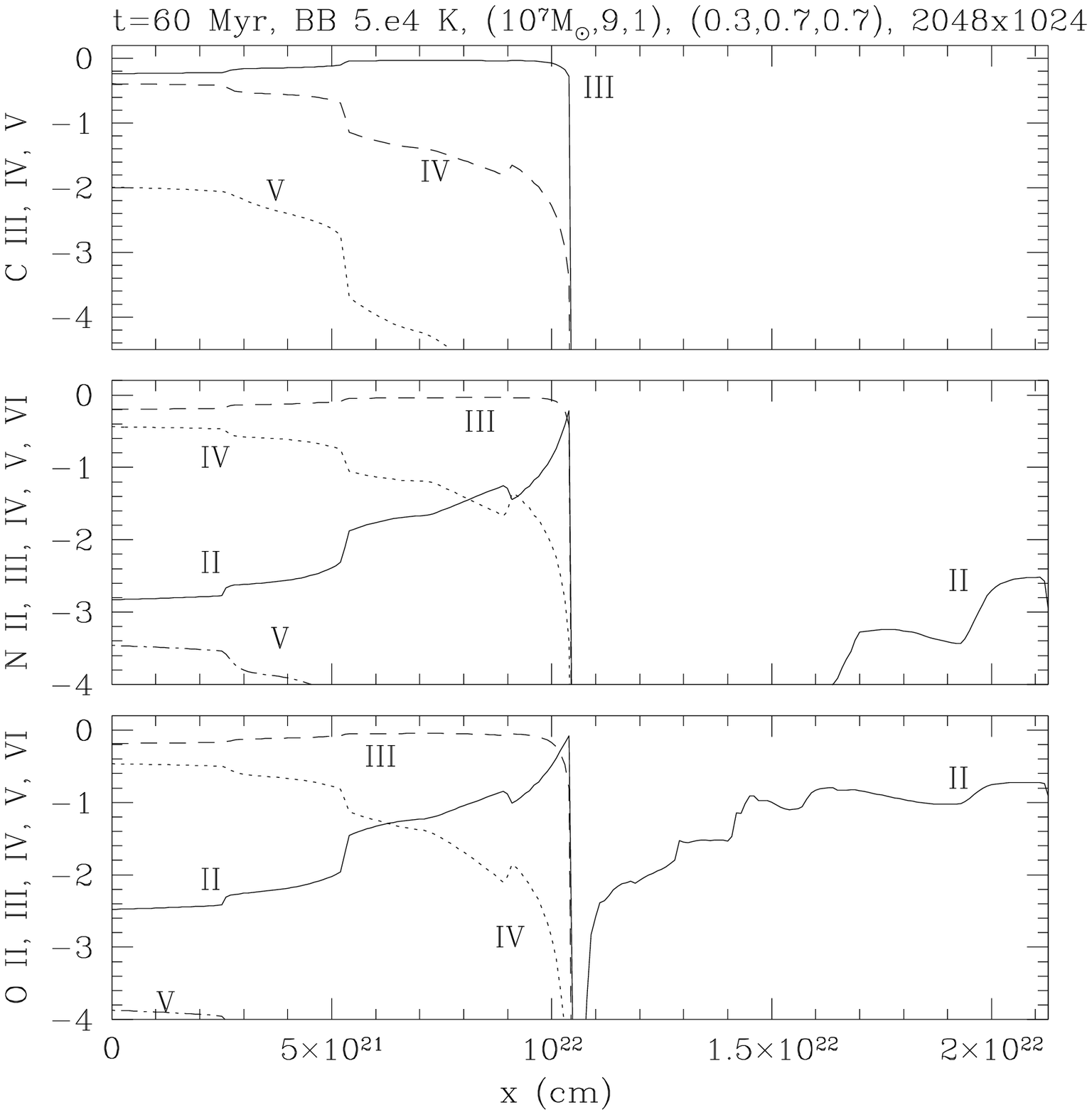}
 \includegraphics[width=3.4in]{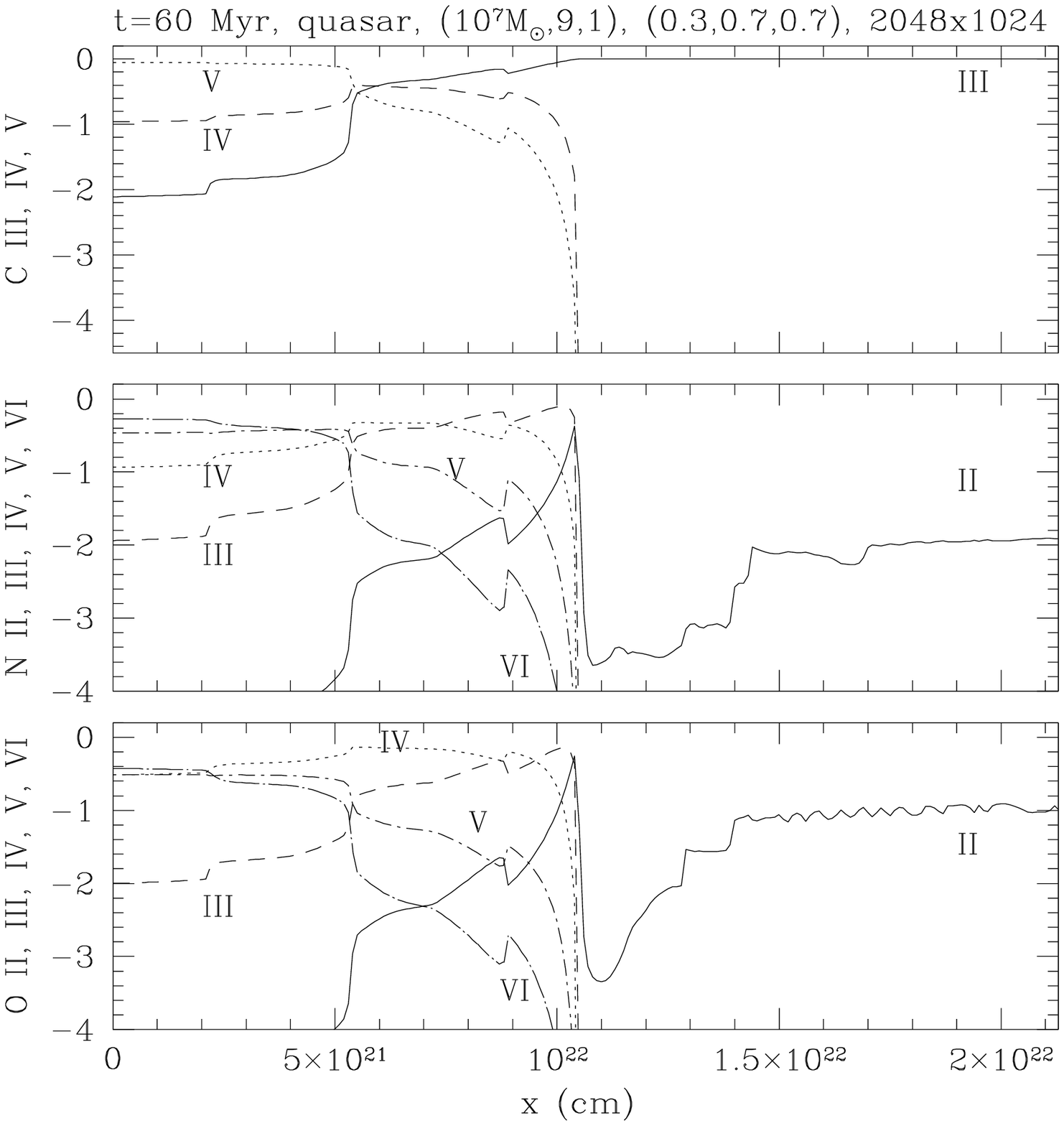}
 \caption{Observational diagnostics II: ionization structure of metals. 
C, N, and O ionic fractions along symmetry
axis at $t=60\rm\,Myr$, for photoevaporating minihalo
(a) (left) BB 5e4 case; (b) (right) QSO case.}
\label{cno_60Myr_QSO}
 \end{figure*}

\begin{figure*}
 \includegraphics[width=3.4in]{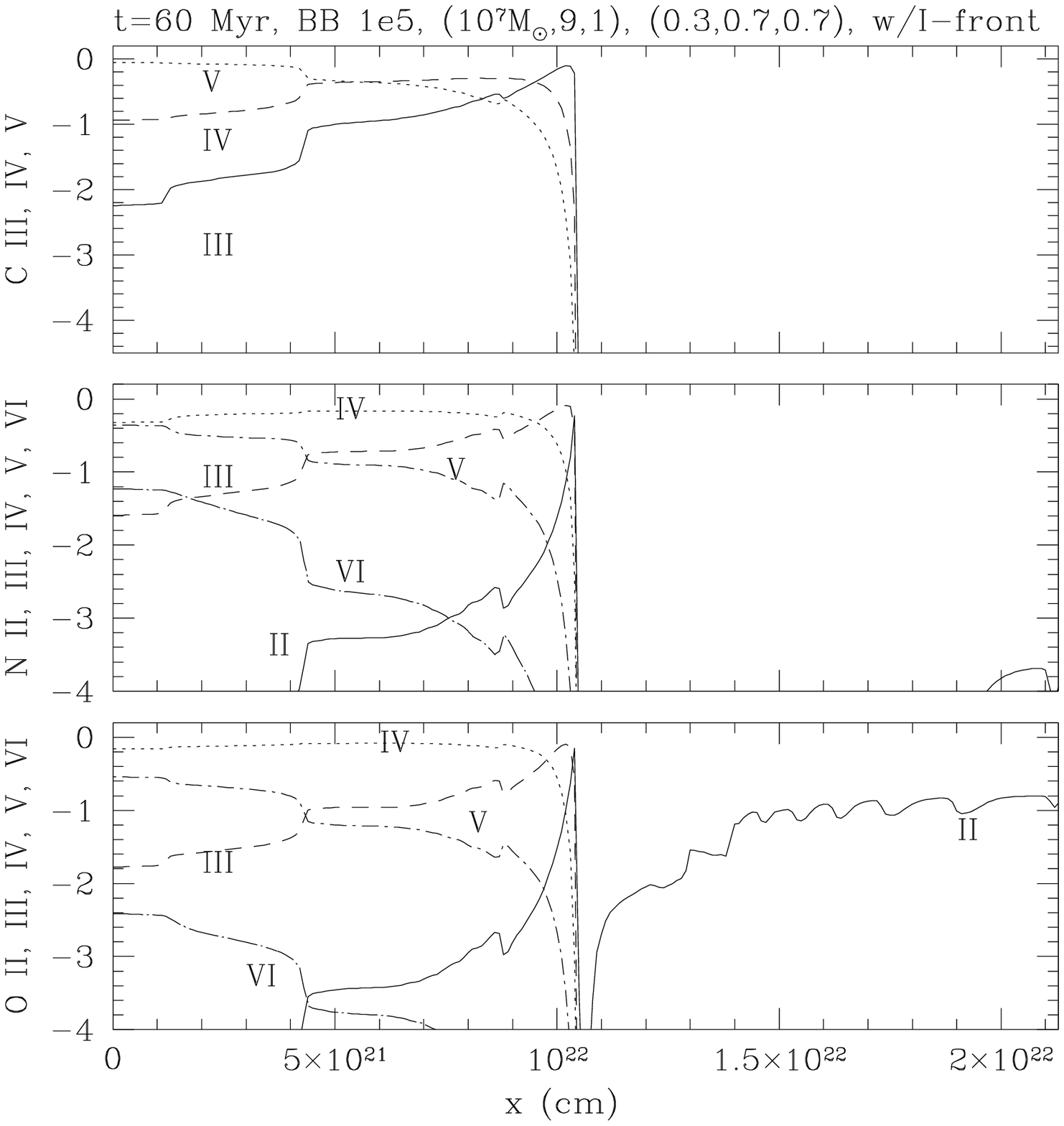}
 \includegraphics[width=3.4in]{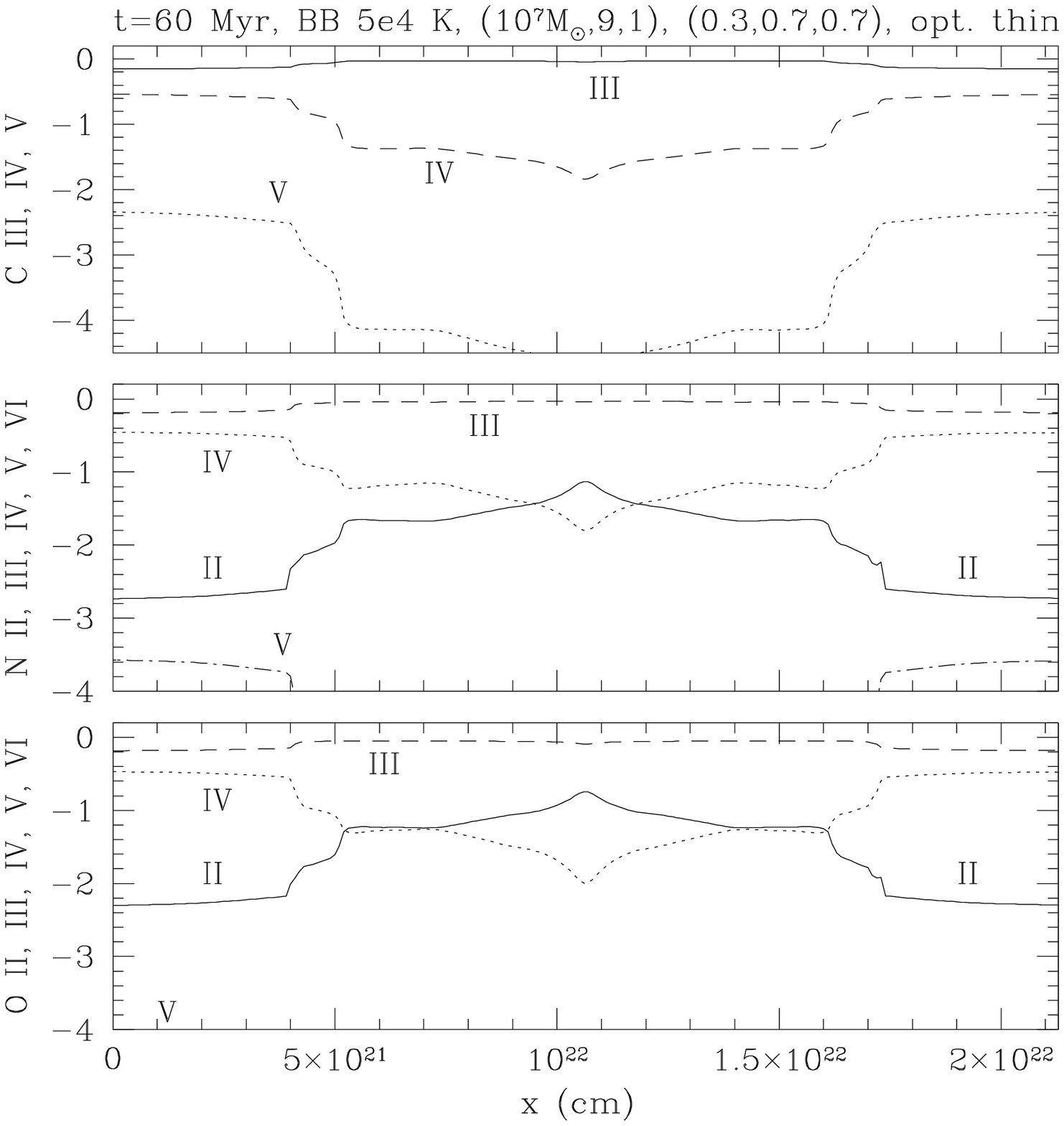}
 \caption{(a) (left) Same as Figure~\ref{cno_60Myr_QSO}, but for BB 1e5 case. 
(b) (right) Same as Figure~\ref{cno_60Myr_QSO}, but for optically-thin BB 5e4 case.}
\label{cno_60Myr_BB1e5}
 \end{figure*}

Figures~\ref{cno_60Myr_QSO} and \ref{cno_60Myr_BB1e5} show 
the spatial variation of the relative abundances of
C, N, and O ions along the symmetry axis at $t=60$~Myrs.
While the QSO case shows the presence at 60~Myrs
of low as well as high ionization stages of the metals, up to C~V, N~VI and
O~VI on the source side of the minihalo and CIII on the neutral side,
the softer spectrum of the BB 5e4 case yields less highly ionized gas 
both on the ionized side of the I-front (C~III, N~III, O~III) and the 
neutral side (C~II, N~I, O~I and II). The BB 1e5 case is 
intermediate between the other two cases, ionizing carbon up to C~V, but 
ionizing nitrogen and oxygen mostly to N~IV and O~IV, respectively, with 
smaller fractions of  N~V and O~V and much smaller, although not negligible 
fractions of N~VI and O~VI than in the QSO case. In the optically-thin
approximation
case in Figure~\ref{cno_60Myr_BB1e5} (right panels), similar 
high ionization stages
are reached for the ionized gas exposed to
the same photoionizing spectrum as in the results above which took
proper account of optical depth, but the low ionization stages of neutral
or partially ionized gas are entirely missing. In the optically-thin
results, all the gas is ionized
instantly, no neutral region of minihalo or
shadow exists and the flow expands symmetrically in all directions,
which would lead to quite different spectral 
signatures as compared to the realistic optically-thick cases shown above.

\section{Summary and conclusions}
\label{summary_sect}

We have presented the first numerical gas dynamics and radiative transfer
simulations of the encounter between an intergalactic I-front and a
minihalo during cosmic reionization at high redshift,
resulting in the photoevaporation of the minihalo gas. We have studied 
all stages of the photoevaporation process in detail, starting from the 
propagation of the fast, weak,
R-type I-front in the low-density IGM, its structure 
and extent in pressure, temperature and ionization of H, He and 
a possible trace of metals for
different possible source spectra, corresponding to quasars, Pop. III 
and Pop. II stars. We have demonstrated for the first 
time the phenomenon of I-front trapping when the global 
R-type I-front runs into a minihalo along its path
and slows down inside the minihalo
to an R-critical front, after which
the front inside the minihalo transform
from an R-type to a D-type front.
We have shown that the heated and ionized gas
behind the front then expands 
supersonically into the IGM, sweeping it up and shock-heating it. This process 
exposes further layers of the minihalo gas to the ionizing radiation
from the external source, heating 
and evaporating them until all the gas is finally
expelled from the minihalo. For our illustrative case of a $10^7M_\odot$ 
halo, overtaken at $z=9$ by an intergalactic I-front created by
a source of $10^{56}$ ionizing photons per 
second at distance of 1 Mpc (or equivalently, a source of 
$10^{52}\rm sec^{-1}$ at 10 kpc), this process is completed
within $\sim100-150$ Myr, depending on the source 
spectrum.

The results presented here are fully consistent with our earlier simulations
of minihalo photoevaporation, going back to Shapiro, Raga, \& Mellema
(1997, 1998) and continuing through those in 
Shapiro \& Raga (2000a,b; 2001), and
Shapiro (2001), although there are some quantitative differences which
result from improvements we have made since then. However, these
results differ from those which have
appeared in the meantime in the recent literature. As mentioned in \S1.3,
for example, 
\citet{BL99} considered the photoevaporation of minihalos without gas
dynamics and without treating the propagation of the I-front.
Their calculations are similar in spirit to (albeit more sophisticated
than) our analytical ISL approximation in \S\ref{strom_sect}.
The result of their
static approximation is that the larger minihalos 
($\geq {\rm few}\times10^5-10^6\,M_\odot$, depending on the redshift and the 
source spectrum) were predicted not to evaporate completely
but rather to leave a significant
fraction of the gas (20\%-40\% for QSO-type power-law 
spectrum for $10^7M_\odot$ halo
at $z=8-20$) still gravitationally bound to the minihalo. Our illustrative
simulations demonstrate
that when dynamical evolution is included this is not the correct outcome and 
all the gas is evaporated from the minihalos, leaving behind only a dark matter
halo almost completely devoid of gas. In a companion paper, 
we will extend these 
studies to show that this conclusion holds for all minihalos,
independent of their mass and redshift 
of collapse.

We have also shown here that a hydrodynamical treatment of the
photoevaporation of minihalos which fails to include radiative transfer
(i.e. the zero optical depth approximation) is also not adequate
for describing this problem. Such an approximation
yields a spherically-symmetric outflow and uniformly high ionization
structure which differ greatly from the more realistic results reported
here which take radiative transfer into account. As a consequence,
any observable features, such as absorption lines, would not be 
predicted correctly by this approximation. The difference between more
realistic simulations like ours, which include radiative transfer,
and those which neglect optical depth can have profound 
consequences for the theory of cosmic reionization, as well.

\citet{HAM01} pointed out the potential importance of minihalos as sinks of
ionizing photons during reionization, due to the increased
recombination rate inside these high-density regions. 
In \S2.5, we described how these authors estimated, 
in the optically-thin limit, the number of ionizing photons per
atom $\xi=n_\gamma/n_a$ needed 
to evaporate a minihalo. They concluded from this
that the clumping due to minihalos can easily
raise the number of photons required to complete reionization by an 
order of magnitude or more
compared to estimates that ignored the minihalos.  
This prompted them to identify a photon budget problem when
comparing the requirements for reionizing the universe with
the available photon supply.
In this paper, we have
used our self-consistent gas dynamics and radiative transfer
simulations to calculate $\xi$ under more realistic assumptions. We 
have shown that the effect of
ignoring radiative transfer and its feedback on the gas dynamics
is to significantly overestimate 
$\xi$. For the illustrative cases shown here, we find that
$\xi$ was overestimated 
by a factor as large as $30-50$, depending on the ionizing spectrum.
In terms of the efficiency factor in equations (\ref{xi_ham}) and 
(\ref{xi_numeric}), this means that we have found values of 
$f\sim1/30$ to $1/50$, depending on the spectrum of the ionizing sources.
Optically-thin estimates also do not account for differences due to
differences in
the external source spectra. We have shown that photoevaporation by Pop. III 
stars is significantly more efficient in terms of the net photon consumption
by this process than is photoevaporation by
QSO or Pop. II stellar sources, even when all sources emit the same number
of H ionizing photons per unit time. These effects
may help alleviate the photon budget 
problem posed by \citet{HAM01}, especially if Pop. III sources dominated
reionization. 

Although we find here that $\xi$ is smaller than previously estimated,
the effects of minihalos as screens and of their evaporation as a sink of
ionizing photons during reionization may still be considerable. For
one, we find $\xi\sim5$, which is still large compared to unity. In
addition, this value applies only to the particular illustrative case of a
$10^7M_\odot$ minihalo exposed to a source with flux $F_0=1$
at $z=9$. Only after we also calculate the values of $\xi$ per
minihalo for the full range of minihalo
masses and redshifts and of source fluxes which can occur
during reionization can we determine the full impact of the minihalos
on that process. We will report those results in a companion paper to follow.
However, we can already see that, if the reionization epoch was extended
in time, as described in \S1.2, with the final overlap of
isolated H II regions occurring at $z<9$, then the neutral regions
which were ionized during this late phase had a significant fraction
($\ga30\%$) of their baryons collapsed into minihalos. In that case,
minihalo photoevaporation by the global I-fronts which reionized the universe
is likely to have dominated this final ``overlap'' phase. 

Apart from the destructive effects of intergalactic I-fronts on
minihalo gas content described here, it has also been suggested
that the same I-fronts might, instead, have a positive feedback effect
on the minihalos rather than photoevaporating them entirely.
It has been proposed, for example,
that under certain conditions, external ionizing radiation
can cause the implosion of the minihalo gas, leading to 
the formation of globular clusters \citep{C01}. We have not observed such an
effect in our simulations.

Finally, we note that intervening minihalos are expected to be
ubiquitous along the line of sight to high redshift sources (Shapiro 2001).
With photoevaporation times $t_{\rm ev}\simgreat100\,\rm Myr$, this
process can continue down to redshifts significantly below $z=10$.
For stellar sources in the $\Lambda$CDM model, our simulations
show that photoevaporation of a $10^7M_\odot$ minihalo
which begins at $z_{\rm initial}=9$ can take 150 Myr to finish,
at $z_{\rm final}=7$, during which time such minihalos can survive without
merging into larger halos. Observations of the absorption spectra of high 
redshift sources like those which reionized the universe should reveal the 
presence of these photoevaporative 
flows and provide a useful diagnostic of the reionization process.

\section*{Acknowledgments}

We thank Garrelt Mellema for sharing his microphysics solver with us, 
for many useful discussions and for a careful reading of an
early version of this manuscript. We thank Hugo Martel for
discussion and collaboration on various aspects of the role of 
minihalos which directly impacts the work reported here.
We thank Marcelo Alvarez for his help in the visualization of
our simulation results.
This research was supported by NSF grant INT-0003682 from 
the International Research Fellowship Program and the Office of 
Multidisciplinary Activities of the Directorate for Mathematical and
Physical Sciences, the Research and Training Network
``The Physics of the Intergalactic Medium'' set up by the European
       Community under the contract HPRN-CT2000-00126 RG29185
NASA ATP Grant NAG5-10825, and Texas
Advanced Research Program Grant 3658-0624-1999.

\appendix

\section{A Time-Dependent Model for Virialized
Halos with Cosmological Infall}

The initial conditions for our simulations are given by the TIS profile
for both the gas and the dark matter density for the virialized halo, 
with zero bulk velocity, as described by
\citet{IS01}, and a matching spherical,
self-similar infall profile as described by
\citet{B85} for the density and infall 
velocity outside the halo, appropriately generalized to 
the case of a low-density 
background universe
models (Iliev \& Shapiro 2001, Appendix A). The match of
the post-shock gas in this infall solution to the TIS is discussed in
detail in \S~7.2 of \cite{SIR99}. This match is made possible by the 
fact that the TIS radius $r_t$ is almost coincident with the shock 
radius of the self-similar infall solution $r_S$, with the latter 
larger than the former by only 1.8\%.\footnote{In this Appendix, spherical
symmetry applies throughout, so we use the variable ``$r$'' to mean the
spherical radius here.}
In order to match the two 
solutions seamlessly, we continue both the density and the velocity
profiles of the infall solution down to $r_t$. 
For the velocity profile outside the TIS outer radius we use the pre-shock
infall profile of the self-similar solution. 
Following the notation of \citet{B85}, we define the dimensionless radius 
$\lambda\equiv{r}/{r_{\rm ta}(t)}$. Here $r_{\rm ta}$ is the turnaround radius
of the shell which is just reaching maximum expansion at time $t$, whose 
time-dependence is given by $r_{\rm ta}(t)=r_{\rm ta}(t_i)(t/t_i)^{8/9}$, 
where $t_i$ is our initial time. Due to self-similarity of the solution every 
feature in radius occurs at some fixed value of $\lambda$. For example, the 
location of the shock is given by $\lambda_S=0.338976$, and the time-varying 
shock radius 
must follow $r_S(t)=r_{\rm ta}(t)\lambda_S$. Similarly, the radial parameters 
of the TIS profile, like core radius $r_0$ and $r_t$, follow the same time 
dependence with $\lambda=(\lambda_0,\lambda_t)=(0.0113175,0.3327339)$, 
respectively. 

The physical velocity $v$ at a given radius 
$r\geq r_t$ t a time $t$ is given by the following parametric solution:
\be
\lambda(\theta)=\sin^2\frac{\theta}2
	\left(\frac{\theta-\sin\theta}{\pi}\right)^{-8/9}
\ee
and
\be
v=\frac rt\frac{\sin\theta(\theta-\sin\theta)}{(1-\cos\theta)^2}
	=\frac{r_0(t)}{\lambda_0t}V(\lambda),
\label{vel_bert}
\ee
where $0\leq\theta\leq2\pi$. For the TIS density profile we use the 
fitting formula to the exact numerical results
given in Appendix B of \cite{IS01}.
The resulting initial 
condition for our simulations is shown in Figure~\ref{init_lam0.7}.

In order to evolve the initial dark matter profile with time 
self-consistently, we assume that the mass of the 
TIS grows over time as a result of
the infall, and our initial solution for the TIS gives way 
to a sequence of equilibrium TIS models, where the TIS is specified 
at each time by the mass $M(t)$ and redshift $z_{\rm coll}$ at that time.
Below we give the time-dependences of all relevant quantities.

The time-dependence of $M(t)$ corresponds to that of the sphere bounded 
by the accretion shock in the Bertschinger solution. 
The central density $\rho_0$ of our TIS solution is just proportional to 
the mean background density at the same epoch. At the early times of
interest to us here, this dependence is simply 
$\rho_0(t)=\rho_0(t_i)(t/t_i)^{-2}$. With this scaling we can write the 
growth of the total mass with time as

\ba
M_{\rm TIS}(t)\!\!\!\!&=&\!\!\!\!
M_{\rm TIS}(t_i)\left[\frac{r_{\rm ta}^3(t)}{r_{\rm ta}^3(t_i)}
	\right]\left[\frac{\rho_b(t)}{\rho_b(t_i)}\right]
\nonumber\\&&
=M_{\rm TIS}(t_i)\left(\frac{t}{t_i}\right)^{2/3}.
\ea
In fact, for any radius which grows in time self-similarly 
(i.e. keeping its $\lambda$-value constant) the mass interior to this 
radius must grow according to
\be
m(\lambda,t)=m(\lambda,t_i)\left(\frac tt_i\right)^{2/3}.
\ee

The gravitational acceleration, $g$, at radius $r$ of this assumed matter 
distribution is given simply in terms of the mass $m(r)$ interior to this
radius: $g=Gm(r)/r^2$. For fixed $\lambda$, $m(\lambda,t)\propto t^{2/3}$, so 
we need only determine $m(r,t)=m[\lambda(r,t),t]$ at fixed $r$ for 
different times. This is given by
\be
m(\lambda,t)=M(\lambda)\frac{4\pi\rho_b(t_i)r_S^3}{3\lambda_S^3}
\left(\frac{t}{t_i}\right)^{2/3},
\ee
where
\be
\lambda(r,t)=\frac{\lambda_0}{r_0(t_i)}\left(\frac{t}{t_i}\right)^{-8/9}r.
\ee

(A) For $\lambda\geq\lambda_t=0.332734$, the infall mass profile is 
\be
M(\lambda)=-\frac92\lambda^2D(\lambda)\left[V(\lambda)-\frac89\lambda\right],
\ee
where $V(\lambda)$ was defined in equation~(\ref{vel_bert}) above 
and 
\be
D(\lambda)=\frac{9(\theta-\sin\theta)^2}{16\sin^6(\theta/2)
	\left\{1+3[1-3V(\lambda)/(2\lambda)]\right\}}
\ee
\citep{B85}.

(B) For $\lambda\leq\lambda_t$ the TIS mass profile is derived by
enforcing hydrostatic equilibrium inside the TIS, using the density
profile fitting formula and the isothermal virial temperature to
solve for the gravitational acceleration needed to balance the pressure
force: 
\ba
M(\lambda)\!\!\!\!&=&\!\!\!\!
6\lambda_0\left(\frac{\rho_0}{\rho_b}\right)\lambda^2
\nonumber\\&&\times
\frac{\xi\left[A/(a^2+\xi^2)^2-B/(b^2+\xi^2)^2\right]}
{\left[A/(a^2+\xi^2)-B/(b^2+\xi^2)\right]},
\ea
where $(A,a^2,B,b^2)=(21.38,9.08,19.81,14.62)$, $\xi=r/r_0$, $r_0$ is the
TIS core radius, and $\rho_b$ is the mean background density.

\end{document}